\documentclass[structabstract]{aa}
\usepackage{color}
\usepackage{graphicx}

\usepackage{natbib}
\usepackage{multirow}
\bibpunct{(}{)}{;}{a}{}{,}

\usepackage{txfonts}
\usepackage{hyperref}

\begin{document}

   \title{Chemical study of two starless cores in the B213/L1495 filament}

   \author{
   L. Moral-Almansa\inst{1}
   \and
     A.~Fuente\inst{1}
    \and
    M. Rodr{\'{\i}}guez-Baras\inst{2}
    \and
   T. Alonso-Albi\inst{3}
   \and    
    G. Esplugues\inst{3}
    \and
    D. Navarro-Almaida\inst{1}
    \and
    P. Rivi\'{e}re-Marichalar\inst{3}
    \and
    B. Tercero\inst{3,} \inst{4}
    \and
    A. Asensio Ramos\inst{5,} \inst{6}
    \and
    C. Westendorp Plaza\inst{5,} \inst{6}    
  }

\institute{ Centro de Astrobiolog\'{\i}a (CSIC-INTA), Ctra. de Ajalvir, km 4, Torrej\'on de Ardoz, 28850, Madrid, Spain
\and
European Space Agency (ESA), European Space Astronomy Centre (ESAC), Camino Bajo del Castillo s/n, 28692 Villanueva de la Cañada, Madrid, Spain
\and
Observatorio Astron\'omico Nacional (OAN), Alfonso XII, 3,  28014, Madrid, Spain
\and
Observatorio de Yebes (IGN), Cerro de la Palera s/n, 19141 Yebes, Guadalajara, Spain
\and
Instituto de Astrof\'{\i}sica de Canarias (IAC), Avda V\'{\i}a L\'{a}ctea s/n, E-38200 La Laguna, Tenerife, Spain
\and
Departamento de Astrof\'{\i}sica, Universidad de La Laguna, E-38205 La Laguna, Tenerife, Spain
}

 \abstract
{The chemical evolution of pre-stellar cores during their transition to a protostellar stage is not yet fully understood. Detailed chemical characterizations of these sources are needed to better define their chemistry during star formation.}
{Our goal is to characterize the chemistry of the starless cores C2 and C16 in the B213/L1495 filament of the Taurus Molecular Cloud, and to understand how it relates to the environmental conditions and the evolutionary state of the cores. }
 {We made use of two complete spectral surveys at 7 mm of these sources, carried out using the Yebes 40-m telescope. Derived molecular abundances were compared with those of other sources in different evolutionary stages and with values computed by chemical models.}
 {Including isotopologs, 22 molecules were detected in B213-C2, and 25 in B213-C16. The derived rotational temperatures have values of between $\sim$ 5 K and $\sim$ 9 K. A comparison of the two sources shows lower abundances in C2, except for l-C$_3$H and HOCO$^+$, which have similar values in both cores. Model results indicate that both cores are best fit assuming early-time chemistry, and point to C2 being in a more advanced evolutionary stage, as it presents a higher molecular hydrogen density and sulfur depletion, and a lower cosmic-ray ionization rate. Our chemical modeling successfully accounts for the abundances of most molecules, including complex organic molecules and long cyanopolynes (HC$_5$N, HC$_7$N), but fails to reproduce those of the carbon chains CCS and C$_3$O.}
  {Chemical differences between C2 and C16 could stem from the evolutionary stage of the cores, with C2 being closer to the pre-stellar phase. Both cores are better fit assuming early-time chemistry of $t \sim 0.1$ Myr. The more intense UV radiation in the northern region of B213 could account for the high abundances of l-C$_3$H and HOCO$^+$ in C2.} 
  
   
 \maketitle  
 \nolinenumbers
 

\section{Introduction}
\label{Sec: Introduction}
The \textit{Herschel} Space Telescope has transformed our understanding of star-forming regions. Images of giant molecular clouds and dark cloud complexes have revealed spectacular networks of filamentary structures where stars are born \citep{2010A&A...518L.102A}. Now we think that interstellar filaments are present throughout the Milky Way and are the preferred sites for star formation. These filaments funnel interstellar gas and dust into increasingly dense concentrations. These concentrations then contract and fragment, leading to gravitationally bound pre-stellar cores that will eventually form stars \citep{2010A&A...518L.102A, Molinari2010, Juvela2012}.

Gas chemistry plays a pivotal role in the process of star formation by regulating critical processes, such as gas cooling and the ionization fraction. Molecular filaments have been observed to fragment into dense cores due to the cooling of the gas by molecules \citep{GoldsmithLanger1978}, thereby reducing the thermal support relative to self-gravity. The ionization fraction exerts a critical influence on the coupling of magnetic fields with the gas, thereby driving the dissipation of turbulence and angular momentum transfer. Consequently, it plays a crucial role in the cloud collapse (isolated vs. clustered star formation) and the dynamics of accretion discs (see \citealp{Padovani2013, Zhao2021}). The gas ionization fraction and molecular abundances are contingent upon the elemental depletion factors, as was demonstrated in \citet{Caselli1998}. In particular, carbon (C) is the primary electron donor in the cloud surface (A$_v$ < 4 mag), and sulfur (S) is the primary donor in the range of A$_V$ = 3.7 to 7 magnitudes, which encompasses a significant fraction of the molecular cloud mass, given its lower ionization potential and its status as a non-depleted element \citep{Goicoechea2006}. Therefore, sulfur depletion constitutes a valuable piece of information for our understanding of the grain composition and evolution.

Dense cores set the initial conditions for the process of star formation (see, e.g., \citealp{Shu1987, Bergin2007}). Gas and solids undergo a continuous evolution from their early stages in molecular clouds to their incorporation into a growing planet through a proto-planetary disk. Although this evolution takes place over the course of several million years, it is now believed that the chemical composition of the disk is largely influenced by the physical and chemical conditions present in the progenitor molecular cloud \citep{Visser2009, Oberg2011, Navarro2024}. In fact, some of the molecules detected in planet-forming disks and comets are thought to be inherited from the pre-stellar phase \citep{Mumma2011, Oberg2023}. The physical and chemical properties of cold, dense cores and their evolution into a collapsing core must be characterized
in order to be able to predict the properties of the stars and planetary systems that will be formed in their centers.
Despite significant advances in recent decades, with several studies on the chemistry of both starless \citep[e.g.][]{Bacmann2012, Vastel2018, Lattanzi2020, Spezzano} and protostellar cores  \citep[e.g.][]{Hirota2010,Fuente2016, Araki2017, Agundez2019, Esplugues2022}, the chemical evolution of dense cores during their transition from a starless to a protostellar stage is not yet fully understood. Particularly, it is still uncertain whether changes in chemical composition are tied to their evolutionary stage or environmental factors. Detailed chemical characterizations of pre-stellar and protostellar objects are needed in order to expand the molecular census in these objects and to better define the chemistry of dense cores during star formation.

In this work, we present a $\lambda$ =  7 mm line survey toward two starless cores, B213-C2 and B213-C16, located in the B213/L1495 filament of the Taurus Molecular Cloud, which were observed with the Yebes 40 m telescope in the $31.3-50.6$ GHz frequency range. This is a rather unexplored window that offers the possibility to study large molecules such as carbon chains, which would be more difficult to detect at higher frequencies due to the high energy associated with the transitions lying at $3$, $2$, or $1.3$ mm. To better characterize the physical conditions of the gas, these surveys have been combined in Sect. $\ref{Sec: Chemical modelling}$ with molecular abundances obtained within the IRAM  30 m Large Program “Gas phase Elemental abundances in Molecular
CloudS” (GEMS)  \citep{Fuente2019}. The observed abundances are then compared with the values computed by a chemical model. Moreover, we take advantage of this wealth of molecular data to 
further confirm our estimation of the sulfur depletion in this filament \citep{Fuente2023}. The comparison of the two sources provides an insight into the relationship between the chemical composition and evolutionary stage of low-mass dense cores, as both starless cores lie within the same filament, reducing the impact of environmental differences.

\section{The sources}
\label{Sec: The sources}
The Taurus molecular cloud, at a distance of 145 pc \citep{Yan}, is considered to be a prototype of a low-mass star-forming region. It has been the target of previous chemical, cloud evolution, and star formation studies \citep[e.g.][]{ Ungerechts, Mizuno, 2008ApJ...680..428G, Spezzano, Fuente2023}, and has been widely mapped in CO \citep{Cernicharo1987, Onishi1996, Narayanan2008} and visual extinction \citep{Cambresy1999, Padoan2002,Schmalzl2010}. The B213/L1495 filament constitutes one of the longest and most prominent filaments within the cloud, extending for over 10 pc \citep{TafallaHacar2015}. It has been largely studied in the (sub)millimeter range \citep[e.g.][]{Palmeirim, Hacar, Marsh2014, TafallaHacar2015, Bracco2017, Shimajiri}. The density of stars decreases from north to south, which is suggestive of a different dynamical and chemical age along the filament \citep{RodriguezBaras2021, Esplugues2022}. The magnetic field lines at the boundary of the B213 filament are oriented perpendicular to its long axis \citep{2008ApJ...680..428G}. The presence of low-density striations parallel to the magnetic field suggests a process of mass accretion into this part of the B213/L1495 complex \citep{2008ApJ...680..428G, Palmeirim, Shimajiri}.

 Several dense cores can be found within this filament, some of them  associated with young stellar objects (YSOs) of different ages \citep{Luhman2009, Rebull2010}, while others show no sign of star formation. This dense core population has been studied using species such as NH$_{3}$, H$^{13}$CO$^{+}$, N$_{2}$H$^{+}$, and SO \citep{Benson1989, Onishi2002, Tatematsu2004, Hacar, Punanova2018}. To better understand the chemical composition of dense cores and how it relates to active star formation, we have centered our study around two starless cores, one of them located close to the most active star-forming region of the filament and the other one in a rather quiescent region. These cores are $\#$2 and  $\#$16 in the nomenclature of \citet{Onishi2002} and \citet{Hacar}. Hereafter, we refer to these cores as B213-C2 and B213-C16. The coordinates of these clumps are  shown in Table \ref{Table: C2 and C16 coordinates}.

 \begin{table}[h!]
\caption{List of sources.}
\label{Table: C2 and C16 coordinates}
\centering
\begin{tabular}{l l l   }
\hline
\hline
\\
Core & \multicolumn{2}{c}{Coordinates} \\
        &  RA(J2000) & Dec(J2000)  \\
\hline

B213-C2  & 04$^{\text{h}}$:17$^{\text{m}}$:50.60$^{\text{s}}$ & +27$^{\circ}$:56$'$:01.0$''$ \\
B213-C16  & 04$^{\text{h}}$:21$^{\text{m}}$:21.00$^{\text{s}}$  & +27$^{\circ}$:00$'$:09.0$''$ \\
\hline
\end{tabular}
\end{table}

\section{Observations}
\label{Sec: Observations}
\begin{figure*}
\centering
 \resizebox{\hsize}{!}{\includegraphics[width=\textwidth, trim={0cm 4.2cm 0cm 0cm}, clip]
 {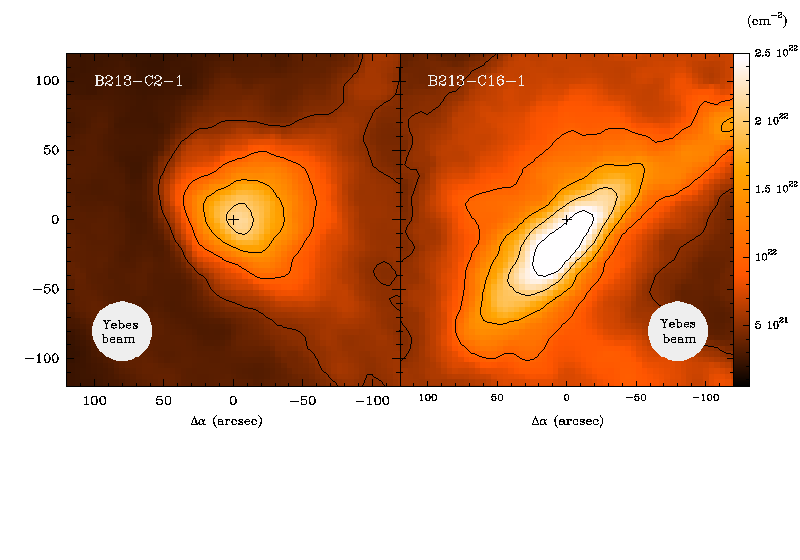}}

\caption{B213-C2 and B213-C16 molecular hydrogen column density maps derived by \citet{Palmeirim}, reconstructed at an angular resolution of 18.2$''$. Contour levels are
 $(5, 10, 15 \ \text{and} \ 20) \times 10^{21} \ \text{cm}^{-2}$ in the left panel and 
 $(5, 10, 15, 20 \ \text{and} \ 25) \times 10^{21} \ \text{cm}^{-2}$ in the right panel. The beam of the Yebes telescope in the Q band  (HPBW$\approx$42.5$"$) is plotted in each panel.
}
\label{Fig: B213  molecular hydrogen maps}

\end{figure*}

The observations were carried out using the Yebes 40-m radio telescope at Yebes (Guadalajara, Spain) (project 20A006). The 7 mm NANOCOSMOS high-electron-mobility transistor (HEMT) receiver and the fast Fourier-transform spectrometers (FFTSs) with $8 \times 2.5$ GHz bands per linear polarization were used, covering the frequency range of 31.3-50.6 GHz and an instantaneous bandwith of 18 GHz with a spectral resolution of 38 kHz \citep{Tercero2021}. The observing mode was frequency switching to optimize telescope efficiency, eliminating the need for an off position to subtract the signal from the sky. Two spectral setups were performed at a slightly different central frequency to identify spurious signals and to cover the full Q band bandwidth. The size of the Yebes telescope's beam in Q band is shown in each panel of Fig. \ref{Fig: B213  molecular hydrogen maps}.

Data reduction was conducted using the CLASS-GILDAS software\footnote{\url{https://www.iram.fr/IRAMFR/GILDAS}}. The Cologne Database for Molecular Spectroscopy\footnote{\url{https://cdms.astro.uni-koeln.de/classic/entries/}}  \citep{2005JMoSt.742..215M} and the JPL Molecular Spectroscopy Catalog \footnote{\url{https://spec.jpl.nasa.gov/ftp/pub/catalog/catform.html/}} \citep{1998JQSRT..60..883P} were used to assign the detected lines to known molecular rotational transitions. Only emission features whose peak intensity is above three times the root mean square (rms) noise level and whose line width is larger than one channel width, were considered as detections. 

\begin{figure*}[h]
\centering

\includegraphics[height=0.2\textheight, width=\textwidth]{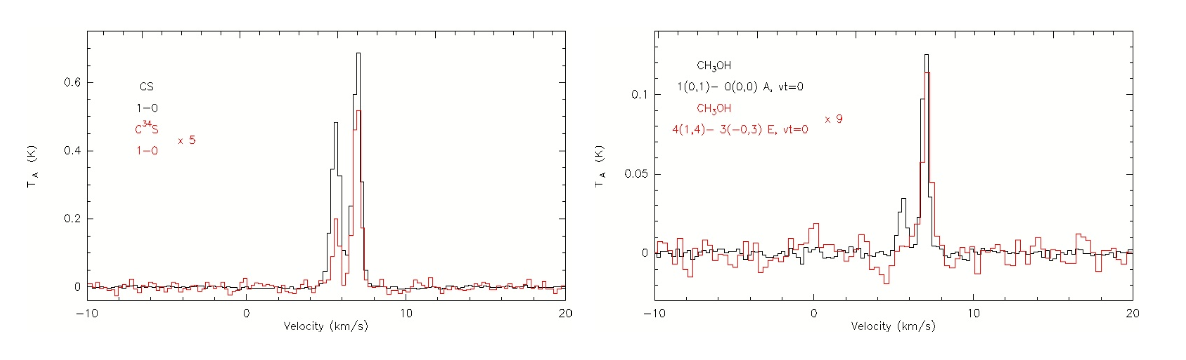}

\caption{\textit{Left}: Velocity components observed in CS overlapped with the two velocity components observed in C$^{34}$S for the detected transitions in B213-C2. \textit{Right}: Superposition of the emission lines detected in B213-C2 for C$\mathrm{H}_{3}$OH with  both one and two visible velocity components.}
\label{Fig: velocity components}

\end{figure*}

\section{Data analysis  and results}
\label{Sec: Data analysis  and results}

We have detected a total of 122 emission lines: 56 of them corresponding to C2, and 66 to C16. The observed lines come from 22 different species for C2, and 25 for C16, including isotopologs. The observed molecules include O-bearing molecules, hydrocarbons, and N-bearing and S-bearing molecules. The detected species are listed in Table \ref{Table: Detected species}. 

\begin{table}[h!]
\caption{Detected species in each source, number of observed lines, and minimum and maximum upper energies of the detected transitions. }
\label{Table: Detected species}
\small
\centering

\begin{tabular}{l l l l l l}

\hline
\hline
\\
Species& Source & Lines & $\text{E}_{up, min}$ & $\text{E}_{up, max}$ \\
           &             &       &        (K)&     (K) \\
\hline
\multirow{2}{*}{CH$_{3}$CHO}& C2&6 &2.77  & 5.12 \\
& C16&5  &2.77 & 5.20   \\
\hline
\multirow{2}{*}{CH$_{3}$OH} &C2 &3  &2.32 &  15.45  \\
&C16 & 3 & 2.32 &28.79    \\
\hline
\multirow{2}{*}{CS} &C2  &1  &2.35 &2.35    \\
& C16 &1  &2.35 & 2.35    \\
\hline
\multirow{2}{*}{H$_{2}$CS}  & C2&1  & 1.65 & 1.65   \\
 & C16 & 1 & 1.65 & 1.65   \\
\hline
\multirow{2}{*}{HCNS} & C2& 0 &-- & --   \\
 &C16&1  & 5.90 & 5.90   \\
\hline
\multirow{2}{*}{HSCN}& C2&0  &-- &--    \\
& C16& 2 & 3.30 &   5.50 \\
\hline
\multirow{2}{*}{HCS$^{+}$}&C2&1  &2.05 & 2.05   \\
& C16&1  &2.05 & 2.05   \\
\hline
\multirow{2}{*}{OCS}&C2& 2 & 3.50 & 5.84   \\
&C16&2   & 3.50 & 5.84  \\
\hline
\multirow{2}{*}{SO$_{2}$}& C2&1  & 29.20 & 29.20   \\
& C16& 0 &-- &--    \\
\hline
\multirow{2}{*}{CCS} &C2 & 4 &3.23 & 12.94   \\
&C16 &4  &3.23 & 12.94   \\
\hline
\multirow{2}{*}{C$_{3}$S}& C2& 3 &5.83 &   9.99 \\
& C16 & 3 &5.83 &   9.99 \\
\hline
\multirow{2}{*}{CH$_{3}$CCH}& C2&1  & 2.46 &2.46    \\
& C16&1  & 2.46 &2.46   \\
\hline
\multirow{2}{*}{CH$_{3}$CN}&C2&1  & 2.65& 2.65    \\
&C16& 1 &2.65 & 2.65   \\
\hline
\multirow{2}{*}{HC$_{5}$N} &C2&7  & 9.97 &  21.85  \\
&C16 &7  & 9.97 &  21.85 \\
\hline
\multirow{2}{*}{HC$_{7}$N}&C2 & 7 & 21.98 & 40.11   \\
&C16 &11  &21.98 &40.11    \\
\hline
\multirow{2}{*}{c-C$_{3}$H$_{2}$}&C2&2  & 8.67 & 18.17  \\
&C16&1  & 8.67 & 8.67   \\
\hline
\multirow{2}{*}{l-C$_{3}$H}&C2& 5 & 1.57 & 1.57   \\
&C16&5  &1.57 & 1.57   \\
\hline
\multirow{2}{*}{HOCO$^{+}$}&C2& 2 & 3.08& 40.41   \\
&C16&1  &3.08 &3.08    \\
\hline
\multirow{2}{*}{C$_{3}$O}&C2& 2 &4.62 & 6.93    \\
&C16&2&4.62 & 6.93    \\
\hline
\multirow{2}{*}{CH$_{2}$CHCN}&C2& 1 &8.83 & 8.83    \\
& C16&5  &4.55 &24.03    \\
\hline
\multirow{2}{*}{l-C$_{3}$H$_{2}$}&C2&3  &2.99 & 16.38    \\
&C16&3  &2.99 & 16.38    \\
\hline
\multirow{2}{*}{$^{13}$CS}&C2&1  &2.22 & 2.22    \\
& C16&1  &2.22 & 2.22   \\
\hline
\multirow{2}{*}{C$^{34}$S}&C2& 1 & 2.31&   2.31 \\
&C16& 1 & 2.31&   2.31 \\
\hline
\multirow{2}{*}{C$^{33}$S}&C2& 0 &-- &--    \\
&C16& 1 &2.33 & 2.33    \\
\hline
\multirow{2}{*}{CC$^{34}$S}& C2& 1 &3.17 &  3.17 \\
&C16&2  &3.17 & 5.30  \\
\hline
\multirow{2}{*}{CCC$^{34}$S}&C2&0  &-- &--    \\
&C16&1  & 5.68 &   5.68  \\
\hline

\end{tabular}
\end{table}

\subsection{Line profiles}
\label{Subsec: Line profiles}

We fit a Gaussian function to each observed line profile making use of the CLASS program. From these Gaussian profiles, we derived the peak velocity ($v_{LSR}$), line width, line intensity (antenna temperature, $T^{\ast}_{A}$, corrected for atmospheric absorption and for antenna ohmic and spillover losses), and equivalent width. To convert the antenna temperature to the main-beam brightness temperature ($T_{MB}$), the following expression was used:
\begin{equation}
T_{MB}=(\eta_{eff}/\eta_{MB})\times T^{\ast}_{A},
\label{eq: TMB}
\end{equation}
where $\eta_{eff}$ is the forward efficiency and $\eta_{MB}$ is the main-beam efficiency\footnote{\url{https://rt40m.oan.es/rt40m_en.php}}. The results are shown in Tables \ref{Line parameters of the detected molecules in B213-C2}–\ref{Line parameters of the detected molecules in B213-C16}.

There are some species in B213-C2 whose transition lines present a double peak emission. These two velocity components, visible in a few rotational transitions, are displaced by approximately 1 $\text{km} \ \text{s}^{-1}$, as is illustrated in both panels of Fig. \ref{Fig: velocity components}. These two-peak profiles had been noticed in previous observations of this filament \citep[e.g.][]{1986A&A...164..349D,2008ApJ...680..428G,2012ApJ...756...12L,Hacar}. The right panel of Fig. \ref{Fig: velocity components} shows the comparison between the observed line profiles for two different transitions of C$\mathrm{H}_{3}$OH. This comparison reveals that one of the velocity components present in the two-peaked transitions aligns in velocity with the component observed across all single-peaked transitions. This suggests that, unlike other dense cores such as L483 \citep{Agundez2019} or B335 \citep{2023A&A...678A.199E}, where the presence of multiple velocity peaks is due to self-absorption, the two-peak profiles observed in B213-C2 are most probably an indication of kinematically distinct components.
We fit a Gaussian profile to each velocity peak in two-peaked transitions. However, because these two velocity components are only visible in a few low-energy transitions corresponding to abundant molecules, for subsequent calculations we exclusively considered the component visible across all observed transitions, at approximately 7 $\text{km} \ \text{s}^{-1}$. The Gaussian fits obtained for both sources can be found in \href{https://zenodo.org/records/18076841}{Zenodo}.

\subsection{Column densities: Rotational diagrams}
\label{Subsec: Rotational diagrams}
For molecules with multiple detected transitions, we determined a rotational temperature ($T_{rot}$) and column density ($N_{tot}$) using the rotational diagram method \citep{1999ApJ...517..209G}. For this, we assumed a single rotational temperature for all energy levels and local thermodynamic equilibrium (LTE). For optically thin emission, assuming that the source fills the beam, the column density, $N_{u}$, of the upper level is given by
\begin{equation} 
\frac{N_{u}}{g_{u}}=\frac{8k\pi}{hc^{3}} \times \frac{\nu_{ul}^{2}}{A_{ul}g_{u}} \times \int T_\text{MB}\text{dv},
\label{eq: N_u}
\end{equation}
where $g_{u}$ is the statistical weight of the upper level, $\nu_{ul}$ is the frequency of the transition, $\int T_\text{MB}\text{dv}$ is the integrated line intensity and $A_{ul}$ is the Einstein coefficient for spontaneous emission. Making use of this expression and applying Boltzmann's equation, the following expression was derived:
\begin{equation}
\text{ln} \left(\frac{N_{u}}{g_{u}} \right)=\text{ln} \left( \frac{N_{tot}}{Q} \right)-\frac{E_{u}}{kT_{rot}},
\label{eq: rotational diagram}
\end{equation}
where $E_{u}/k$ is the energy of the upper level and $Q$ is the partition function at $T_{rot}$, given by

\begin{equation}
\label{eq: partition function}
Q=\sum_{i}g_{i}e^{-E_{i}/kT_{rot}}
.\end{equation}

\begin{figure*}

\centering

 \centering
\begin{minipage}[t]{.186\linewidth}
\includegraphics[height=0.15\textheight, width=\linewidth]{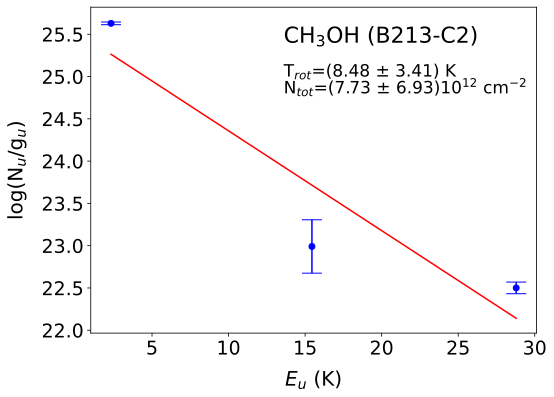}
\end{minipage}
\begin{minipage}[t]{.186\linewidth}
\includegraphics[height=0.15\textheight, width=\linewidth]{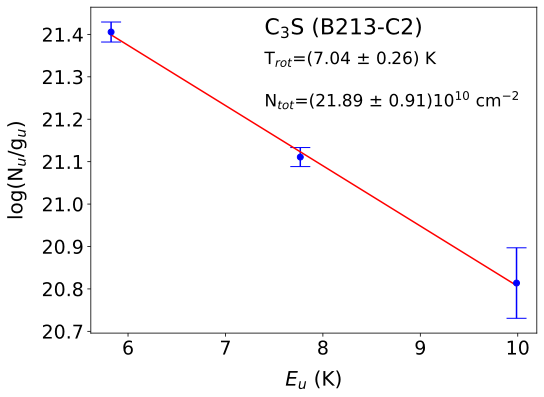}
\end{minipage}
\begin{minipage}[t]{.186\linewidth}
\includegraphics[height=0.15\textheight, width=\linewidth]{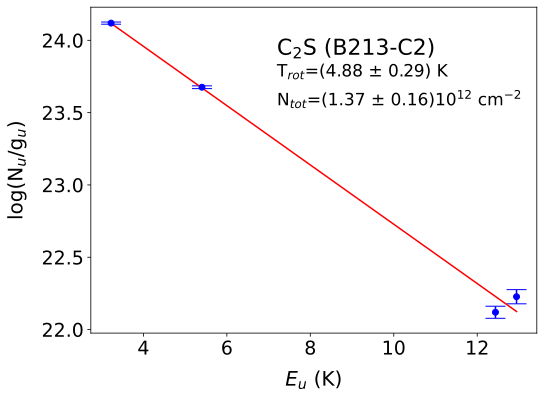}
\end{minipage}
\begin{minipage}[b]{.186\linewidth}
\includegraphics[height=0.15\textheight, width=\linewidth]{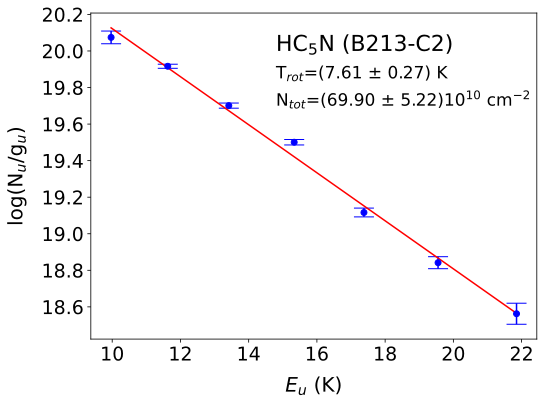}
\end{minipage}
\begin{minipage}[b]{.186\linewidth}
\includegraphics[height=0.15\textheight, width=\linewidth]{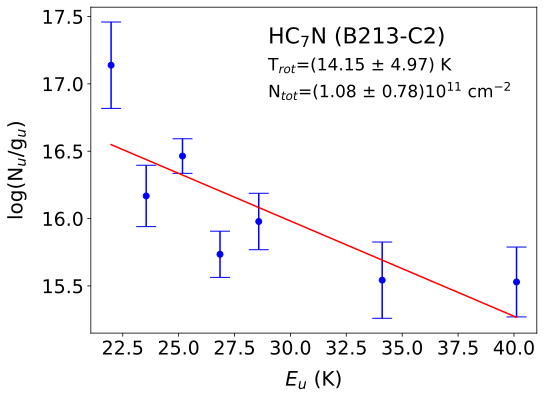}
\end{minipage}

\caption{Rotational diagrams for the detected molecules in B213-C2. The calculated values for the rotational temperature, $T_{rot}$, and column density, $N_{tot}$, are indicated for each molecule. }
\label{Fig: Rotational diagrams C2}
\end{figure*}

\begin{figure*}[h!]

\centering

\begin{minipage}[t]{.2325\linewidth}
\includegraphics[height=0.15\textheight, width=\linewidth]{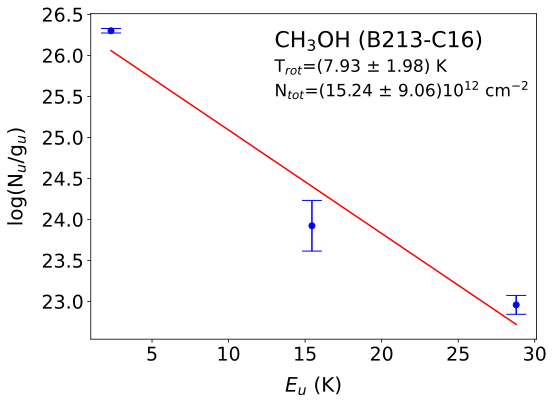}
\end{minipage}
\begin{minipage}[t]{.2325\linewidth}
\includegraphics[height=0.15\textheight, width=\linewidth]{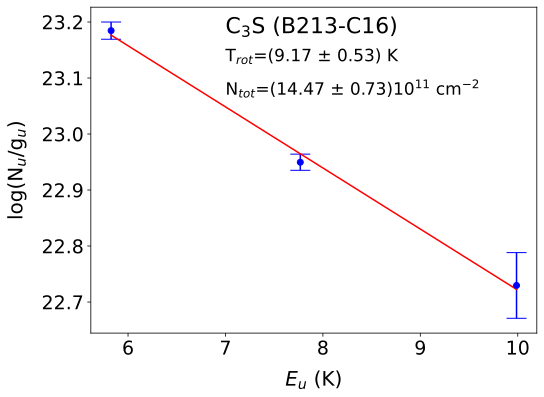}
\end{minipage}
\begin{minipage}[t]{.2325\linewidth}
\includegraphics[height=0.15\textheight, width=\linewidth]{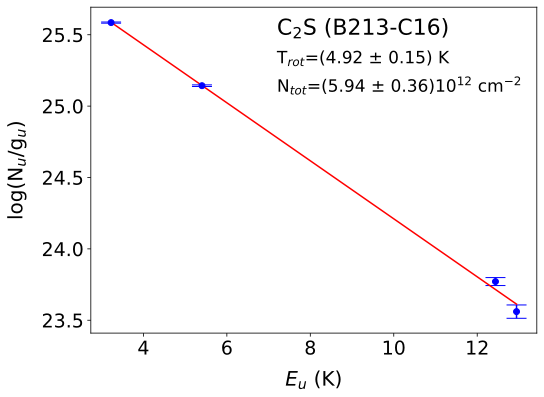}
\end{minipage}
\begin{minipage}[b]{.2325\linewidth}
\includegraphics[height=0.15\textheight, width=\linewidth]{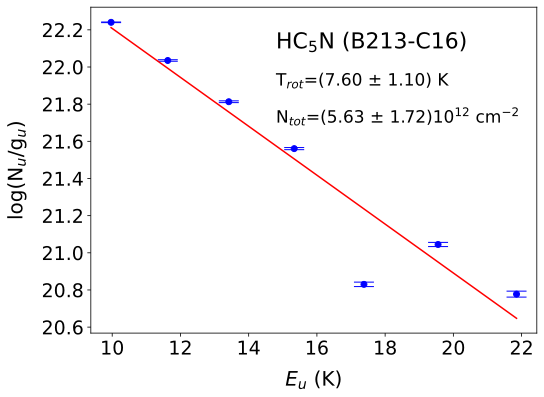}
\end{minipage}
\
\\

\begin{minipage}[b]{.2325\linewidth}
\includegraphics[height=0.15\textheight, width=\linewidth]{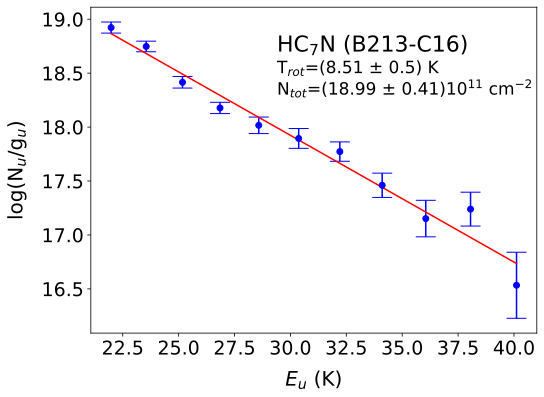}
\end{minipage}
\begin{minipage}[b]{.2325\linewidth}
\includegraphics[height=0.15\textheight, width=\linewidth]{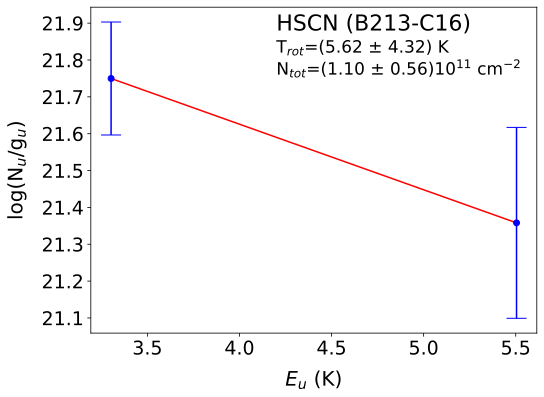}
\end{minipage}
\begin{minipage}[b]{.2325\linewidth}
\includegraphics[height=0.15\textheight, width=\linewidth]{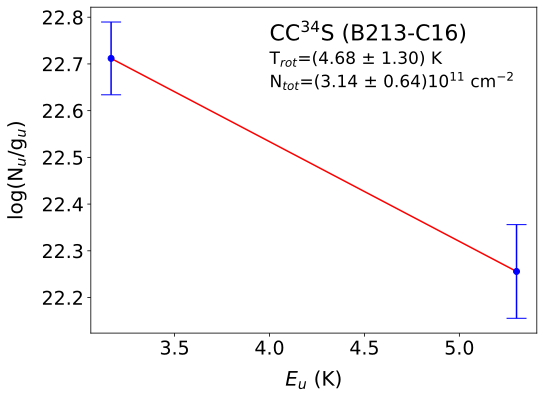}
\end{minipage}
\begin{minipage}[b]{.2325\linewidth}
\includegraphics[height=0.15\textheight, width=\linewidth]{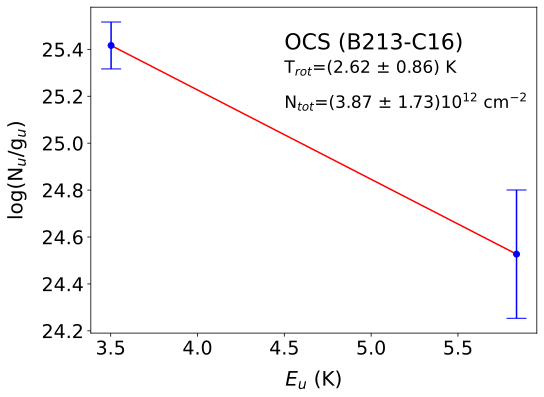}
\end{minipage}

\caption{Rotational diagrams for the detected molecules in B213-C16. The calculated values for the rotational temperature, $T_{rot}$, and column density, $N_{tot}$, are indicated for each molecule. }
\label{Fig: Rotational diagrams C16}
\end{figure*}

Considering Eq. (\ref{eq: rotational diagram}) as a linear equation, the rotational temperature, $T_{rot}$, and total column density, $N_{tot}$, can be derived from the slope and intercept of the regression line, respectively. The values of $\text{ln}(Q)$ for each molecule at $T_{rot}$ were derived by linear interpolation from the data collected in CDMS\footnote{\url{https://cdms.astro.uni-koeln.de/classic/entries/partition_function.html}}.  The resulting rotational diagrams are shown in Figs. \ref{Fig: Rotational diagrams C2} and \ref{Fig: Rotational diagrams C16}. The derived $T_{rot}$ and $N_{tot}$ values are listed in Table \ref{Table: Column densities and abundances}. The uncertainties shown in Table \ref{Table: Column densities and abundances} were obtained from the uncertainties in the least squares fit of the rotational diagrams.

The calculated rotational temperatures for both cores take values in the range of $4.68 \lesssim T_{rot} \lesssim 9.17 \ \text{K}$, with mean values and standard deviations of $T_{rot, \ C2}=8.43 \pm 3.46$ K and $T_{rot, \ C16}=6.38 \pm 2.27$ K. Two notable exceptions are OCS in C16 and HC$_{7}$N in C2, with rotational temperatures of $2.62$ K and $14.15$ K, respectively. The low $T_{rot}$ derived for OCS hints at emission from a moderate- or low-density region, with the gas being subthermally excited. On the other hand, even though the derived value for HC$_{7}$N is higher than expected for a cold dense core, considering the elevated uncertainty computed for this rotational temperature, this measure is still compatible with a rotational temperature of $\sim 10 \ \text{K}$.
The rest of the calculated values are coherent with emission arising from a cold region. Previous ammonia mappings of Taurus cores derived a median temperature of 9.5 K \citep{Jijina1999, Seo2015}.

In those cases in which it was not possible to apply rotational diagrams, i.e., molecules with a single detected transition or molecules with multiple detected transitions but very similar upper level energies, we assigned each of these molecules a rotational temperature of 8 K, which is consistent with the $T_{rot}$ values previously derived by rotational diagrams. Column densities were then determined from Eqs. \ref{eq: N_u} and \ref{eq: rotational diagram}.

To estimate the impact of the uncertainties in the rotational temperature, we also calculated the total column densities and abundances assuming a fixed temperature of 8 K for those species with rotational diagrams (species in Figs. \ref{Fig: Rotational diagrams C2} and \ref{Fig: Rotational diagrams C16}). The calculated values are shown in Table \ref{Table: Column densities and abundances}. As can be derived from the table, the values obtained following both methods are compatible in the majority of cases.

The upper limits of the column densities were determined for the undetected species. 
The $3\sigma$ upper limit was derived from the rms ($\sigma$), the velocity resolution of the spectrum ($\Delta v$), and the line width ($\Delta V$), assuming $\Delta V=1 \ \text{km} \ \text{s}^{-1}$. The upper limit to the equivalent width is given by
\begin{equation} 
\int T_\text{MB}\text{dv} <  3 \sigma \times \sqrt{\Delta v \times \Delta V }
.\end{equation}

\subsection{Abundances}
\label{Subsec: abundances}
Fractional abundances relative to H$_{2}$ were calculated from the column densities previously obtained, using the expression $N$(X)/$N$(H$_{2}$) and the H$_{2}$ column density and dust temperature maps of B213 (see Fig. \ref{Fig: B213  molecular hydrogen maps}), obtained by \citet{Palmeirim} derived from the Herschel Gould Belt Survey \citep{2010A&A...518L.102A} and Planck data \citep{2010A&A...518L..88B} at an angular resolution of 18.2$''$. The total column density of H$_{2}$, $N$(H$_{2}$), is related to the visual extinction by the expression $N(\text{H}_{2})=10^{21} \ A_{\rm v}$ \citep{Bohlin1978}, with $A_{\rm v}$ given by the extinction maps. Abundances were derived considering $N$(H$_{2}$)$_{C2}$$=2.09 \times 10^{22}$ cm$^{-2}$ and  $N$(H$_{2}$)$_{C16}$$=2.48 \times 10^{22}$ cm$^{-2}$. The calculated abundances can be found in Table \ref{Table: Column densities and abundances}.

For B213-C2, the derived fractional abundances take values of between $3.04 \times 10^{-12}$ and $3.35 \times 10^{-9}$, with CC$^{34}$S being the least abundant molecule and CS the most abundant. For C16, abundances take values of between $4.44 \times 10^{-12}$ and $2.59 \times 10^{-9}$ with the least and most abundant molecules being, respectively, HSCN and CH$_{3}$CCH.

\section{Discussion}
\label{Sec: Discussion}

\subsection{Chemical comparison: C2 and C16}
\label{Subsec: Column densities and abundances}
\begin{figure}[h!]

\centering
 \resizebox{\hsize}{!}{\includegraphics{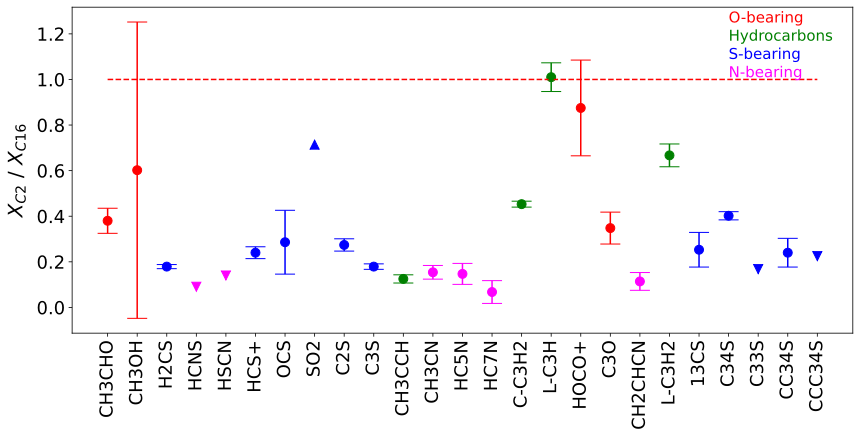}}

\caption{Fractional abundance ratios between B213-C2 and B213-C16, for all the detected species. Lower limits of the ratios, corresponding to molecules detected in B213-C16 but not in B213-C2, are marked with downward triangles. Conversely, upper limits of the ratios, for molecules detected in B213-C2 but not in B213-C16, are indicated with upward triangles.}

\label{Fig: C2 C16 abundance comparison}
\end{figure}

One of the goals of this paper is to explore the chemical evolution of starless cores in their process to become a collapsing core. Previous works suggest that the cores located in the northern part of B213, such as B213-C2, are more evolved than those located in the south \citep{Spezzano,Esplugues2022}. The wealth of species detected in our 7~mm survey allows us to perform a more complete chemical comparison.
Figure \ref{Fig: C2 C16 abundance comparison} shows fractional abundance ratios between B213-C2 and B213-C16. As is seen in the figure, almost all the detected molecules are significantly more abundant in B213-C16. This would imply a higher molecular depletion in B213-C2, as is expected in a more evolved core. The only exceptions to this rule are l-C$_3$H and HOCO$^+$.

The decrease in the density of stars from north to south makes B213-C2, located in the northern part of the filament, more likely to be affected by the presence of low-mass stars located nearby. The cones of the outflows of these stars carve out the medium, allowing the interstellar radiation field to penetrate the molecular cloud \citep{Spezzano}. Although the inner parts of the cloud are shielded from UV radiation by the dust, the subsequent exposure to UV radiation in the outer regions could contribute to reduce abundances for some species and an increase in the abundances of certain molecules such as HOCO$^+$, related with irradiated ices \citep{Minh1991}. In the same line, the radical l-C$_3$H is known to present enhanced abundances in photodissociation regions \citep{Pety2012, Guzman2015}.
As can be seen in Fig. \ref{Fig: C2 C16 abundance comparison}, these molecules display similar fractional abundances in both sources.

Another possible cause of the chemical differentiation observed between C16 and C2 could be the accretion of background molecular gas. \citet{2008ApJ...680..428G} and \citet{Palmeirim} found that the B213 region of the B213/L1495 filament is surrounded by a large number of low-density striations that are oriented approximately perpendicular to the main filament and parallel to the magnetic field. Blueshifted and redshifted components in both $^{12}$CO(1-0) and $^{13}$CO(1-0) emission are also visible to the southwest and northeast of the B213 filament. This morphology is suggestive of mass accretion along the field lines into the filament. \citet{Shimajiri} proposed that the filament was initially formed by large-scale compression of HI gas and is now growing in mass as a result of the gravitational accretion of molecular gas of the ambient cloud. It is possible that this accretion of ambient material into this part of the B213/L1495 filament along the magnetic field lines is enhancing the molecular abundances in the embedded cores.
However, these low-density striations are not present in the B7-B10 region \citep[see Fig. 2 of][]{Hacar} of the B213/L1495 filament, where C2 is located. \citet{Chapman2011} indicated that above core 7 (see Fig. \ref{Fig: B213  molecular hydrogen maps}), where the filament presents a sharp turn to the north, the magnetic field switches from being perpendicular to the filament to becoming parallel. In the presence of a magnetic field, motions of ionized gas encounter a significant magnetic resistance perpendicular to the field lines \citep{Nagai1998} and, under these conditions, the accretion of background material is mainly parallel to the field lines.

Finally, as was commented on above, the lower abundances observed in C2 could stem from the evolutionary state of the cores, with B213-C2 constituting a more evolved object, closer to the pre-stellar core phase. A higher density would cause the depletion of some molecules, which would condense out onto dust grain surfaces. This would be reflected in reduced abundances in B213-C2. These factors are not exclusive, and the observed differences in molecular abundances between the two sources are most likely the result of their addition.

\subsection{Chemical comparison with other sources}
\label{Subsec: Chemical comparison with other sources}

\begin{figure*}[h!]

\centering
\begin{minipage}[t]{.322\linewidth}
\includegraphics[height=0.18\textheight, width=\linewidth]{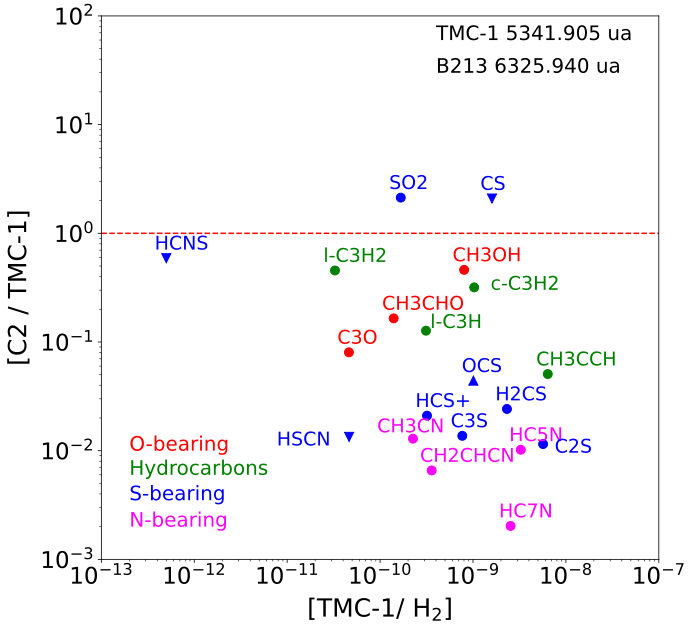}
\end{minipage}
\begin{minipage}[t]{.322\linewidth}
\includegraphics[height=0.18\textheight, width=\linewidth]{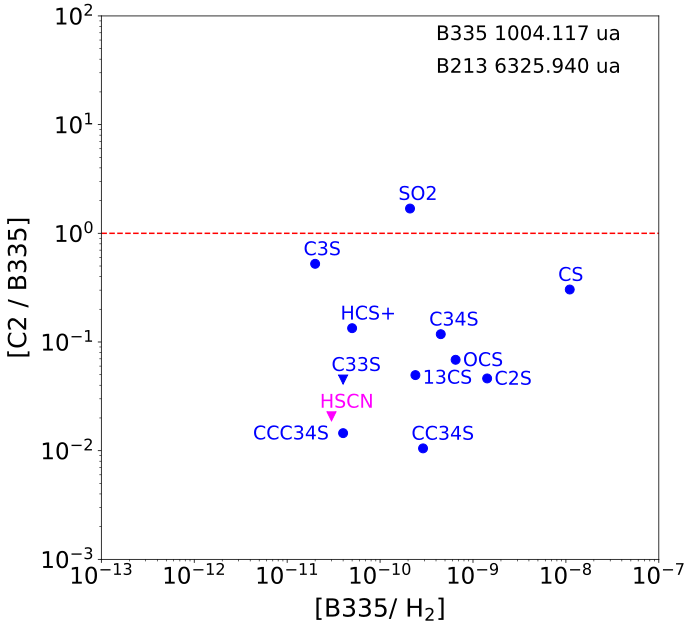}
\end{minipage}
\begin{minipage}[t]{.322\linewidth}
\includegraphics[height=0.18\textheight, width=\linewidth]{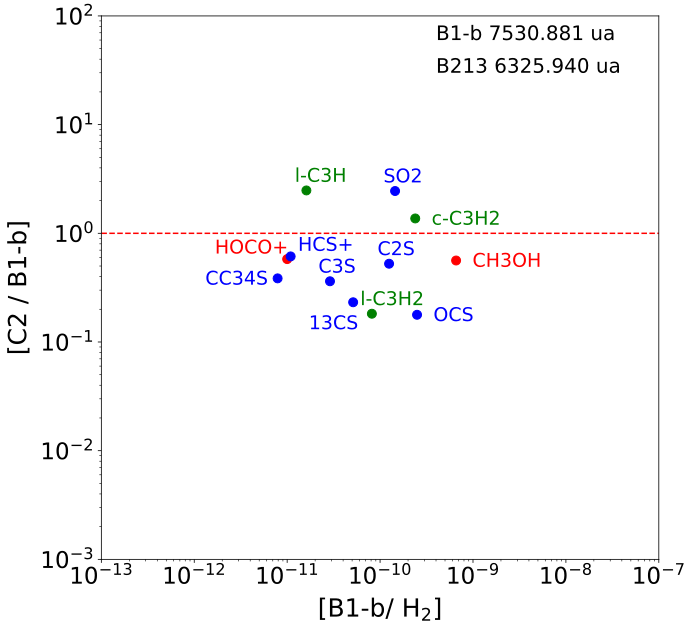}
\end{minipage}

\begin{minipage}[b]{.322\linewidth}
\includegraphics[height=0.18\textheight, width=\linewidth]{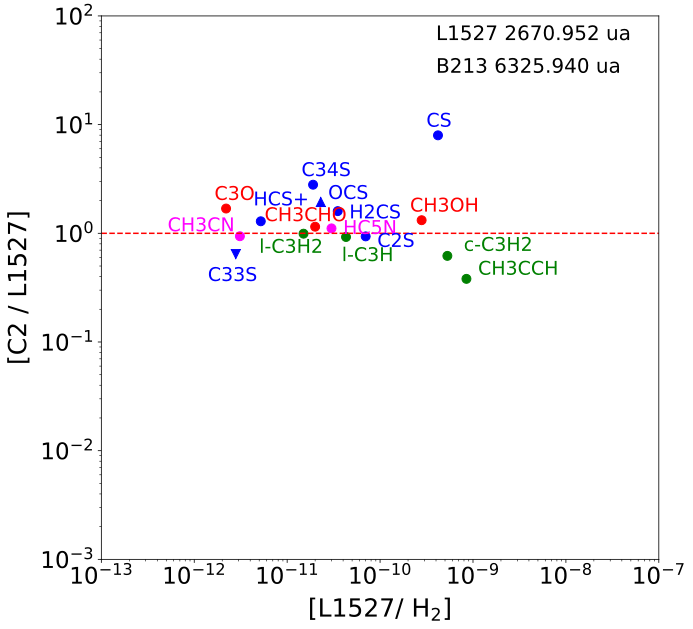}
\end{minipage}
\begin{minipage}[b]{.322\linewidth}
\includegraphics[height=0.18\textheight, width=\linewidth]{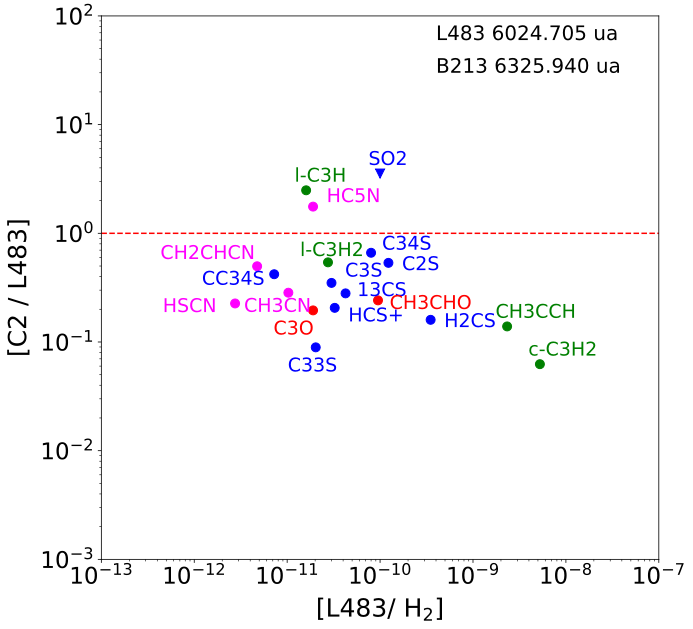}
\end{minipage}

\caption{Fractional abundance ratios between B213-C2 and five reference sources: TMC-1, B335, B1-b, L1527, and L483.  Upper and lower limits are indicated with downward and upward triangles, respectively. L483: \citet{Agundez2019}.  TMC-1: \citet{Gratier2016}; \citet{Cernicharo2024}; \citet{Wakelam2013}. L1527: \citet{Yoshida2019}. B335: \citet{2023A&A...678A.199E}. B1-b: \citet{Fuente2016}; \citet{Loison2017}; \citet{WidicusWeaver}. N(H$_{2}$) values used to obtain fractional abundances are $7.60 \times 10^{22} \ \text{cm}^{-2}$ for B1-b \citep{Daniel2013}, and $1.80 \times 10^{22} \ \text{cm}^{-2}$  for TMC-1 \citep{Palmeirim, RodriguezBaras2021}. Physical scales for each source are indicated in the upper left corner.} 

\label{Fig: C2 and other sources abundance comparison}
\end{figure*}

\begin{figure*}[h!]
\centering

\begin{minipage}[t]{.322\linewidth}
\includegraphics[height=0.18\textheight, width=\linewidth]{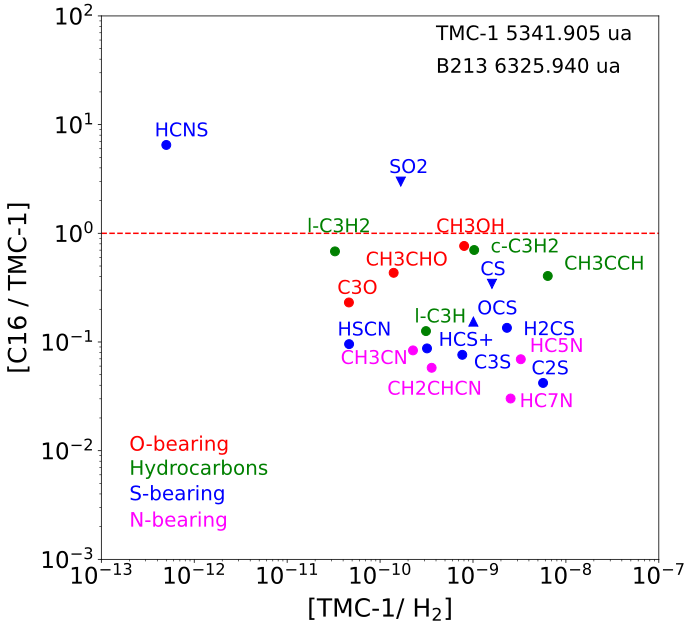}
\end{minipage}
\begin{minipage}[t]{.322\linewidth}
\includegraphics[height=0.18\textheight, width=\linewidth]{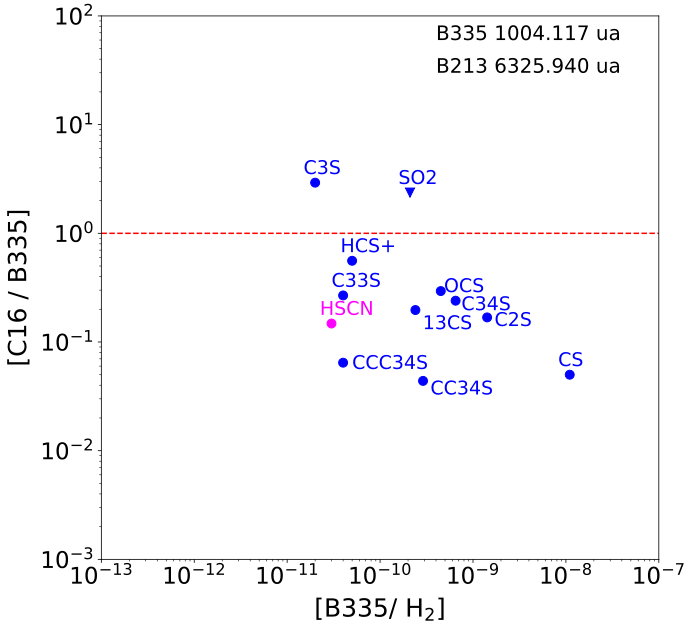}
\end{minipage}
\begin{minipage}[t]{.322\linewidth}
\includegraphics[height=0.18\textheight, width=\linewidth]{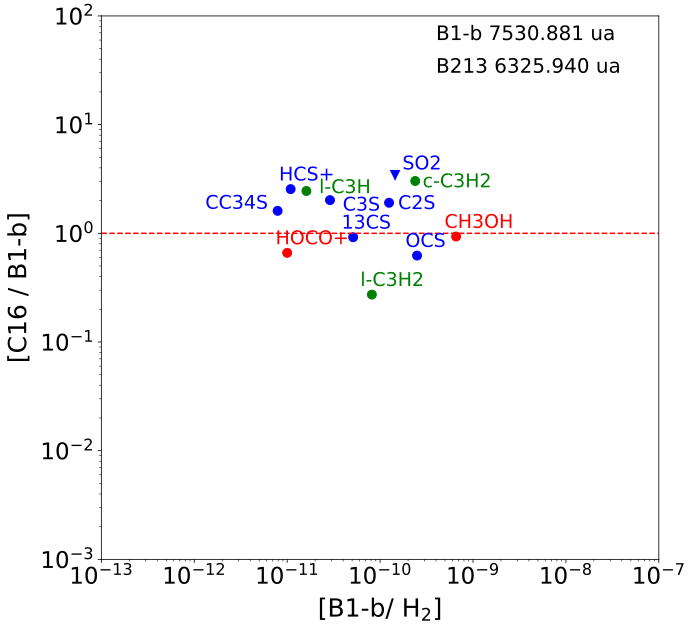}
\end{minipage}

\begin{minipage}[b]{.322\linewidth}
\includegraphics[height=0.18\textheight, width=\linewidth]{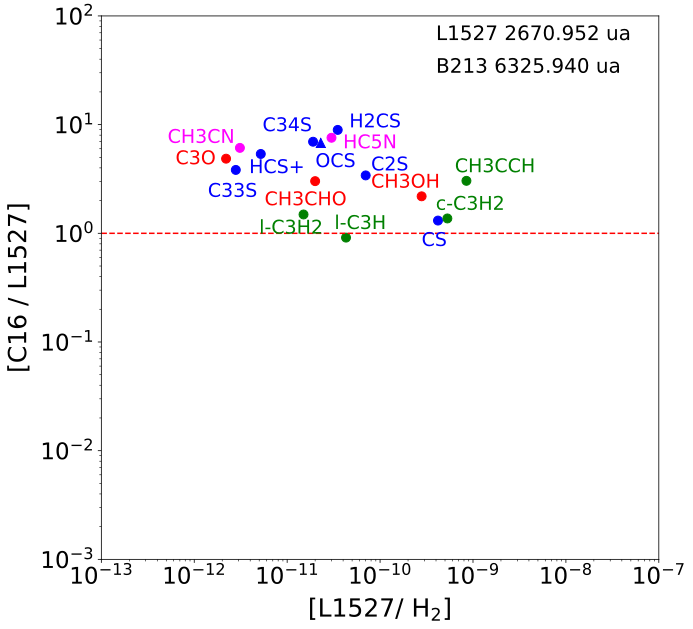}
\end{minipage}
\begin{minipage}[b]{.322\linewidth}
\includegraphics[height=0.18\textheight, width=\linewidth]{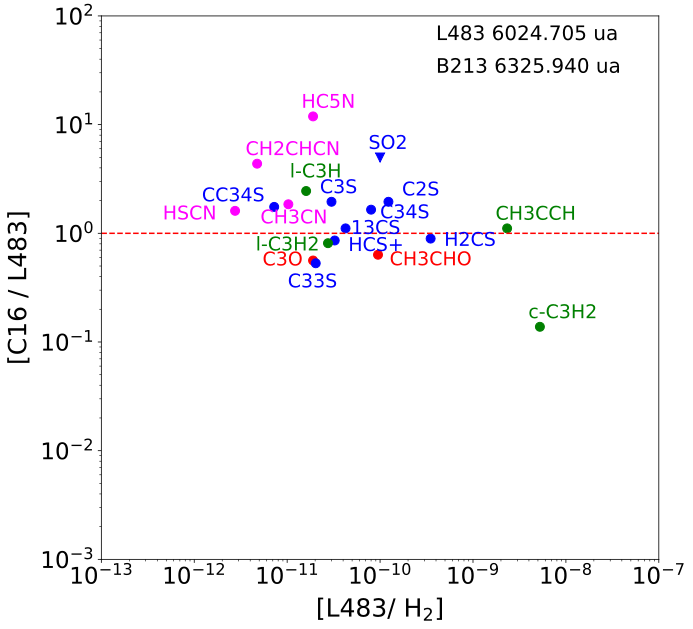}
\end{minipage}

\caption{Fractional abundance ratios between B213-C16 and five reference sources: TMC-1, B335, B1-b, L1527 and L483. Upper and lower limits are indicated with downward and upward triangles, respectively. L483: \citet{Agundez2019}.  TMC-1: \citet{Gratier2016}; \citet{Cernicharo2024}; \citet{Wakelam2013}. L1527: \citet{Yoshida2019}. B335: \citet{2023A&A...678A.199E}. B1-b: \citet{Fuente2016}; \citet{Loison2017}; \citet{WidicusWeaver}. N(H$_{2}$) values used to obtain fractional abundances are $7.60 \times 10^{22} \ \text{cm}^{-2}$ for B1-b \citep{Daniel2013}, and $1.80 \times 10^{22} \ \text{cm}^{-2}$  for TMC-1 \citep{Palmeirim, RodriguezBaras2021}. Physical scales for each source are indicated in the upper left corner.} 
\label{Fig: C16 and other sources abundance comparison}
\end{figure*}

Once a protostar is formed, it heats the surrounding material, increasing dust temperature, and releasing the icy molecules to the gas phase. At this early stage, the cold envelope is still infalling and its chemical composition remains similar to the parent core. Figures \ref{Fig: C2 and other sources abundance comparison} and \ref{Fig: C16 and other sources abundance comparison} show fractional abundances relative to H$_{2}$ in C2 and C16 compared with those in five reference sources representative of the earliest stages of the formation of a low-mass star: TMC-1, $\text{B}\, 335$, B1-b, L1527, and L483.

The considered sources present different evolutionary stages, including both starless cores and cores with indications of star formation. Taurus Molecular Cloud-1 (TMC-1) is a cold dense cloud with no signs of star formation, located at the centre of the Taurus Molecular Cloud at a distance of $140.2 ^{+1.3}_{-1.3}$ pc \citep{Galli2019}. Barnard 335 ($\text{B}\,335$) is a prototypical example of a large, isolated dust globule \citep{Keene1980}, which is associated with an embedded far-infrared source, IRAS 19347+0727 \citep{Keene1983}. The presence of saturated complex organic molecules (COMs) identifies $\text{B}\,335$ as a hot corino \citep{Cazaux2003, Imai2016}. The source Barnard 1b (B1-b) is a dark cloud that hosts two young protostellar objects, B1b-N and B1b-S \citep{Huang2013}. The high HCS$^+$/CS ratio for $\text{B}\,335$ \citep{2023A&A...678A.199E} compared to the value observed in B1-b \citep{Fuente2016} suggests an early evolutionary stage of $\text{B}\,335$ compared to B1-b, as rate coefficients for ion-CS reactions increase at low temperatures \citep{Clary1985}. On the other hand, the source L483 is an optical dark cloud with an embedded infrared source, which is classified as a Class 0/I object \citep{Tafalla2000, Agundez2019}. Finally, the dense core L1527 contains a solar-type Class 0/I protostar, IRAS04368+2557 \citep{Yoshida2019}. Therefore, the sources can be approximately ordered in an evolutionary sequence as $$\text{TMC-1}  \rightarrow \text{$\text{B}\,335$} \rightarrow \text{B1-b} \rightarrow \text{L1527} \rightarrow \text{L483},$$
with starless cores B213-C2 and B213-C16 being in an evolutionary state between TMC-1 and $\text{B}\,335$. The order of the plots in Figs. \ref{Fig: C2 and other sources abundance comparison} and \ref{Fig: C16 and other sources abundance comparison} follows this evolutionary sequence.

For TMC-1, column densities are mostly taken from the line survey of the cyanopolyyne peak position of TMC-1 (TMC-1(CP)) carried out with the Nobeyama 45 m telescope \citep{Kaifu2004}, revised by \citet{Gratier2016}, and also from the QUIJOTE line survey of TMC-1 (CP) \citep{Cernicharo2024} and from \citet{Wakelam2013}. In the case of the source $\text{B}\,335$, abundances are based on a study of its sulfur chemistry through observations in the spectral range $\lambda = 7, 3$ and 2 mm, conducted with the Yebes-40 m and IRAM-30 m telescopes \citep{2023A&A...678A.199E}.  B1-b data is based on the chemical studied carried out by \citet{Fuente2016}. Abundances for L1527 were taken from  a  $\lambda= 3$ mm line survey conducted with the Nobeyama-45 m telescope \citep{Yoshida2019}. Finally, for the source L483, colum densities were taken from the  $\lambda = 3$ mm line survey carried out with the IRAM-30 m telescope in the 80–116 GHz frequency range \citep{Agundez2019}. Most of the abundances plotted in Figs. \ref{Fig: C2 and other sources abundance comparison} and \ref{Fig: C16 and other sources abundance comparison} have been measured at similar spatial scales ($3000 \ \text{au} -6000 \ \text{au}$), which makes their comparison reasonable. However, $\text{B}\,335$ and L1527 present smaller spatial scales.

As can be seen in Figs. \ref{Fig: C2 and other sources abundance comparison} and \ref{Fig: C16 and other sources abundance comparison}, TMC-1 presents the highest abundances of all the compared sources. This is to be expected, as previous studies have shown that TMC-1 is a particularly molecule-rich dark cloud, especially in carbon chains \citep{Wakelam2013}.
A high cosmic-ray ionization rate, $\zeta_{H_{2}}$, has been suggested as an explanation for the richness of TMC-1 in this type of molecule \citep{Wakelam2013}. This target is considered to be a prototype for the starless core stage prior to the collapse.

The source $\text{B}\,335$ presents a clear chemical differentiation with respect to  C2 and C16. The overabundance of sulfur-bearing species is particularly pronounced when comparing abundances of sulfur carbon chains. Recent studies on sulfur chemistry in $\text{B}\,335$ have identified it as an especially rich source in this type of molecule, unlike other Class 0 sources \citep{2023A&A...678A.199E}. $\text{B}\,335$ presents an early evolutionary stage comparable to the ages of pre-stellar condensations \citep{Caselli2012}. This could be the result of the isolated source nature of $\text{B}\,335$ compared to other protostars formed in dense molecular clouds. Accretion of diffuse cloud material has also been proposed as a possible cause of the different chemistry observed in $\text{B}\,335$ \citep{2023A&A...678A.199E}. In any case, the lower molecular abundances observed in both C2 and C16 with respect to $\text{B}\,335$ are likely a result of the chemical richness of this peculiar source.

The upper right panels of Figs. \ref{Fig: C2 and other sources abundance comparison} and \ref{Fig: C16 and other sources abundance comparison} show that the abundances of hydrocarbons and oxygen-bearing molecules in B1-b are of the same order as in C2 and C16. However, marked chemical differentiation is observed for S-bearing molecules. This difference could be explained in terms of freeze-out, as the sources present varying degrees of sulfur depletion. Previous works based on methanol observations suggest a chemical differentiation between the northern and southern cores of the B213 filament \citep{Spezzano, Punanova2022}. A recent study of the H$_{2}$CS deuterated compounds \citep{Esplugues2022} showed that the cores located in the northern part of the filament are more chemically evolved than the southern cores. Moreover, based on observations of NS, \citet{Hily2022} proposed a correlation between sulfur depletion and chemical age, with more evolved cores presenting a higher sulfur depletion. On the other hand, the source B1-b presents a moderate sulfur depletion ($[\text{S} /\text{H}] \sim 6.0 \times 10^{-7}$), which could be the consequence of the star formation activity in the region with multiple outflows in the surroundings or the result of the rapid collapse of B1-b \citep{Fuente2016}. 
These differences in sulfur depletion could explain the higher abundances of S-bearing molecules observed in B1-b with respect to C2, located in the northern part of the B213 filament, and the lower abundances in B1-b with respect to C16, located south of the filament. The bottom panels of Figs. \ref{Fig: C2 and other sources abundance comparison} and \ref{Fig: C16 and other sources abundance comparison} show that, in general, L1527 presents lower molecular abundances than C2 and C16, with the exception of hydrocarbons, which take higher or similar values to those observed in C2. Abundances in L483 are overabundant with respect to C2 and underabundant when comparing with C16. Among carbon-chain molecules, the C$_{n}$H$_{m}$ species seem especially abundant in both L1527 and L483, presenting similar values to those observed in C16, and even higher in the case of L483. A longer lifetime than the other carbon-chain molecules has been proposed as a possible cause of the high abundances of these molecules. Otherwise, the C$_{n}$H$_{m}$ molecules may be regenerated near the protostar, as both L1527 and L483 are warm carbon-chain chemistry (WCCC) sources \citep{Sakai2008, Oya2017}. In the scenario of the production of unsaturated carbon chains, often referred to as WCCC (\citealt{Sakai2013}), this regeneration process would be driven by the evaporation of CH$_{4}$ due to protostellar activity, as CH$_{4}$ has a relatively low sublimation temperature ($\sim$ 30 K). The higher temperatures in the vicinity of the protostar can cause the C$_{n}$H$_{m}$ to evaporate from dust grains, with the subsequent carbon-chain production in the gas phase. However, under the physical conditions prevailing in systems that result in WCCC, the formation of HC$_{n}$N is slower than that of C$_{n}$H$_{m}$, due to slow neutral-neutral reactions, which would suppress the abundances of HC$_{n}$N and other N-bearing species \citep{Turner1998, Sakai2008, Yoshida2019}. As can be seen in the bottom panels of Fig.  \ref{Fig: C2 and other sources abundance comparison}, the abundance of HC$_{5}$N in both L1527 and L483, takes similar values to the ones observed in C2.

\section{Comparison with chemical models}
\label{Sec: Chemical modelling}
To better constrain the physical and chemical properties of both C2 and C16, we compared the observational data to a set of chemical models. These models were computed by a neural emulator of the astrochemical code NAUTILUS 1.1 based on conditional neural fields \citep{Asensio2024}. This emulator reproduces Nautilus predictions with a precision well below $0.1$ dex, allowing us to carry  out a broad exploration of the space parameter in a reasonable time.

NAUTILUS 1.1 \citep{Raud2016} is a three-phase model that considers gas, grain surface, and grain mantle phases, along with their interactions. Given a set of physical and chemical input parameters, it solves the kinetic equations for the gas-phase and the solid species at the surface of interstellar dust grains, calculating the evolution with time of chemical abundances. The use of neural networks to simulate the behaviour of chemical models considerably reduces the computation time, allowing a more in-depth exploration of the parameter space and reducing the uncertainties for all the considered species. 

\subsection{Parameter space}
\label{Subsec: Parameter space}
We ran a set of models with typical physical conditions in dark clouds, which were selected randomly in the intervals defined in Table \ref{Table: Parameter space}. All the parameters except for the temperature, which is linearly sampled, are sampled uniformly in log space as they mostly span several orders of magnitude. The visual extinction takes the fixed values 10.5 and 12.4 for C2 and C16, respectively \citep{Palmeirim, RodriguezBaras2021}.

 \begin{table}[h!]
\caption{Chemical model parameters.}
\label{Table: Parameter space}
\centering
\begin{tabular}{l l  }
\hline
\hline
\\
C2 & \\
$t$  & $10^{4}-10^{7}$ yr \\
$n_{H} $  & $(0.6-2)\times10^{5} $ cm$^{-3}$\\
$A_{\rm v}$ & 10.5 mag \\
$\zeta_{H_{2}}$  & $(0.1-5)\times 10^{-16}$ s$^{-1}$ \\
$[$S / H$]$ & $(0.0075-1.5) \times 10^{-5}$\\
$\chi_{UV}$ (Draine) & $1-50$ \\
$T$ & $10-14$ K\\

\hline
\\
C16 & \\
$t$  & $10^{4}-10^{7}$ yr \\
$n_{H} $  & $(0.4-2)\times10^{5} $ cm$^{-3}$ \\
$A_{\rm v}$ & 12.4 mag \\
$\zeta_{H_{2}}$  &  $(0.1-5)\times 10^{-16}$ s$^{-1}$\\
$[$S / H$]$ & $(0.0075-1.5) \times 10^{-5}$\\
$\chi_{UV}$ (Draine) & $1-50$ \\
$T$ & $10-14$ K \\
\hline

\end{tabular}
\end{table}

We considered chemical ages, $t$, between early chemistry (0.01 Myr) and steady-state chemistry (10 Myr). The sulfur elemental abundance [S/H] varied between the cases of no depletion, [S/H]$=1.5\times 10^{-5}$, and high-depletion, [S/H]$=7.5\times 10^{-8}$. These values correspond to different estimates of sulfur depletion in star-forming regions \citep{Esplugues2014,Vastel2018, Navarro2020, Navarro2021}. The cosmic-ray ionization rate, $\zeta_{H_{2}}$, was allowed to take values between $\zeta_{H_{2}}=10^{-17}$ s$^{-1}$, found in dense and evolved cores \citep{Caselli2002}, and $\zeta_{H_{2}}=5 \times 10^{-16}$ s$^{-1}$, expected in diffuse molecular gas \citep{NeufeldWolfire2017}. We considered $\chi_{UV}=50$ as the upper limit for the UV field strength in units of Draine field \citep{Draine1978}, based on the values derived toward photon-dominated regions such as the Horsehead nebula \citep{Pety2005, Goicoechea2006, Riviere2019}.
\subsection{Methodology}
\label{subsec: Methodology}

\begin{figure*}[h]

\begin{minipage}[t]{.52\linewidth}
\includegraphics[width=\linewidth]{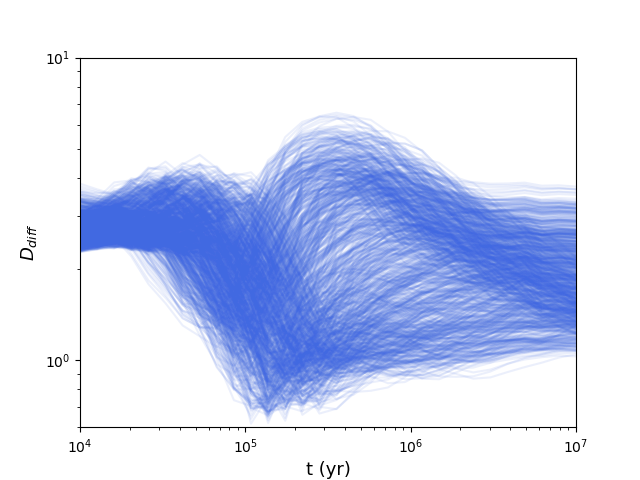}
\end{minipage}
\begin{minipage}[b]{.52\linewidth}
\includegraphics[width=\linewidth]{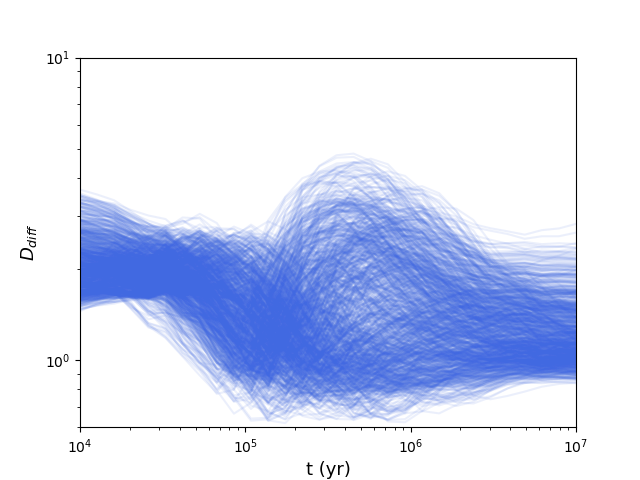}
\end{minipage}

\caption{$D_{diff} \ (t)$ for a set of $10^{3}$ models considering the physical conditions listed for B213-C2 (left) and B213-C16 (right).}
\label{Fig: best t values}

\end{figure*}

\begin{figure*}[h]

\begin{minipage}[t]{.52\linewidth}
\includegraphics[width=\linewidth]{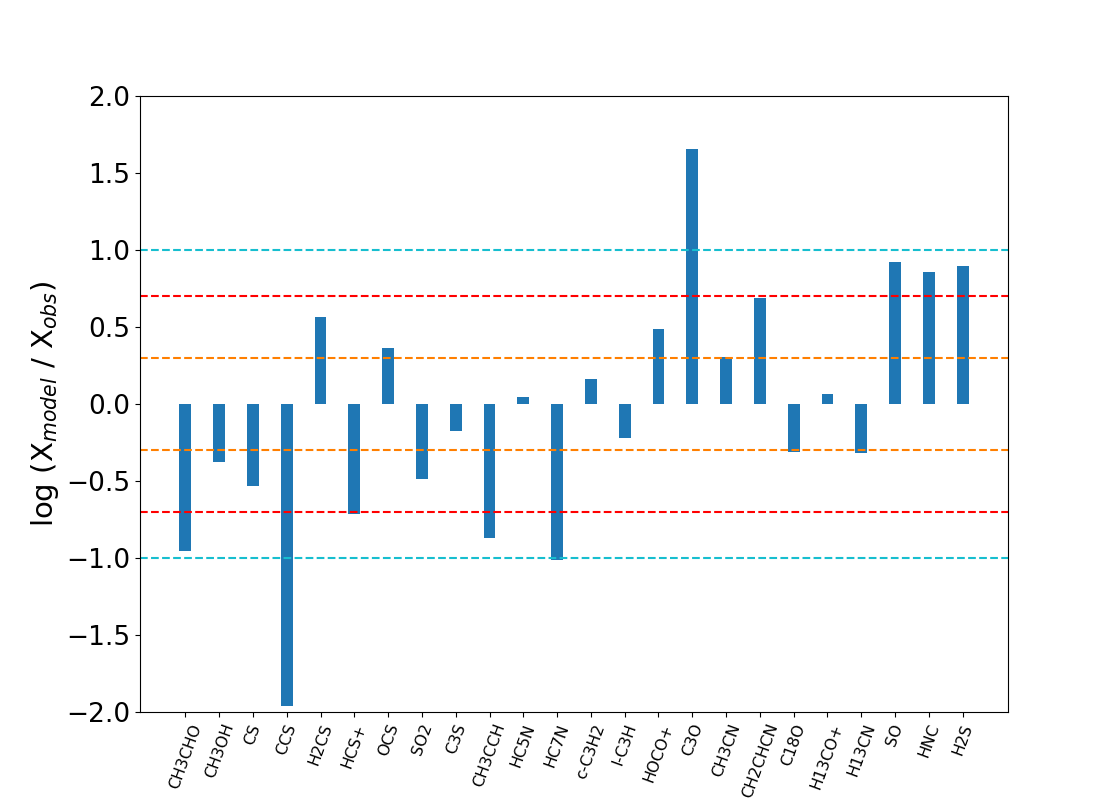}
\end{minipage}
\begin{minipage}[b]{.52\linewidth}
\includegraphics[width=\linewidth]{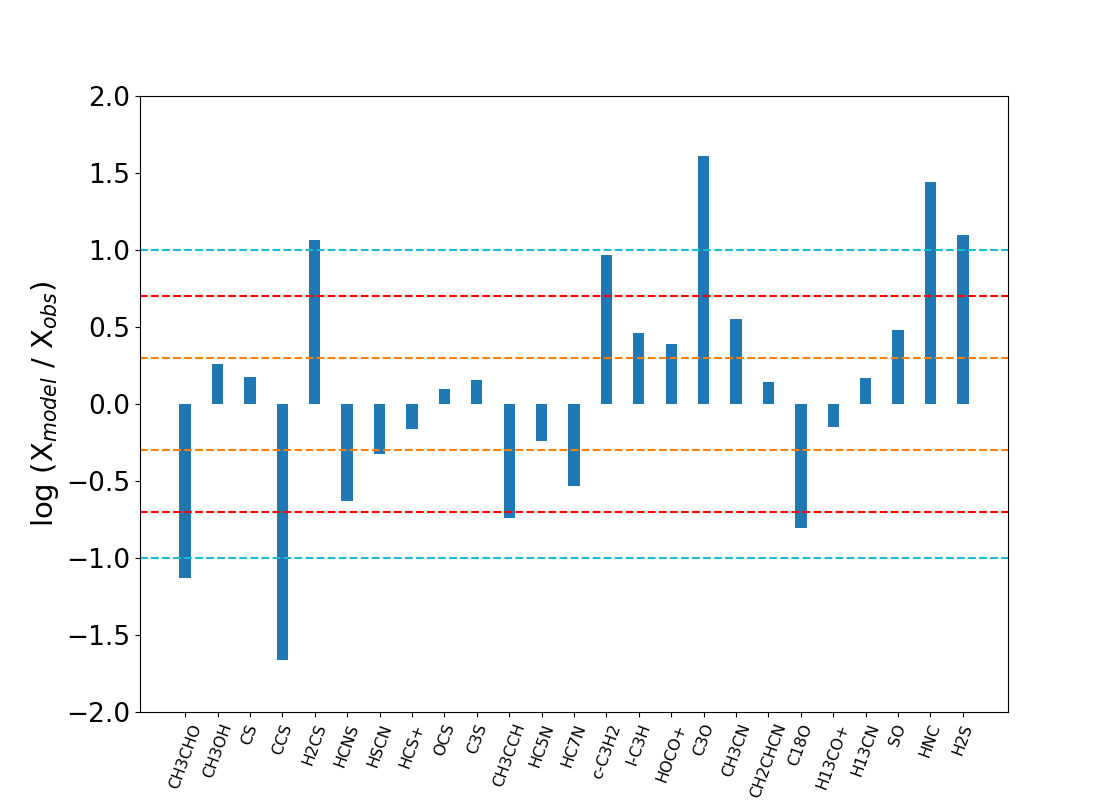}
\end{minipage}

\caption{Ratios between the best-fitting model prediction and the observed abundances for B213-C2 (left) and B213-C16 (right), indicated with blue bars. Discrepancies between model and observations of a factor 10, 5 and 2 are indicated with dashed blue, red, and orange lines, respectively.}
\label{Fig: comparison with models}
\end{figure*}

Our aim is to determine the values of $t$, $n_{H}$, $\zeta_{H_{2}}$, [S/H], $\chi_{UV}$, and $T$ that best fit the observational data for each core. We need to select a parameter in order to describe the goodness of each model. The standard chi-squared $\chi^{2}$ is not adequate to estimate errors when the molecular abundances included in the fit differ from the observed abundances by several orders of magnitude. For this reason, we used the parameter $D_{diff}$, defined as

\begin{equation}
D_{diff} \ (t)=\frac{1}{n_{obs}}\sum_{i}[log_{10}(X^{i}_{mod})(t)-log_{10}(X^{i}_{obs})]^{2}
\label{eq: distance of disagreement}
,\end{equation}
where $n_{obs}$ is the number of detected species in the source, $X^{i}_{mod} \ (t)$ is the model predicted abundance for the $i$ species at a time, $t$, and $X^{i}_{obs}$ is the abundance derived from the observations. This parameter is the square of the “disagreement parameter” that has been previously used in astrochemistry by
different authors \citep[see e.g.,][]{Wakelam2006, Vastel2018, Fuente2023, Taillard2025}.

First, we ran the models allowing the parameter $t$ to vary among a range of discrete values in the interval $10^{4}-10^{7}$ Myr, with the purpose of determining the evolution time that best reproduces the observational data. We then ran a new set of models fixing $t$ to the best-fitting value previously determined. Reasonable intervals for the rest of the parameters were obtained from a set of best-fitting models including those whose $D_{diff}$ parameter differs from the minimum $D_{diff}$ value by less than $1\%$.

In our fitting, we used the molecular abundances derived in this paper as well as those of C$^{18}$O, H$^{13}$CO$^+$, H$^{13}$CN, SO, HNC, and H$_{2}$S derived by \cite{RodriguezBaras2021} using the database of the IRAM LP GEMS database\footnote{Database available in https://iram-institute.org/science-portal/proposals/lp/completed/lp021-gas-phase-elemental-abundances-in-molecular-clouds/}. The abundances for CS, C$^{18}$O, H$^{13}$CO$^{+}$, and H$^{13}$CN were derived from those of C$^{34}$S, CO, HCO$^{+}$, and HCN, considering the isotopic ratios $^{32}$S/$^{34}$S$=22.5$, $^{12}$C/$^{13}$C$=60$, and $^{16}$O/$^{18}$O$=550$ \citep{Wilson1994, Savage2002, Gratier2016}.

\subsection{Comparison with observations}
\label{subsec: Comparison with observations: fractional abundances in B213-C2}

\begin{table*}[ht!]
\tiny
\centering
\caption{Parameters corresponding to the best-fitting models.}
\label{Table: Best models}
\centering
\begin{tabular}{ l l l l l l l l l }
\hline
\hline
\\
\multicolumn{9}{c}{\text{$D_{diff, \ min}$}} \\
\\
\hline
Source & $t$ (Myr) & $A_{\rm v}$ (mag) & $n_{H} $ (cm$^{-3}$) & $\zeta_{H_{2}}$ (s$^{-1}$)  & $[$S / H$]$ & $\chi_{UV}$ (Draine) & $T$ (K) & $D_{diff}$ \\
\hline
\\
C2 & $0.11$ & 10.5 & $1.99 \times 10^{5}$  & $1.48\times 10^{-17}$ & $1.78 \times 10^{-7}$ & $2.70$ & $13.64$ & $0.600$\\
C16 & $0.17$ & 12.4 & $1.63 \times 10^{5}$ & $1.17\times 10^{-16}$ & $3.49 \times 10^{-7}$ & $6.90$ & $13.97$ & $0.605$ \\
\hline
\\
\multicolumn{9}{c}{\text{$D_{diff}\in [D_{diff, \ min}, \ 1.01 \times D_{diff, \ min}]$ }} \\
\\
\hline
Source & $t$ (Myr) & $A_{\rm v}$ (mag) & $n_{H} $ (cm$^{-3}$) & $\zeta_{H_{2}}$ (s$^{-1}$)  & $[$S / H$]$ & $\chi_{UV}$ (Draine) & $T$ (K) & $D_{diff}$ \\
\hline
\\
C2 & $0.11$ & 10.5 & $(1.94-1.99) \times 10^{5}$  & $(1.33-1.63) \times 10^{-17}$ & $(1.48-2.36) \times 10^{-7}$ & $1.61-11.19$ & $13.64-13.99$ & $0.600-0.606$ \\
C16 & $0.17$ & 12.4 & $(1.29-1.69) \times 10^{5}$ & $(1.15-1.25) \times 10^{-16}$ & $(2.71-3.49) \times 10^{-7}$ & $1.25-6.90$ & $13.78-13.98$ & $0.605-0.607$ \\
\hline

\end{tabular}
\end{table*}

Our first aim is to determine the chemical age, $t$, that best reproduces the observational data for each source. Figure \ref{Fig: best t values} shows the $D_{diff}$ parameter as a function of time for a set of $10^{3}$ models generated with the physical conditions listed in Table \ref{Table: Parameter space}. From both panels of Fig. \ref{Fig: best t values}, we have deduced that the observed abundances are best fit assuming early time chemistry ($t \sim 0.1$ Myr), which is consistent with the idea that this filament is still accreting background material of the ambient cloud, keeping the gas chemistry far from the steady state \citep{Fuente2023}. 

To constrain the values for the rest of the parameters, we ran another set of models, fixing the chemical age, $t$, to the best-fitting value for each core. Figure~\ref{Fig: comparison with models} shows the comparison between the observed abundances and the predictions of the  best-fitting models shown in Table~\ref{Table: Best models}. Horizontal dashed lines indicate discrepancies by factors of 2, 5, and 10. Most molecules are well reproduced within a factor of $<$5, including CH$_3$OH, and other COMs, such as cyanopolynes. However, a few are poorly reproduced by the model with discrepancies of more than one order of magnitude. In particular, the abundances for HNC, H$_2$S, and the carbon chains C$_2$S, and C$_3$O are poorly reproduced in the two studied cores. The molecules HNC and H$_2$S are over-reproduced by a factor of 10. In the case of HNC, the abundance derived from the GEMS database is uncertain since the observed line, HNC J=1$\rightarrow$0, is expected to be optically thick in dense cores. \citet{Tasa2025} presented new observations of  HN$^{13}$C  in C2 and C16. Assuming HN$^{12}$C/ HN$^{13}$C=60, we estimate HNC abundances a factor of $\sim$10 higher than those used in the fitting, in better agreement with model predictions. Opacity effects could also be significant in the case of H$_2$S. Taking the N(H$_2^{34}$S) column density derived by \citet{RodriguezBaras2023} toward C2, and assuming $^{32}$S/$^{34}$S$=22.5$, we obtain a H$_2$S abundance of H$_2$S $\sim$2.7 times higher than the value previously obtained by \citet{RodriguezBaras2021}.  \citet{RodriguezBaras2023} did not detect H$_2^{34}$S toward C16. Even taking into account an uncertainty of a factor of $\sim$3 in the H$_2$S abundance, model predictions remain higher than the observed values. The cases of C$_2$S and C$_3$O  cannot be attributed to opacity effects, according to the observed CCS/CC$^{34}$S and CCCS/CCC$^{34}$S column density ratios measured in this work. The poor agreement of the model predictions with observations is more likely related to uncertainties in the chemistry of these species. Our data suggests that although our state-of-the-art chemical models are doing a good job at predicting some COMs and nitrogen-bearing species, the description of sulfur chemistry and especially the formation of the sulfur and oxygen carbon chains needs to be improved. 

To test the robustness of our modeling, we repeated the fitting assuming different chemical times. As is shown in Fig. \ref{Fig: best t values}, the best-fitting models for C2 reach $D_{diff}$ values within a 1$\%$ difference with respect to $D_{diff, \ min}$ for $t \sim 0.1$ Myr and remain within a 5 $\%$ difference for $t$ values in the interval $0.10-0.17$ Myr. For C16, we find larger variation, with the best-fitting models reaching $D_{diff}$ values within  a 1$\%$ difference for $t \sim 0.17$ Myr, and up to a 5$\%$ difference for $t$ values in the interval $0.13-1$ Myr. To evaluate the uncertainties that fixing the time can introduce in the fitting of C16, we repeated it for t= 1 Myr. We found that the main difference between the different times is a decrease in density for longer times. Specifically, we obtain  a density about 3 times lower, $n_{H} \sim 4.0 \times 10^{4}$ cm$^{-3}$, adopting t = 1 Myr.
The variations in the rest of parameters are within the uncertainties. \citet{Tasa2025} estimated densities of n$_{\rm H} \sim$ 2.0 $\times$10$^5$ cm$^{-3}$ toward this core based on a multi-transition analysis of DNC. This result provides further support for our best-fitting model in Table~\ref{Table: Best models} being the more accurate one with the information available until now.

Chemical differences between starless cores can be due to different evolutionary stages, environmental conditions, and also the cloud history. Our fitting results show that C2 is best reproduced by $n_{H}=(1.94-1.99) \times 10^{5}$ cm$^{-3}$, while C16 is best-fit by $n_{H}=(1.29-1.69) \times 10^{5}$ cm$^{-3}$. These higher molecular hydrogen densities in C2 with respect to C16 are consistent with C2 being a denser, more evolved object, closer to the pre-stellar core phase. Regarding the sulfur elemental abundance, it also hints at this possibility, as C2 presents a higher sulfur depletion, which is suggestive of a more advanced evolutionary stage in agreement with \citet{Hily2022}.
Despite being embedded in a harsher environment, with a higher number of low-mass stars located nearby, C2 seems better fit by low cosmic-ray ionization rates ($\zeta_{H_{2}}=(1.33-1.63) \times 10^{-17}$ s$^{-1}$). This could also be related to a higher density in C2, as $\zeta_ {H_{2}}$ is attenuated by an increasing gas column density \citep{Padovani2009}. 

All physical and chemical indicators point to a more evolved stage for C2 than for C16, consistent with \citet{Esplugues2022}. Based on our chemical model, we cannot constrain the environmental UV field since the impinging  radiation is strongly attenuated at high optical depths. The chemical age derived from our observations is similar in the two starless cores, 
contrary to the results based on other evolutionary indicators.  Our 0D chemical model does not account for physical and chemical gradients along the line of sight. It does not account for either the dynamical evolution or turbulence effects, which could affect molecular abundances, especially of sulfurated species \citep{Beitia2024}. More sophisticated models accounting for chemistry and dynamics in a consistent way should be used to gain further insights into the accreting and collapsing history of this complex star-forming filament and the cores within it.

\section{Summary and conclusions}
\label{Sec: Summary and conclusions}

We have carried out a $\lambda= 7$ mm line survey in the 31.3-50.6 GHz frequency range for the starless cores B213-C2 and B213-C16, using the Yebes 40-m telescope. We detected 122 emission lines, 56 of them corresponding to B213-C2 and the remaining 66 to B213-C16. Including isotopologs, 22 species were detected in B213-C2, while 25 were detected in B213-C16. The observed molecules include O-bearing molecules (CH$_{3}$CHO, CH$_{3}$OH, HOCO$^{+}$), hydrocarbons (CH$_{3}$CCH, c-C$_{3}$H$_{2}$, l-C$_{3}$H, l-C$_{3}$H$_{2}$), N-bearing molecules (CH$_{3}$CN, HC$_{5}$N, HC$_{7}$N, CH$_{2}$CHCN), and S-bearing molecules (CS, H$_{2}$CS, HCNS, HSCN, HCS$^{+}$, OCS, SO$_{2}$, CCS, C$_{3}$S, $^{13}$CS, C$^{34}$S, C$^{33}$S, CC$^{34}$S, CCC$^{34}$S).

To determine column densities, we used the rotational diagram method, deriving rotational temperatures mostly in the range of $5 \lesssim T_{rot} \lesssim 9 \ \text{K}$, consistent with emission arising from a cold source. In general, fractional molecular abundances in B213-C2 are lower than in B213-C16. This could be the result of environmental factors such as C2 presenting a higher exposure to UV radiation from nearby low-mass stars, or a rejuvenation of the chemistry in C16 due to the accretion of background molecular gas. The lower molecular abundances in C2 could also stem from the evolutionary state of the cores, with C2 constituting a denser, more evolved object closer to the pre-stellar phase. A higher density would cause the depletion of some molecules, which would condense out onto dust grain surfaces.

The molecular abundances in both cores were also compared with those of five reference sources (TMC-1, $\text{B}\,335$, B1-b, L1527, and L483) with different evolutionary stages. TMC-1 presents the highest abundances of all the compared sources. This is to be expected, as TMC-1 has been shown to be a particularly molecule-rich dark cloud. The source $\text{B}\,335$ presents a clear chemical differentiation with respect to C2 and C16, with considerably higher abundances of sulfur carbon chains. This is probably due to $\text{B}\,335$ being especially rich in this type of molecule, unlike other Class 0 sources. Hydrocarbons and oxygen-bearing molecules in B1-b are of the same order as in C2 and C16. However, marked chemical differentiation is observed for S-bearing molecules. This difference could be explained by the varying degrees of sulfur depletion in this source, with B1-b presenting moderate sulfur depletion ($D_{S}=25$). In general, L1527 presents lower molecular abundances than C2 and C16, while abundances in L483 are overabundant with respect to C2 and underabundant compared to C16.

The comparison with chemical models shows that both cores are best fit with $t \sim 0.1$ Myr. However, the parameters corresponding to the best-fitting models hint at C2 being at a more advanced evolutionary stage, as it presents a higher molecular hydrogen density and sulfur depletion than C16, and a lower cosmic-ray ionization rate. More  complex simulations, coupling dynamics and chemistry,  are required to describe  in detail the evolution of a dense core during its collapse.

\section*{Data availability}
\label{Sec: Data availability}
The Gaussian fits obtained for both sources can be found at \href{https://zenodo.org/records/18076841}{Zenodo}.

  \begin{acknowledgements}
This project has received funding from the European Research Council (ERC) under the European Union$’$s Horizon Europe research and innovation programme ERC-AdG-2022 (GA No. 101096293). Views and opinions expressed are however those of the author(s) only and do not necessarily reflect those of the European Union or the European Research Council Executive Agency. Neither the European Union nor the granting authority can be held responsible for them. AF, MRB, GE, PRM and BT also thank project PID2022-137980NB-I00 funded by the Spanish Ministry of Science and Innovation/State Agency of Research MCIN/AEI/10.13039/501100011033 and by “ERDF A way of making Europe.” BT also thanks project PID2023-147545NB-I00. AAR acknowledges funding from the Agencia Estatal de Investigación del Ministerio de Ciencia, Innovación y Universidades (MCIU/AEI) under grant ``Polarimetric Inference of Magnetic Fields'' and the European Regional
Development Fund (ERDF) with reference PID2022-136563NB-I00/10.13039/501100011033. The authors thank the anonymous referee for their valuable suggestions.

\end{acknowledgements}  
\bibliography{references}

@ARTICLE{Fuente2019,
       author = {{Fuente}, A. and {Navarro}, D.~G. and {Caselli}, P. and {Gerin}, M. and {Kramer}, C. and {Roueff}, E. and {Alonso-Albi}, T. and {Bachiller}, R. and {Cazaux}, S. and {Commercon}, B. and {Friesen}, R. and {Garc{\'\i}a-Burillo}, S. and {Giuliano}, B.~M. and {Goicoechea}, J.~R. and {Gratier}, P. and {Hacar}, A. and {Jim{\'e}nez-Serra}, I. and {Kirk}, J. and {Lattanzi}, V. and {Loison}, J.~C. and {Malinen}, J. and {Marcelino}, N. and {Mart{\'\i}n-Dom{\'e}nech}, R. and {Mu{\~n}oz-Caro}, G. and {Pineda}, J. and {Tafalla}, M. and {Tercero}, B. and {Ward-Thompson}, D. and {Trevi{\~n}o-Morales}, S.~P. and {Rivi{\'e}re-Marichalar}, P. and {Roncero}, O. and {Vidal}, T. and {Ballester}, M.~Y.},
        title = "{Gas phase Elemental abundances in Molecular cloudS (GEMS). I. The prototypical dark cloud TMC 1}",
      journal = {\aap},
         year = 2019,
        month = apr,
       volume = {624},
          eid = {A105},
        pages = {A105},
          doi = {10.1051/0004-6361/201834654},
archivePrefix = {arXiv},
       eprint = {1809.04978},
 primaryClass = {astro-ph.GA},
       adsurl = {https://ui.adsabs.harvard.edu/abs/2019A&A...624A.105F}
}

@ARTICLE{Shu1987,
       author = {{Shu}, Frank H. and {Adams}, Fred C. and {Lizano}, Susana},
        title = "{Star formation in molecular clouds: observation and theory.}",
      journal = {\araa},
         year = 1987,
        month = jan,
       volume = {25},
        pages = {23-81},
          doi = {10.1146/annurev.aa.25.090187.000323},
       adsurl = {https://ui.adsabs.harvard.edu/abs/1987ARA&A..25...23S}
}

@ARTICLE{Bergin2007,
       author = {{Bergin}, Edwin A. and {Tafalla}, Mario},
        title = "{Cold Dark Clouds: The Initial Conditions for Star Formation}",
      journal = {\araa},
         year = 2007,
        month = sep,
       volume = {45},
       number = {1},
        pages = {339-396},
          doi = {10.1146/annurev.astro.45.071206.100404},
archivePrefix = {arXiv},
       eprint = {0705.3765},
 primaryClass = {astro-ph},
       adsurl = {https://ui.adsabs.harvard.edu/abs/2007ARA&A..45..339B}
}

@ARTICLE{Caselli1998,
       author = {{Caselli}, P. and {Walmsley}, C.~M. and {Terzieva}, R. and {Herbst}, Eric},
        title = "{The Ionization Fraction in Dense Cloud Cores}",
      journal = {\apj},
         year = 1998,
        month = may,
       volume = {499},
       number = {1},
        pages = {234-249},
          doi = {10.1086/305624},
       adsurl = {https://ui.adsabs.harvard.edu/abs/1998ApJ...499..234C}
}

@ARTICLE{Zhao2021,
       author = {{Zhao}, Bo and {Caselli}, Paola and {Li}, Zhi-Yun and {Krasnopolsky}, Ruben and {Shang}, Hsien and {Lam}, Ka Ho},
        title = "{The interplay between ambipolar diffusion and Hall effect on magnetic field decoupling and protostellar disc formation}",
      journal = {\mnras},
         year = 2021,
        month = aug,
       volume = {505},
       number = {4},
        pages = {5142-5163},
          doi = {10.1093/mnras/stab1295},
archivePrefix = {arXiv},
       eprint = {2009.07820},
 primaryClass = {astro-ph.SR},
       adsurl = {https://ui.adsabs.harvard.edu/abs/2021MNRAS.505.5142Z}
}

@ARTICLE{Padovani2013,
       author = {{Padovani}, M. and {Hennebelle}, P. and {Galli}, D.},
        title = "{Cosmic-ray ionisation in collapsing clouds}",
      journal = {\aap},
         year = 2013,
        month = dec,
       volume = {560},
          eid = {A114},
        pages = {A114},
          doi = {10.1051/0004-6361/201322407},
archivePrefix = {arXiv},
       eprint = {1310.2158},
 primaryClass = {astro-ph.GA},
       adsurl = {https://ui.adsabs.harvard.edu/abs/2013A&A...560A.114P}
}

@ARTICLE{Agundez2019,
       author = {{Ag{\'u}ndez}, M. and {Marcelino}, N. and {Cernicharo}, J. and
         {Roueff}, E. and {Tafalla}, M.},
        title = "{A sensitive {\ensuremath{\lambda}} 3 mm line survey of L483. A broad view of the chemical composition of a core around a Class 0 object}",
      journal = {\aap},
         year = 2019,
        month = may,
       volume = {625},
          eid = {A147},
        pages = {A147},
          doi = {10.1051/0004-6361/201935164},
archivePrefix = {arXiv},
       eprint = {1904.06565},
 primaryClass = {astro-ph.GA},
       adsurl = {https://ui.adsabs.harvard.edu/abs/2019A&A...625A.147A}
}

@ARTICLE{2005JMoSt.742..215M,
       author = {{M{\"u}ller}, Holger S.~P. and {Schl{\"o}der}, Frank and {Stutzki}, J{\"u}rgen and {Winnewisser}, Gisbert},
        title = "{The Cologne Database for Molecular Spectroscopy, CDMS: a useful tool for astronomers and spectroscopists}",
      journal = {Journal of Molecular Structure},
         year = 2005,
        month = may,
       volume = {742},
       number = {1-3},
        pages = {215-227},
          doi = {10.1016/j.molstruc.2005.01.027},
       adsurl = {https://ui.adsabs.harvard.edu/abs/2005JMoSt.742..215M}
}

@ARTICLE{1998JQSRT..60..883P,
       author = {{Pickett}, H.~M. and {Poynter}, R.~L. and {Cohen}, E.~A. and {Delitsky}, M.~L. and {Pearson}, J.~C. and {M{\"u}ller}, H.~S.~P.},
        title = "{Submillimeter, millimeter and microwave spectral line catalog.}",
      journal = {\jqsrt},
         year = 1998,
        month = nov,
       volume = {60},
       number = {5},
        pages = {883-890},
          doi = {10.1016/S0022-4073(98)00091-0},
       adsurl = {https://ui.adsabs.harvard.edu/abs/1998JQSRT..60..883P}
}

@ARTICLE{2008ApJ...680..428G,
       author = {{Goldsmith}, Paul F. and {Heyer}, Mark and {Narayanan}, Gopal and {Snell}, Ronald and {Li}, Di and {Brunt}, Chris},
        title = "{Large-Scale Structure of the Molecular Gas in Taurus Revealed by High Linear Dynamic Range Spectral Line Mapping}",
      journal = {\apj},
         year = 2008,
        month = jun,
       volume = {680},
       number = {1},
        pages = {428-445},
          doi = {10.1086/587166},
archivePrefix = {arXiv},
       eprint = {0802.2206},
 primaryClass = {astro-ph},
       adsurl = {https://ui.adsabs.harvard.edu/abs/2008ApJ...680..428G}
}

@ARTICLE{2012ApJ...756...12L,
       author = {{Li}, Di and {Goldsmith}, Paul F.},
        title = "{Is the Taurus B213 Region a True Filament?: Observations of Multiple Cyanoacetylene Transitions}",
      journal = {\apj},
         year = 2012,
        month = sep,
       volume = {756},
       number = {1},
          eid = {12},
        pages = {12},
          doi = {10.1088/0004-637X/756/1/12},
archivePrefix = {arXiv},
       eprint = {1207.0044},
 primaryClass = {astro-ph.GA},
       adsurl = {https://ui.adsabs.harvard.edu/abs/2012ApJ...756...12L}
}

@ARTICLE{Hacar,
       author = {{Hacar}, A. and {Tafalla}, M. and {Kauffmann}, J. and {Kov{\'a}cs}, A.},
        title = "{Cores, filaments, and bundles: hierarchical core formation in the L1495/B213 Taurus region}",
      journal = {\aap},
         year = 2013,
        month = jun,
       volume = {554},
          eid = {A55},
        pages = {A55},
          doi = {10.1051/0004-6361/201220090},
archivePrefix = {arXiv},
       eprint = {1303.2118},
 primaryClass = {astro-ph.GA},
       adsurl = {https://ui.adsabs.harvard.edu/abs/2013A&A...554A..55H}
}

@ARTICLE{1986A&A...164..349D,
       author = {{Duvert}, G. and {Cernicharo}, J. and {Baudry}, A.},
        title = "{A molecular survey of three dark clouds in Taurus.}",
      journal = {\aap},
         year = 1986,
        month = aug,
       volume = {164},
        pages = {349-357},
       adsurl = {https://ui.adsabs.harvard.edu/abs/1986A&A...164..349D}
}

@ARTICLE{2023A&A...678A.199E,
       author = {{Esplugues}, G. and {Rodr{\'\i}guez-Baras}, M. and {San Andr{\'e}s}, D. and {Navarro-Almaida}, D. and {Fuente}, A. and {Rivi{\`e}re-Marichalar}, P. and {S{\'a}nchez-Monge}, {\'A}. and {Drozdovskaya}, M.~N. and {Spezzano}, S. and {Caselli}, P.},
        title = "{Evolution of Chemistry in the envelope of HOt corinoS (ECHOS). I. Extremely young sulphur chemistry in the isolated Class 0 object B 335}",
      journal = {\aap},
         year = 2023,
        month = oct,
       volume = {678},
          eid = {A199},
        pages = {A199},
          doi = {10.1051/0004-6361/202346721},
archivePrefix = {arXiv},
       eprint = {2309.01713},
 primaryClass = {astro-ph.GA},
       adsurl = {https://ui.adsabs.harvard.edu/abs/2023A&A...678A.199E}
}

@ARTICLE{1999ApJ...517..209G,
       author = {{Goldsmith}, Paul F. and {Langer}, William D.},
        title = "{Population Diagram Analysis of Molecular Line Emission}",
      journal = {\apj},
         year = 1999,
        month = may,
       volume = {517},
       number = {1},
        pages = {209-225},
          doi = {10.1086/307195},
       adsurl = {https://ui.adsabs.harvard.edu/abs/1999ApJ...517..209G}
}

@ARTICLE{Palmeirim,
       author = {{Palmeirim}, P. and {Andr{\'e}}, Ph. and {Kirk}, J. and {Ward-Thompson}, D. and {Arzoumanian}, D. and {K{\"o}nyves}, V. and {Didelon}, P. and {Schneider}, N. and {Benedettini}, M. and {Bontemps}, S. and {Di Francesco}, J. and {Elia}, D. and {Griffin}, M. and {Hennemann}, M. and {Hill}, T. and {Martin}, P.~G. and {Men'shchikov}, A. and {Molinari}, S. and {Motte}, F. and {Nguyen Luong}, Q. and {Nutter}, D. and {Peretto}, N. and {Pezzuto}, S. and {Roy}, A. and {Rygl}, K.~L.~J. and {Spinoglio}, L. and {White}, G.~L.},
        title = "{Herschel view of the Taurus B211/3 filament and striations: evidence of filamentary growth?}",
      journal = {\aap},
         year = 2013,
        month = feb,
       volume = {550},
          eid = {A38},
        pages = {A38},
          doi = {10.1051/0004-6361/201220500},
archivePrefix = {arXiv},
       eprint = {1211.6360},
 primaryClass = {astro-ph.SR},
       adsurl = {https://ui.adsabs.harvard.edu/abs/2013A&A...550A..38P}
}

@ARTICLE{2010A&A...518L.102A,
       author = {{Andr{\'e}}, Ph. and {Men'shchikov}, A. and {Bontemps}, S. and {K{\"o}nyves}, V. and {Motte}, F. and {Schneider}, N. and {Didelon}, P. and {Minier}, V. and {Saraceno}, P. and {Ward-Thompson}, D. and {di Francesco}, J. and {White}, G. and {Molinari}, S. and {Testi}, L. and {Abergel}, A. and {Griffin}, M. and {Henning}, Th. and {Royer}, P. and {Mer{\'\i}n}, B. and {Vavrek}, R. and {Attard}, M. and {Arzoumanian}, D. and {Wilson}, C.~D. and {Ade}, P. and {Aussel}, H. and {Baluteau}, J. -P. and {Benedettini}, M. and {Bernard}, J. -Ph. and {Blommaert}, J.~A.~D.~L. and {Cambr{\'e}sy}, L. and {Cox}, P. and {di Giorgio}, A. and {Hargrave}, P. and {Hennemann}, M. and {Huang}, M. and {Kirk}, J. and {Krause}, O. and {Launhardt}, R. and {Leeks}, S. and {Le Pennec}, J. and {Li}, J.~Z. and {Martin}, P.~G. and {Maury}, A. and {Olofsson}, G. and {Omont}, A. and {Peretto}, N. and {Pezzuto}, S. and {Prusti}, T. and {Roussel}, H. and {Russeil}, D. and {Sauvage}, M. and {Sibthorpe}, B. and {Sicilia-Aguilar}, A. and {Spinoglio}, L. and {Waelkens}, C. and {Woodcraft}, A. and {Zavagno}, A.},
        title = "{From filamentary clouds to prestellar cores to the stellar IMF: Initial highlights from the Herschel Gould Belt Survey}",
      journal = {\aap},
         year = 2010,
        month = jul,
       volume = {518},
          eid = {L102},
        pages = {L102},
          doi = {10.1051/0004-6361/201014666},
archivePrefix = {arXiv},
       eprint = {1005.2618},
 primaryClass = {astro-ph.GA},
       adsurl = {https://ui.adsabs.harvard.edu/abs/2010A&A...518L.102A}
}

@ARTICLE{RodriguezBaras2021,
       author = {{Rodr{\'\i}guez-Baras}, M. and {Fuente}, A. and {Rivi{\'e}re-Marichalar}, P. and {Navarro-Almaida}, D. and {Caselli}, P. and {Gerin}, M. and {Kramer}, C. and {Roueff}, E. and {Wakelam}, V. and {Esplugues}, G. and {Garc{\'\i}a-Burillo}, S. and {Le Gal}, R. and {Spezzano}, S. and {Alonso-Albi}, T. and {Bachiller}, R. and {Cazaux}, S. and {Commercon}, B. and {Goicoechea}, J.~R. and {Loison}, J.~C. and {Trevi{\~n}o-Morales}, S.~P. and {Roncero}, O. and {Jim{\'e}nez-Serra}, I. and {Laas}, J. and {Hacar}, A. and {Kirk}, J. and {Lattanzi}, V. and {Mart{\'\i}n-Dom{\'e}nech}, R. and {Mu{\~n}oz-Caro}, G. and {Pineda}, J.~E. and {Tercero}, B. and {Ward-Thompson}, D. and {Tafalla}, M. and {Marcelino}, N. and {Malinen}, J. and {Friesen}, R. and {Giuliano}, B.~M.},
        title = "{Gas phase Elemental abundances in Molecular cloudS (GEMS). IV. Observational results and statistical trends}",
      journal = {\aap},
         year = 2021,
        month = apr,
       volume = {648},
          eid = {A120},
        pages = {A120},
          doi = {10.1051/0004-6361/202040112},
archivePrefix = {arXiv},
       eprint = {2102.13153},
 primaryClass = {astro-ph.GA},
       adsurl = {https://ui.adsabs.harvard.edu/abs/2021A&A...648A.120R}
}

@ARTICLE{2010A&A...518L..88B,
       author = {{Bernard}, J. -Ph. and {Paradis}, D. and {Marshall}, D.~J. and {Montier}, L. and {Lagache}, G. and {Paladini}, R. and {Veneziani}, M. and {Brunt}, C.~M. and {Mottram}, J.~C. and {Martin}, P. and {Ristorcelli}, I. and {Noriega-Crespo}, A. and {Compi{\`e}gne}, M. and {Flagey}, N. and {Anderson}, L.~D. and {Popescu}, C.~C. and {Tuffs}, R. and {Reach}, W. and {White}, G. and {Benedettini}, M. and {Calzoletti}, L. and {Digiorgio}, A.~M. and {Faustini}, F. and {Juvela}, M. and {Joblin}, C. and {Joncas}, G. and {Mivilles-Deschenes}, M. -A. and {Olmi}, L. and {Traficante}, A. and {Piacentini}, F. and {Zavagno}, A. and {Molinari}, S.},
        title = "{Dust temperature tracing the ISRF intensity in the Galaxy}",
      journal = {\aap},
         year = 2010,
        month = jul,
       volume = {518},
          eid = {L88},
        pages = {L88},
          doi = {10.1051/0004-6361/201014540},
       adsurl = {https://ui.adsabs.harvard.edu/abs/2010A&A...518L..88B}
}

@ARTICLE{Shimajiri,
       author = {{Shimajiri}, Y. and {Andr{\'e}}, Ph. and {Palmeirim}, P. and {Arzoumanian}, D. and {Bracco}, A. and {K{\"o}nyves}, V. and {Ntormousi}, E. and {Ladjelate}, B.},
        title = "{Probing accretion of ambient cloud material into the Taurus B211/B213 filament}",
      journal = {\aap},
         year = 2019,
        month = mar,
       volume = {623},
          eid = {A16},
        pages = {A16},
          doi = {10.1051/0004-6361/201834399},
archivePrefix = {arXiv},
       eprint = {1811.06240},
 primaryClass = {astro-ph.GA},
       adsurl = {https://ui.adsabs.harvard.edu/abs/2019A&A...623A..16S}
}

@ARTICLE{Yan,
       author = {{Yan}, Qing-Zeng and {Zhang}, Bo and {Xu}, Ye and {Guo}, Sufen and {Macquart}, Jean-Pierre and {Tang}, Zheng-Hong and {Walsh}, Andrew John},
        title = "{Distances to molecular clouds at high galactic latitudes based on Gaia DR2}",
      journal = {\aap},
         year = 2019,
        month = apr,
       volume = {624},
          eid = {A6},
        pages = {A6},
          doi = {10.1051/0004-6361/201834337},
archivePrefix = {arXiv},
       eprint = {1902.02052},
 primaryClass = {astro-ph.GA},
       adsurl = {https://ui.adsabs.harvard.edu/abs/2019A&A...624A...6Y}
}

@ARTICLE{Spezzano,
       author = {{Spezzano}, S. and {Fuente}, A. and {Caselli}, P. and {Vasyunin}, A. and {Navarro-Almaida}, D. and {Rodr{\'\i}guez-Baras}, M. and {Punanova}, A. and {Vastel}, C. and {Wakelam}, V.},
        title = "{Gas phase Elemental abundances in Molecular cloudS (GEMS) V. Methanol in Taurus}",
      journal = {\aap},
         year = 2022,
        month = jan,
       volume = {657},
          eid = {A10},
        pages = {A10},
          doi = {10.1051/0004-6361/202141971},
archivePrefix = {arXiv},
       eprint = {2110.01675},
 primaryClass = {astro-ph.GA},
       adsurl = {https://ui.adsabs.harvard.edu/abs/2022A&A...657A..10S}
}

@ARTICLE{Fuente2023,
       author = {{Fuente}, A. and {Rivi{\`e}re-Marichalar}, P. and {Beitia-Antero}, L. and {Caselli}, P. and {Wakelam}, V. and {Esplugues}, G. and {Rodr{\'\i}guez-Baras}, M. and {Navarro-Almaida}, D. and {Gerin}, M. and {Kramer}, C. and {Bachiller}, R. and {Goicoechea}, J.~R. and {Jim{\'e}nez-Serra}, I. and {Loison}, J.~C. and {Ivlev}, A. and {Mart{\'\i}n-Dom{\'e}nech}, R. and {Spezzano}, S. and {Roncero}, O. and {Mu{\~n}oz-Caro}, G. and {Cazaux}, S. and {Marcelino}, N.},
        title = "{Gas phase Elemental abundances in Molecular cloudS (GEMS). VII. Sulfur elemental abundance}",
      journal = {\aap},
         year = 2023,
        month = feb,
       volume = {670},
          eid = {A114},
        pages = {A114},
          doi = {10.1051/0004-6361/202244843},
archivePrefix = {arXiv},
       eprint = {2212.03742},
 primaryClass = {astro-ph.GA},
       adsurl = {https://ui.adsabs.harvard.edu/abs/2023A&A...670A.114F}
}

@ARTICLE{TafallaHacar2015,
       author = {{Tafalla}, M. and {Hacar}, A.},
        title = "{Chains of dense cores in the Taurus L1495/B213 complex}",
      journal = {\aap},
         year = 2015,
        month = feb,
       volume = {574},
          eid = {A104},
        pages = {A104},
          doi = {10.1051/0004-6361/201424576},
archivePrefix = {arXiv},
       eprint = {1412.1083},
 primaryClass = {astro-ph.GA},
       adsurl = {https://ui.adsabs.harvard.edu/abs/2015A&A...574A.104T}
}

@ARTICLE{Ungerechts,
       author = {{Ungerechts}, H. and {Thaddeus}, P.},
        title = "{A CO Survey of the Dark Nebulae in Perseus, Taurus, and Auriga}",
      journal = {\apjs},
         year = 1987,
        month = mar,
       volume = {63},
        pages = {645},
          doi = {10.1086/191176},
       adsurl = {https://ui.adsabs.harvard.edu/abs/1987ApJS...63..645U}
}

@ARTICLE{Mizuno,
       author = {{Mizuno}, A. and {Onishi}, T. and {Yonekura}, Y. and {Nagahama}, T. and {Ogawa}, H. and {Fukui}, Y.},
        title = "{Overall Distribution of Dense Molecular Gas and Star Formation in the Taurus Cloud Complex}",
      journal = {\apjl},
         year = 1995,
        month = jun,
       volume = {445},
        pages = {L161},
          doi = {10.1086/187914},
       adsurl = {https://ui.adsabs.harvard.edu/abs/1995ApJ...445L.161M}
}

@ARTICLE{Cernicharo1987,
       author = {{Cernicharo}, J. and {Guelin}, M.},
        title = "{The physical and chemical state of HCL2.}",
      journal = {\aap},
         year = 1987,
        month = apr,
       volume = {176},
        pages = {299-316},
       adsurl = {https://ui.adsabs.harvard.edu/abs/1987A&A...176..299C}
}

@ARTICLE{Onishi1996,
       author = {{Onishi}, Toshikazu and {Mizuno}, Akira and {Kawamura}, Akiko and {Ogawa}, Hideo and {Fukui}, Yasuo},
        title = "{A C 18O Survey of Dense Cloud Cores in Taurus: Core Properties}",
      journal = {\apj},
         year = 1996,
        month = jul,
       volume = {465},
        pages = {815},
          doi = {10.1086/177465},
       adsurl = {https://ui.adsabs.harvard.edu/abs/1996ApJ...465..815O}
}

@ARTICLE{Narayanan2008,
       author = {{Narayanan}, Gopal and {Heyer}, Mark H. and {Brunt}, Christopher and {Goldsmith}, Paul F. and {Snell}, Ronald and {Li}, Di},
        title = "{The Five College Radio Astronomy Observatory CO Mapping Survey of the Taurus Molecular Cloud}",
      journal = {\apjs},
         year = 2008,
        month = jul,
       volume = {177},
       number = {1},
        pages = {341-361},
          doi = {10.1086/587786},
archivePrefix = {arXiv},
       eprint = {0802.2556},
 primaryClass = {astro-ph},
       adsurl = {https://ui.adsabs.harvard.edu/abs/2008ApJS..177..341N}
}

@ARTICLE{Cambresy1999,
       author = {{Cambr{\'e}sy}, L.},
        title = "{Mapping of the extinction in giant molecular clouds using optical star counts}",
      journal = {\aap},
         year = 1999,
        month = may,
       volume = {345},
        pages = {965-976},
          doi = {10.48550/arXiv.astro-ph/9903149},
archivePrefix = {arXiv},
       eprint = {astro-ph/9903149},
 primaryClass = {astro-ph},
       adsurl = {https://ui.adsabs.harvard.edu/abs/1999A&A...345..965C}
}

@ARTICLE{Padoan2002,
       author = {{Padoan}, Paolo and {Cambr{\'e}sy}, Laurent and {Langer}, William},
        title = "{Structure Function Scaling of a 2MASS Extinction Map of Taurus}",
      journal = {\apjl},
         year = 2002,
        month = nov,
       volume = {580},
       number = {1},
        pages = {L57-L60},
          doi = {10.1086/345403},
archivePrefix = {arXiv},
       eprint = {astro-ph/0208217},
 primaryClass = {astro-ph},
       adsurl = {https://ui.adsabs.harvard.edu/abs/2002ApJ...580L..57P}
}

@ARTICLE{Schmalzl2010,
       author = {{Schmalzl}, Markus and {Kainulainen}, Jouni and {Quanz}, Sascha P. and {Alves}, Jo{\~a}o and {Goodman}, Alyssa A. and {Henning}, Thomas and {Launhardt}, Ralf and {Pineda}, Jaime E. and {Rom{\'a}n-Z{\'u}{\~n}iga}, Carlos G.},
        title = "{Star Formation in the Taurus Filament L 1495: From Dense Cores to Stars}",
      journal = {\apj},
         year = 2010,
        month = dec,
       volume = {725},
       number = {1},
        pages = {1327-1336},
          doi = {10.1088/0004-637X/725/1/1327},
archivePrefix = {arXiv},
       eprint = {1010.2755},
 primaryClass = {astro-ph.GA},
       adsurl = {https://ui.adsabs.harvard.edu/abs/2010ApJ...725.1327S}
}

@ARTICLE{Yoshida2019,
       author = {{Yoshida}, Kento and {Sakai}, Nami and {Nishimura}, Yuri and {Tokudome}, Tomoya and {Watanabe}, Yoshimasa and {Sakai}, Takeshi and {Takano}, Shuro and {Yamamoto}, Satoshi},
        title = "{An unbiased spectral line survey observation toward the low-mass star-forming region L1527}",
      journal = {\pasj},
         year = 2019,
        month = dec,
       volume = {71},
          eid = {S18},
        pages = {S18},
          doi = {10.1093/pasj/psy136},
archivePrefix = {arXiv},
       eprint = {1901.06546},
 primaryClass = {astro-ph.GA},
       adsurl = {https://ui.adsabs.harvard.edu/abs/2019PASJ...71S..18Y}
}

@ARTICLE{Keene1980,
       author = {{Keene}, J. and {Hildebrand}, R.~H. and {Whitcomb}, S.~E. and {Harper}, D.~A.},
        title = "{Far-infrared observations of the globule B335}",
      journal = {\apjl},
         year = 1980,
        month = aug,
       volume = {240},
        pages = {L43-L46},
          doi = {10.1086/183320},
       adsurl = {https://ui.adsabs.harvard.edu/abs/1980ApJ...240L..43K}
}

@ARTICLE{Keene1983,
       author = {{Keene}, J. and {Davidson}, J.~A. and {Harper}, D.~A. and {Hildebrand}, R.~H. and {Jaffe}, D.~T. and {Loewenstein}, R.~F. and {Low}, F.~J. and {Pernic}, R.},
        title = "{Far-infrared detection of low-luminosity star formation in the BOK globule B 335.}",
      journal = {\apjl},
         year = 1983,
        month = nov,
       volume = {274},
        pages = {L43-L47},
          doi = {10.1086/184147},
       adsurl = {https://ui.adsabs.harvard.edu/abs/1983ApJ...274L..43K}
}

@ARTICLE{Huang2013,
       author = {{Huang}, Yun-Hsin and {Hirano}, Naomi},
        title = "{Probing the Earliest Stage of Protostellar Evolution{\textemdash}Barnard 1-bN and Barnard 1-bS}",
      journal = {\apj},
         year = 2013,
        month = apr,
       volume = {766},
       number = {2},
          eid = {131},
        pages = {131},
          doi = {10.1088/0004-637X/766/2/131},
archivePrefix = {arXiv},
       eprint = {1302.6053},
 primaryClass = {astro-ph.SR},
       adsurl = {https://ui.adsabs.harvard.edu/abs/2013ApJ...766..131H}
}

@ARTICLE{Kaifu2004,
       author = {{Kaifu}, Norio and {Ohishi}, Masatoshi and {Kawaguchi}, Kentarou and {Saito}, Shuji and {Yamamoto}, Satoshi and {Miyaji}, Takeshi and {Miyazawa}, Keisuke and {Ishikawa}, Shin-Ichi and {Noumaru}, Chiaki and {Harasawa}, Sumiko and {Okuda}, Michiko and {Suzuki}, Hiroko},
        title = "{A 8.8--50GHz Complete Spectral Line Survey toward TMC-1 I. Survey Data}",
      journal = {\pasj},
         year = 2004,
        month = feb,
       volume = {56},
        pages = {69-173},
          doi = {10.1093/pasj/56.1.69},
       adsurl = {https://ui.adsabs.harvard.edu/abs/2004PASJ...56...69K}
}

@ARTICLE{Gratier2016,
       author = {{Gratier}, P. and {Majumdar}, L. and {Ohishi}, M. and {Roueff}, E. and {Loison}, J.~C. and {Hickson}, K.~M. and {Wakelam}, V.},
        title = "{A New Reference Chemical Composition for TMC-1}",
      journal = {\apjs},
         year = 2016,
        month = aug,
       volume = {225},
       number = {2},
          eid = {25},
        pages = {25},
          doi = {10.3847/0067-0049/225/2/25},
archivePrefix = {arXiv},
       eprint = {1610.00524},
 primaryClass = {astro-ph.GA},
       adsurl = {https://ui.adsabs.harvard.edu/abs/2016ApJS..225...25G}
}

@ARTICLE{Fuente2016,
       author = {{Fuente}, A. and {Cernicharo}, J. and {Roueff}, E. and {Gerin}, M. and {Pety}, J. and {Marcelino}, N. and {Bachiller}, R. and {Lefloch}, B. and {Roncero}, O. and {Aguado}, A.},
        title = "{Ionization fraction and the enhanced sulfur chemistry in Barnard 1}",
      journal = {\aap},
         year = 2016,
        month = sep,
       volume = {593},
          eid = {A94},
        pages = {A94},
          doi = {10.1051/0004-6361/201628285},
archivePrefix = {arXiv},
       eprint = {1605.04724},
 primaryClass = {astro-ph.GA},
       adsurl = {https://ui.adsabs.harvard.edu/abs/2016A&A...593A..94F}
}

@ARTICLE{Wakelam2013,
       author = {{Ag{\'u}ndez}, Marcelino and {Wakelam}, Valentine},
        title = "{Chemistry of Dark Clouds: Databases, Networks, and Models}",
      journal = {Chemical Reviews},
         year = 2013,
        month = dec,
       volume = {113},
       number = {12},
        pages = {8710-8737},
          doi = {10.1021/cr4001176},
archivePrefix = {arXiv},
       eprint = {1310.3651},
 primaryClass = {astro-ph.GA},
       adsurl = {https://ui.adsabs.harvard.edu/abs/2013ChRv..113.8710A}
}

@ARTICLE{Cernicharo2024,
       author = {{Cernicharo}, J. and {Ag{\'u}ndez}, M. and {Cabezas}, C. and {Tercero}, B. and {Fuentetaja}, R. and {Marcelino}, N. and {de Vicente}, P.},
        title = "{Discovery of thiofulminic acid with the QUIJOTE line survey: A study of the isomers of HNCS and HNCO in TMC-1}",
      journal = {\aap},
         year = 2024,
        month = feb,
       volume = {682},
          eid = {L4},
        pages = {L4},
          doi = {10.1051/0004-6361/202349105},
archivePrefix = {arXiv},
       eprint = {2401.11785},
 primaryClass = {astro-ph.GA},
       adsurl = {https://ui.adsabs.harvard.edu/abs/2024A&A...682L...4C}
}

@ARTICLE{Loison2017,
       author = {{Loison}, Jean-Christophe and {Ag{\'u}ndez}, Marcelino and {Wakelam}, Valentine and {Roueff}, Evelyne and {Gratier}, Pierre and {Marcelino}, N{\'u}ria and {Reyes}, Dianailys Nu{\~n}ez and {Cernicharo}, Jos{\'e} and {Gerin}, Maryvonne},
        title = "{The interstellar chemistry of C$_{3}$H and C$_{3}$H$_{2}$ isomers}",
      journal = {\mnras},
         year = 2017,
        month = oct,
       volume = {470},
       number = {4},
        pages = {4075-4088},
          doi = {10.1093/mnras/stx1265},
archivePrefix = {arXiv},
       eprint = {1707.07926},
 primaryClass = {astro-ph.GA},
       adsurl = {https://ui.adsabs.harvard.edu/abs/2017MNRAS.470.4075L}
}

@ARTICLE{WidicusWeaver,
       author = {{Widicus Weaver}, Susanna L. and {Laas}, Jacob C. and {Zou}, Luyao and {Kroll}, Jay A. and {Rad}, Mary L. and {Hays}, Brian M. and {Sanders}, James L. and {Lis}, Dariusz C. and {Cross}, Trevor N. and {Wehres}, Nadine and {McGuire}, Brett A. and {Sumner}, Matthew C.},
        title = "{Deep, Broadband Spectral Line Surveys of Molecule-rich Interstellar Clouds}",
      journal = {\apjs},
         year = 2017,
        month = sep,
       volume = {232},
       number = {1},
          eid = {3},
        pages = {3},
          doi = {10.3847/1538-4365/aa8098},
       adsurl = {https://ui.adsabs.harvard.edu/abs/2017ApJS..232....3W}
}

@ARTICLE{Daniel2013,
       author = {{Daniel}, F. and {G{\'e}rin}, M. and {Roueff}, E. and {Cernicharo}, J. and {Marcelino}, N. and {Lique}, F. and {Lis}, D.~C. and {Teyssier}, D. and {Biver}, N. and {Bockel{\'e}e-Morvan}, D.},
        title = "{Nitrogen isotopic ratios in Barnard 1: a consistent study of the N$_{2}$H$^{+}$, NH$_{3}$, CN, HCN, and HNC isotopologues}",
      journal = {\aap},
         year = 2013,
        month = dec,
       volume = {560},
          eid = {A3},
        pages = {A3},
          doi = {10.1051/0004-6361/201321939},
archivePrefix = {arXiv},
       eprint = {1309.5782},
 primaryClass = {astro-ph.GA},
       adsurl = {https://ui.adsabs.harvard.edu/abs/2013A&A...560A...3D}
}

@ARTICLE{Cazaux2003,
       author = {{Cazaux}, S. and {Tielens}, A.~G.~G.~M. and {Ceccarelli}, C. and {Castets}, A. and {Wakelam}, V. and {Caux}, E. and {Parise}, B. and {Teyssier}, D.},
        title = "{The Hot Core around the Low-mass Protostar IRAS 16293-2422: Scoundrels Rule!}",
      journal = {\apjl},
         year = 2003,
        month = aug,
       volume = {593},
       number = {1},
        pages = {L51-L55},
          doi = {10.1086/378038},
       adsurl = {https://ui.adsabs.harvard.edu/abs/2003ApJ...593L..51C}
}

@ARTICLE{Imai2016,
       author = {{Imai}, Muneaki and {Sakai}, Nami and {Oya}, Yoko and {L{\'o}pez-Sepulcre}, Ana and {Watanabe}, Yoshimasa and {Ceccarelli}, Cecilia and {Lefloch}, Bertrand and {Caux}, Emmanuel and {Vastel}, Charlotte and {Kahane}, Claudine and {Sakai}, Takeshi and {Hirota}, Tomoya and {Aikawa}, Yuri and {Yamamoto}, Satoshi},
        title = "{Discovery of a Hot Corino in the Bok Globule B335}",
      journal = {\apjl},
         year = 2016,
        month = oct,
       volume = {830},
       number = {2},
          eid = {L37},
        pages = {L37},
          doi = {10.3847/2041-8205/830/2/L37},
archivePrefix = {arXiv},
       eprint = {1610.03942},
 primaryClass = {astro-ph.SR},
       adsurl = {https://ui.adsabs.harvard.edu/abs/2016ApJ...830L..37I}
}

@ARTICLE{Tafalla2000,
       author = {{Tafalla}, M. and {Myers}, P.~C. and {Mardones}, D. and {Bachiller}, R.},
        title = "{L483: a protostar in transition from Class 0 to Class I}",
      journal = {\aap},
         year = 2000,
        month = jul,
       volume = {359},
        pages = {967-976},
          doi = {10.48550/arXiv.astro-ph/0005525},
archivePrefix = {arXiv},
       eprint = {astro-ph/0005525},
 primaryClass = {astro-ph},
       adsurl = {https://ui.adsabs.harvard.edu/abs/2000A&A...359..967T}
}

@ARTICLE{Punanova2022,
       author = {{Punanova}, Anna and {Vasyunin}, Anton and {Caselli}, Paola and {Howard}, Alexander and {Spezzano}, Silvia and {Shirley}, Yancy and {Scibelli}, Samantha and {Harju}, Jorma},
        title = "{Methanol Mapping in Cold Cores: Testing Model Predictions}",
      journal = {\apj},
         year = 2022,
        month = mar,
       volume = {927},
       number = {2},
          eid = {213},
        pages = {213},
          doi = {10.3847/1538-4357/ac4e7d},
archivePrefix = {arXiv},
       eprint = {2112.04538},
 primaryClass = {astro-ph.GA},
       adsurl = {https://ui.adsabs.harvard.edu/abs/2022ApJ...927..213P}
}

@ARTICLE{Esplugues2022,
       author = {{Esplugues}, G. and {Fuente}, A. and {Navarro-Almaida}, D. and {Rodr{\'\i}guez-Baras}, M. and {Majumdar}, L. and {Caselli}, P. and {Wakelam}, V. and {Roueff}, E. and {Bachiller}, R. and {Spezzano}, S. and {Rivi{\`e}re-Marichalar}, P. and {Mart{\'\i}n-Dom{\'e}nech}, R. and {Mu{\~n}oz Caro}, G.~M.},
        title = "{Gas phase Elemental abundances in Molecular cloudS (GEMS). VI. A sulphur journey across star-forming regions: study of thioformaldehyde emission}",
      journal = {\aap},
         year = 2022,
        month = jun,
       volume = {662},
          eid = {A52},
        pages = {A52},
          doi = {10.1051/0004-6361/202142936},
archivePrefix = {arXiv},
       eprint = {2204.02645},
 primaryClass = {astro-ph.SR},
       adsurl = {https://ui.adsabs.harvard.edu/abs/2022A&A...662A..52E}
}

@ARTICLE{Hily2022,
       author = {{Hily-Blant}, P. and {Pineau des For{\^e}ts}, G. and {Faure}, A. and {Lique}, F.},
        title = "{Sulfur gas-phase abundance in dense cores}",
      journal = {\aap},
         year = 2022,
        month = feb,
       volume = {658},
          eid = {A168},
        pages = {A168},
          doi = {10.1051/0004-6361/201936498},
archivePrefix = {arXiv},
       eprint = {2112.01076},
 primaryClass = {astro-ph.GA},
       adsurl = {https://ui.adsabs.harvard.edu/abs/2022A&A...658A.168H}
}

@ARTICLE{Clary1985,
       author = {{Clary}, D.~C. and {Smith}, D. and {Adams}, N.~G.},
        title = "{Temperature dependence of rate coefficients for reactions of ions with dipolar molecules}",
      journal = {Chemical Physics Letters},
         year = 1985,
        month = sep,
       volume = {119},
       number = {4},
        pages = {320-326},
          doi = {10.1016/0009-2614(85)80425-5},
       adsurl = {https://ui.adsabs.harvard.edu/abs/1985CPL...119..320C}
}

@ARTICLE{Caselli2012,
       author = {{Caselli}, Paola and {Ceccarelli}, Cecilia},
        title = "{Our astrochemical heritage}",
      journal = {\aapr},
         year = 2012,
        month = oct,
       volume = {20},
          eid = {56},
        pages = {56},
          doi = {10.1007/s00159-012-0056-x},
archivePrefix = {arXiv},
       eprint = {1210.6368},
 primaryClass = {astro-ph.GA},
       adsurl = {https://ui.adsabs.harvard.edu/abs/2012A&ARv..20...56C}
}

@ARTICLE{Sakai2008,
       author = {{Sakai}, Nami and {Sakai}, Takeshi and {Hirota}, Tomoya and {Yamamoto}, Satoshi},
        title = "{Abundant Carbon-Chain Molecules toward the Low-Mass Protostar IRAS 04368+2557 in L1527}",
      journal = {\apj},
         year = 2008,
        month = jan,
       volume = {672},
       number = {1},
        pages = {371-381},
          doi = {10.1086/523635},
       adsurl = {https://ui.adsabs.harvard.edu/abs/2008ApJ...672..371S}
}

@ARTICLE{Turner1998,
       author = {{Turner}, B.~E. and {Lee}, Ho-Hsin and {Herbst}, Eric},
        title = "{The Physics and Chemistry of Small Translucent Molecular Clouds. IX. Acetylenic Chemistry}",
      journal = {\apjs},
         year = 1998,
        month = mar,
       volume = {115},
       number = {1},
        pages = {91-118},
          doi = {10.1086/313077},
       adsurl = {https://ui.adsabs.harvard.edu/abs/1998ApJS..115...91T}
}

@ARTICLE{Navarro2020,
       author = {{Navarro-Almaida}, D. and {Le Gal}, R. and {Fuente}, A. and {Rivi{\`e}re-Marichalar}, P. and {Wakelam}, V. and {Cazaux}, S. and {Caselli}, P. and {Laas}, J.~C. and {Alonso-Albi}, T. and {Loison}, J.~C. and {Gerin}, M. and {Kramer}, C. and {Roueff}, E. and {Bachiller}, R. and {Commer{\c{c}}on}, B. and {Friesen}, R. and {Garc{\'\i}a-Burillo}, S. and {Goicoechea}, J.~R. and {Giuliano}, B.~M. and {Jim{\'e}nez-Serra}, I. and {Kirk}, J.~M. and {Lattanzi}, V. and {Malinen}, J. and {Marcelino}, N. and {Mart{\'\i}n-Dom{\`e}nech}, R. and {Mu{\~n}oz Caro}, G.~M. and {Pineda}, J. and {Tercero}, B. and {Trevi{\~n}o-Morales}, S.~P. and {Roncero}, O. and {Hacar}, A. and {Tafalla}, M. and {Ward-Thompson}, D.},
        title = "{Gas phase Elemental abundances in Molecular cloudS (GEMS). II. On the quest for the sulphur reservoir in molecular clouds: the H$_{2}$S case}",
      journal = {\aap},
         year = 2020,
        month = may,
       volume = {637},
          eid = {A39},
        pages = {A39},
          doi = {10.1051/0004-6361/201937180},
archivePrefix = {arXiv},
       eprint = {2004.03475},
 primaryClass = {astro-ph.GA},
       adsurl = {https://ui.adsabs.harvard.edu/abs/2020A&A...637A..39N}
}

@ARTICLE{Jijina1999,
       author = {{Jijina}, J. and {Myers}, P.~C. and {Adams}, Fred C.},
        title = "{Dense Cores Mapped in Ammonia: A Database}",
      journal = {\apjs},
         year = 1999,
        month = nov,
       volume = {125},
       number = {1},
        pages = {161-236},
          doi = {10.1086/313268},
       adsurl = {https://ui.adsabs.harvard.edu/abs/1999ApJS..125..161J}
}

@ARTICLE{Wakelam2006,
       author = {{Wakelam}, V. and {Herbst}, E. and {Selsis}, F.},
        title = "{The effect of uncertainties on chemical models of dark clouds}",
      journal = {\aap},
         year = 2006,
        month = may,
       volume = {451},
       number = {2},
        pages = {551-562},
          doi = {10.1051/0004-6361:20054682},
archivePrefix = {arXiv},
       eprint = {astro-ph/0601611},
 primaryClass = {astro-ph},
       adsurl = {https://ui.adsabs.harvard.edu/abs/2006A&A...451..551W}
}

@ARTICLE{Vastel2018,
       author = {{Vastel}, Charlotte and {Qu{\'e}nard}, D. and {Le Gal}, R. and {Wakelam}, V. and {Andrianasolo}, A. and {Caselli}, P. and {Vidal}, T. and {Ceccarelli}, C. and {Lefloch}, B. and {Bachiller}, R.},
        title = "{Sulphur chemistry in the L1544 pre-stellar core}",
      journal = {\mnras},
         year = 2018,
        month = aug,
       volume = {478},
       number = {4},
        pages = {5514-5532},
          doi = {10.1093/mnras/sty1336},
archivePrefix = {arXiv},
       eprint = {1806.01102},
 primaryClass = {astro-ph.GA},
       adsurl = {https://ui.adsabs.harvard.edu/abs/2018MNRAS.478.5514V}
}

@ARTICLE{Marsh2014,
       author = {{Marsh}, K.~A. and {Griffin}, M.~J. and {Palmeirim}, P. and {Andr{\'e}}, Ph. and {Kirk}, J. and {Stamatellos}, D. and {Ward-Thompson}, D. and {Roy}, A. and {Bontemps}, S. and {di Francesco}, J. and {Elia}, D. and {Hill}, T. and {K{\"o}nyves}, V. and {Motte}, F. and {Nguyen-Luong}, Q. and {Peretto}, N. and {Pezzuto}, S. and {Rivera-Ingraham}, A. and {Schneider}, N. and {Spinoglio}, L. and {White}, G.},
        title = "{Properties of starless and prestellar cores in Taurus revealed by Herschel: SPIRE/PACS imaging}",
      journal = {\mnras},
         year = 2014,
        month = apr,
       volume = {439},
       number = {4},
        pages = {3683-3693},
          doi = {10.1093/mnras/stu219},
archivePrefix = {arXiv},
       eprint = {1401.7871},
 primaryClass = {astro-ph.GA},
       adsurl = {https://ui.adsabs.harvard.edu/abs/2014MNRAS.439.3683M}
}

@ARTICLE{Bracco2017,
       author = {{Bracco}, A. and {Palmeirim}, P. and {Andr{\'e}}, Ph. and {Adam}, R. and {Ade}, P. and {Bacmann}, A. and {Beelen}, A. and {Beno{\^\i}t}, A. and {Bideaud}, A. and {Billot}, N. and {Bourrion}, O. and {Calvo}, M. and {Catalano}, A. and {Coiffard}, G. and {Comis}, B. and {D'Addabbo}, A. and {D{\'e}sert}, F. -X. and {Didelon}, P. and {Doyle}, S. and {Goupy}, J. and {K{\"o}nyves}, V. and {Kramer}, C. and {Lagache}, G. and {Leclercq}, S. and {Mac{\'\i}as-P{\'e}rez}, J.~F. and {Maury}, A. and {Mauskopf}, P. and {Mayet}, F. and {Monfardini}, A. and {Motte}, F. and {Pajot}, F. and {Pascale}, E. and {Peretto}, N. and {Perotto}, L. and {Pisano}, G. and {Ponthieu}, N. and {Rev{\'e}ret}, V. and {Rigby}, A. and {Ritacco}, A. and {Rodriguez}, L. and {Romero}, C. and {Roy}, A. and {Ruppin}, F. and {Schuster}, K. and {Sievers}, A. and {Triqueneaux}, S. and {Tucker}, C. and {Zylka}, R.},
        title = "{Probing changes of dust properties along a chain of solar-type prestellar and protostellar cores in Taurus with NIKA}",
      journal = {\aap},
         year = 2017,
        month = aug,
       volume = {604},
          eid = {A52},
        pages = {A52},
          doi = {10.1051/0004-6361/201731117},
archivePrefix = {arXiv},
       eprint = {1706.08407},
 primaryClass = {astro-ph.GA},
       adsurl = {https://ui.adsabs.harvard.edu/abs/2017A&A...604A..52B}
}

@ARTICLE{Luhman2009,
       author = {{Luhman}, K.~L. and {Mamajek}, E.~E. and {Allen}, P.~R. and {Cruz}, K.~L.},
        title = "{An Infrared/X-Ray Survey for New Members of the Taurus Star-Forming Region}",
      journal = {\apj},
         year = 2009,
        month = sep,
       volume = {703},
       number = {1},
        pages = {399-419},
          doi = {10.1088/0004-637X/703/1/399},
archivePrefix = {arXiv},
       eprint = {0911.5451},
 primaryClass = {astro-ph.GA},
       adsurl = {https://ui.adsabs.harvard.edu/abs/2009ApJ...703..399L}
}

@ARTICLE{Rebull2010,
       author = {{Rebull}, L.~M. and {Padgett}, D.~L. and {McCabe}, C. -E. and {Hillenbrand}, L.~A. and {Stapelfeldt}, K.~R. and {Noriega-Crespo}, A. and {Carey}, S.~J. and {Brooke}, T. and {Huard}, T. and {Terebey}, S. and {Audard}, M. and {Monin}, J. -L. and {Fukagawa}, M. and {G{\"u}del}, M. and {Knapp}, G.~R. and {Menard}, F. and {Allen}, L.~E. and {Angione}, J.~R. and {Baldovin-Saavedra}, C. and {Bouvier}, J. and {Briggs}, K. and {Dougados}, C. and {Evans}, N.~J. and {Flagey}, N. and {Guieu}, S. and {Grosso}, N. and {Glauser}, A.~M. and {Harvey}, P. and {Hines}, D. and {Latter}, W.~B. and {Skinner}, S.~L. and {Strom}, S. and {Tromp}, J. and {Wolf}, S.},
        title = "{The Taurus Spitzer Survey: New Candidate Taurus Members Selected Using Sensitive Mid-Infrared Photometry}",
      journal = {\apjs},
         year = 2010,
        month = feb,
       volume = {186},
       number = {2},
        pages = {259-307},
          doi = {10.1088/0067-0049/186/2/259},
archivePrefix = {arXiv},
       eprint = {0911.3176},
 primaryClass = {astro-ph.SR},
       adsurl = {https://ui.adsabs.harvard.edu/abs/2010ApJS..186..259R}
}

@ARTICLE{Seo2015,
       author = {{Seo}, Young Min and {Shirley}, Yancy L. and {Goldsmith}, Paul and {Ward-Thompson}, Derek and {Kirk}, Jason M. and {Schmalzl}, Markus and {Lee}, Jeong-Eun and {Friesen}, Rachel and {Langston}, Glen and {Masters}, Joe and {Garwood}, Robert W.},
        title = "{An Ammonia Spectral Map of the L1495-B218 Filaments in the Taurus Molecular Cloud. I. Physical Properties of Filaments and Dense Cores}",
      journal = {\apj},
         year = 2015,
        month = jun,
       volume = {805},
       number = {2},
          eid = {185},
        pages = {185},
          doi = {10.1088/0004-637X/805/2/185},
archivePrefix = {arXiv},
       eprint = {1503.05179},
 primaryClass = {astro-ph.GA},
       adsurl = {https://ui.adsabs.harvard.edu/abs/2015ApJ...805..185S}
}

@ARTICLE{Tercero2021,
       author = {{Tercero}, F. and {L{\'o}pez-P{\'e}rez}, J.~A. and {Gallego}, J.~D. and {Beltr{\'a}n}, F. and {Garc{\'\i}a}, O. and {Patino-Esteban}, M. and {L{\'o}pez-Fern{\'a}ndez}, I. and {G{\'o}mez-Molina}, G. and {Diez}, M. and {Garc{\'\i}a-Carre{\~n}o}, P. and {Malo}, I. and {Amils}, R. and {Serna}, J.~M. and {Albo}, C. and {Hern{\'a}ndez}, J.~M. and {Vaquero}, B. and {Gonz{\'a}lez-Garc{\'\i}a}, J. and {Barbas}, L. and {L{\'o}pez-Fern{\'a}ndez}, J.~A. and {Bujarrabal}, V. and {G{\'o}mez-Garrido}, M. and {Pardo}, J.~R. and {Santander-Garc{\'\i}a}, M. and {Tercero}, B. and {Cernicharo}, J. and {de Vicente}, P.},
        title = "{Yebes 40 m radio telescope and the broad band Nanocosmos receivers at 7 mm and 3 mm for line surveys}",
      journal = {\aap},
         year = 2021,
        month = jan,
       volume = {645},
          eid = {A37},
        pages = {A37},
          doi = {10.1051/0004-6361/202038701},
archivePrefix = {arXiv},
       eprint = {2010.16224},
 primaryClass = {astro-ph.IM},
       adsurl = {https://ui.adsabs.harvard.edu/abs/2021A&A...645A..37T}
}

@ARTICLE{Benson1989,
       author = {{Benson}, P.~J. and {Myers}, P.~C.},
        title = "{A Survey for Dense Cores in Dark Clouds}",
      journal = {\apjs},
         year = 1989,
        month = sep,
       volume = {71},
        pages = {89},
          doi = {10.1086/191365},
       adsurl = {https://ui.adsabs.harvard.edu/abs/1989ApJS...71...89B}
}

@ARTICLE{Onishi2002,
       author = {{Onishi}, Toshikazu and {Mizuno}, Akira and {Kawamura}, Akiko and {Tachihara}, Kengo and {Fukui}, Yasuo},
        title = "{A Complete Search for Dense Cloud Cores in Taurus}",
      journal = {\apj},
         year = 2002,
        month = aug,
       volume = {575},
       number = {2},
        pages = {950-973},
          doi = {10.1086/341347},
       adsurl = {https://ui.adsabs.harvard.edu/abs/2002ApJ...575..950O}
}

@ARTICLE{Tatematsu2004,
       author = {{Tatematsu}, Ken'ichi and {Umemoto}, Tomofumi and {Kandori}, Ryo and {Sekimoto}, Yutaro},
        title = "{N$_{2}$H$^{+}$ Observations of Molecular Cloud Cores in Taurus}",
      journal = {\apj},
         year = 2004,
        month = may,
       volume = {606},
       number = {1},
        pages = {333-340},
          doi = {10.1086/382862},
archivePrefix = {arXiv},
       eprint = {astro-ph/0401584},
 primaryClass = {astro-ph},
       adsurl = {https://ui.adsabs.harvard.edu/abs/2004ApJ...606..333T}
}

@ARTICLE{Punanova2018,
       author = {{Punanova}, A. and {Caselli}, P. and {Pineda}, J.~E. and {Pon}, A. and {Tafalla}, M. and {Hacar}, A. and {Bizzocchi}, L.},
        title = "{Kinematics of dense gas in the L1495 filament}",
      journal = {\aap},
         year = 2018,
        month = sep,
       volume = {617},
          eid = {A27},
        pages = {A27},
          doi = {10.1051/0004-6361/201731159},
archivePrefix = {arXiv},
       eprint = {1806.03354},
 primaryClass = {astro-ph.GA},
       adsurl = {https://ui.adsabs.harvard.edu/abs/2018A&A...617A..27P}
}

@ARTICLE{Minh1991,
       author = {{Minh}, Y.~C. and {Brewer}, M.~K. and {Irvine}, W.~M. and {Friberg}, P. and {Johansson}, L.~E.~B.},
        title = "{Abundance and chemistry of interstellar HOCO+.}",
      journal = {\aap},
         year = 1991,
        month = apr,
       volume = {244},
        pages = {470},
       adsurl = {https://ui.adsabs.harvard.edu/abs/1991A&A...244..470M}
}

@ARTICLE{Chapman2011,
       author = {{Chapman}, Nicholas L. and {Goldsmith}, Paul F. and {Pineda}, Jorge L. and {Clemens}, D.~P. and {Li}, Di and {Kr{\v{c}}o}, Marko},
        title = "{The Magnetic Field in Taurus Probed by Infrared Polarization}",
      journal = {\apj},
         year = 2011,
        month = nov,
       volume = {741},
       number = {1},
          eid = {21},
        pages = {21},
          doi = {10.1088/0004-637X/741/1/21},
archivePrefix = {arXiv},
       eprint = {1108.0410},
 primaryClass = {astro-ph.GA},
       adsurl = {https://ui.adsabs.harvard.edu/abs/2011ApJ...741...21C}
}

@ARTICLE{Nagai1998,
       author = {{Nagai}, Tomoya and {Inutsuka}, Shu-ichiro and {Miyama}, Shoken M.},
        title = "{An Origin of Filamentary Structure in Molecular Clouds}",
      journal = {\apj},
         year = 1998,
        month = oct,
       volume = {506},
       number = {1},
        pages = {306-322},
          doi = {10.1086/306249},
       adsurl = {https://ui.adsabs.harvard.edu/abs/1998ApJ...506..306N}
}

@ARTICLE{Bacmann2012,
       author = {{Bacmann}, A. and {Taquet}, V. and {Faure}, A. and {Kahane}, C. and {Ceccarelli}, C.},
        title = "{Detection of complex organic molecules in a prestellar core: a new challenge for astrochemical models}",
      journal = {\aap},
         year = 2012,
        month = may,
       volume = {541},
          eid = {L12},
        pages = {L12},
          doi = {10.1051/0004-6361/201219207},
       adsurl = {https://ui.adsabs.harvard.edu/abs/2012A&A...541L..12B}
}

@ARTICLE{Lattanzi2020,
       author = {{Lattanzi}, Valerio and {Bizzocchi}, Luca and {Vasyunin}, Anton I. and {Harju}, Jorma and {Giuliano}, Barbara M. and {Vastel}, Charlotte and {Caselli}, Paola},
        title = "{Molecular complexity in pre-stellar cores: a 3 mm-band study of L183 and L1544}",
      journal = {\aap},
         year = 2020,
        month = jan,
       volume = {633},
          eid = {A118},
        pages = {A118},
          doi = {10.1051/0004-6361/201936884},
       adsurl = {https://ui.adsabs.harvard.edu/abs/2020A&A...633A.118L}
}

@ARTICLE{Araki2017,
       author = {{Araki}, Mitsunori and {Takano}, Shuro and {Sakai}, Nami and {Yamamoto}, Satoshi and {Oyama}, Takahiro and {Kuze}, Nobuhiko and {Tsukiyama}, Koichi},
        title = "{Long Carbon Chains in the Warm Carbon-chain-chemistry Source L1527: First Detection of C$_{7}$H in Molecular Clouds}",
      journal = {\apj},
         year = 2017,
        month = sep,
       volume = {847},
       number = {1},
          eid = {51},
        pages = {51},
          doi = {10.3847/1538-4357/aa8637},
       adsurl = {https://ui.adsabs.harvard.edu/abs/2017ApJ...847...51A}
}

@ARTICLE{Hirota2010,
       author = {{Hirota}, Tomoya and {Sakai}, Nami and {Yamamoto}, Satoshi},
        title = "{Depletion of CCS in a Candidate Warm-carbon-chain-chemistry Source L483}",
      journal = {\apj},
         year = 2010,
        month = sep,
       volume = {720},
       number = {2},
        pages = {1370-1373},
          doi = {10.1088/0004-637X/720/2/1370},
archivePrefix = {arXiv},
       eprint = {1007.1066},
 primaryClass = {astro-ph.SR},
       adsurl = {https://ui.adsabs.harvard.edu/abs/2010ApJ...720.1370H}
}

@ARTICLE{Asensio2024,
       author = {{Asensio Ramos}, A. and {Westendorp Plaza}, C. and {Navarro-Almaida}, D. and {Rivi{\`e}re-Marichalar}, P. and {Wakelam}, V. and {Fuente}, A.},
        title = "{A fast neural emulator for interstellar chemistry}",
      journal = {\mnras},
         year = 2024,
        month = jul,
       volume = {531},
       number = {4},
        pages = {4930-4943},
          doi = {10.1093/mnras/stae1432},
archivePrefix = {arXiv},
       eprint = {2406.02387},
 primaryClass = {astro-ph.IM},
       adsurl = {https://ui.adsabs.harvard.edu/abs/2024MNRAS.531.4930A}
}

@ARTICLE{Raud2016,
       author = {{Ruaud}, Maxime and {Wakelam}, Valentine and {Hersant}, Franck},
        title = "{Gas and grain chemical composition in cold cores as predicted by the Nautilus three-phase model}",
      journal = {\mnras},
         year = 2016,
        month = jul,
       volume = {459},
       number = {4},
        pages = {3756-3767},
          doi = {10.1093/mnras/stw887},
archivePrefix = {arXiv},
       eprint = {1604.05216},
 primaryClass = {astro-ph.GA},
       adsurl = {https://ui.adsabs.harvard.edu/abs/2016MNRAS.459.3756R}
}

@ARTICLE{Wilson1994,
       author = {{Wilson}, T.~L. and {Rood}, R.},
        title = "{Abundances in the Interstellar Medium}",
      journal = {\araa},
         year = 1994,
        month = jan,
       volume = {32},
        pages = {191-226},
          doi = {10.1146/annurev.aa.32.090194.001203},
       adsurl = {https://ui.adsabs.harvard.edu/abs/1994ARA&A..32..191W}
}

@ARTICLE{Savage2002,
       author = {{Savage}, C. and {Apponi}, A.~J. and {Ziurys}, L.~M. and {Wyckoff}, S.},
        title = "{Galactic $^{12}$C/$^{13}$C Ratios from Millimeter-Wave Observations of Interstellar CN}",
      journal = {\apj},
         year = 2002,
        month = oct,
       volume = {578},
       number = {1},
        pages = {211-223},
          doi = {10.1086/342468},
       adsurl = {https://ui.adsabs.harvard.edu/abs/2002ApJ...578..211S}
}

@ARTICLE{Esplugues2014,
       author = {{Esplugues}, G.~B. and {Viti}, S. and {Goicoechea}, J.~R. and {Cernicharo}, J.},
        title = "{Modelling the sulphur chemistry evolution in Orion KL}",
      journal = {\aap},
         year = 2014,
        month = jul,
       volume = {567},
          eid = {A95},
        pages = {A95},
          doi = {10.1051/0004-6361/201323010},
archivePrefix = {arXiv},
       eprint = {1406.2278},
 primaryClass = {astro-ph.SR},
       adsurl = {https://ui.adsabs.harvard.edu/abs/2014A&A...567A..95E}
}

@ARTICLE{Navarro2021,
       author = {{Navarro-Almaida}, D. and {Fuente}, A. and {Majumdar}, L. and {Wakelam}, V. and {Caselli}, P. and {Rivi{\`e}re-Marichalar}, P. and {Trevi{\~n}o-Morales}, S.~P. and {Cazaux}, S. and {Jim{\'e}nez-Serra}, I. and {Kramer}, C. and {Chac{\'o}n-Tanarro}, A. and {Kirk}, J.~M. and {Ward-Thompson}, D. and {Tafalla}, M.},
        title = "{Evolutionary view through the starless cores in Taurus. Deuteration in TMC 1-C and TMC 1-CP}",
      journal = {\aap},
         year = 2021,
        month = sep,
       volume = {653},
          eid = {A15},
        pages = {A15},
          doi = {10.1051/0004-6361/202140820},
archivePrefix = {arXiv},
       eprint = {2107.00423},
 primaryClass = {astro-ph.GA},
       adsurl = {https://ui.adsabs.harvard.edu/abs/2021A&A...653A..15N}
}

@ARTICLE{Caselli2002,
       author = {{Caselli}, P. and {Walmsley}, C.~M. and {Zucconi}, A. and {Tafalla}, M. and {Dore}, L. and {Myers}, P.~C.},
        title = "{Molecular Ions in L1544. II. The Ionization Degree}",
      journal = {\apj},
         year = 2002,
        month = jan,
       volume = {565},
       number = {1},
        pages = {344-358},
          doi = {10.1086/324302},
archivePrefix = {arXiv},
       eprint = {astro-ph/0109023},
 primaryClass = {astro-ph},
       adsurl = {https://ui.adsabs.harvard.edu/abs/2002ApJ...565..344C}
}

@ARTICLE{NeufeldWolfire2017,
       author = {{Neufeld}, David A. and {Wolfire}, Mark G.},
        title = "{The Cosmic-Ray Ionization Rate in the Galactic Disk, as Determined from Observations of Molecular Ions}",
      journal = {\apj},
         year = 2017,
        month = aug,
       volume = {845},
       number = {2},
          eid = {163},
        pages = {163},
          doi = {10.3847/1538-4357/aa6d68},
archivePrefix = {arXiv},
       eprint = {1704.03877},
 primaryClass = {astro-ph.GA},
       adsurl = {https://ui.adsabs.harvard.edu/abs/2017ApJ...845..163N}
}

@ARTICLE{Pety2005,
       author = {{Pety}, J. and {Teyssier}, D. and {Foss{\'e}}, D. and {Gerin}, M. and {Roueff}, E. and {Abergel}, A. and {Habart}, E. and {Cernicharo}, J.},
        title = "{Are PAHs precursors of small hydrocarbons in photo-dissociation regions? The Horsehead case}",
      journal = {\aap},
         year = 2005,
        month = jun,
       volume = {435},
       number = {3},
        pages = {885-899},
          doi = {10.1051/0004-6361:20041170},
archivePrefix = {arXiv},
       eprint = {astro-ph/0501339},
 primaryClass = {astro-ph},
       adsurl = {https://ui.adsabs.harvard.edu/abs/2005A&A...435..885P}
}

@ARTICLE{Goicoechea2006,
       author = {{Goicoechea}, J.~R. and {Pety}, J. and {Gerin}, M. and {Teyssier}, D. and {Roueff}, E. and {Hily-Blant}, P. and {Baek}, S.},
        title = "{Low sulfur depletion in the Horsehead PDR}",
      journal = {\aap},
         year = 2006,
        month = sep,
       volume = {456},
       number = {2},
        pages = {565-580},
          doi = {10.1051/0004-6361:20065260},
archivePrefix = {arXiv},
       eprint = {astro-ph/0605716},
 primaryClass = {astro-ph},
       adsurl = {https://ui.adsabs.harvard.edu/abs/2006A&A...456..565G}
}

@ARTICLE{Riviere2019,
       author = {{Rivi{\`e}re-Marichalar}, P. and {Fuente}, A. and {Goicoechea}, J.~R. and {Pety}, J. and {Le Gal}, R. and {Gratier}, P. and {Guzm{\'a}n}, V. and {Roueff}, E. and {Loison}, J.~C. and {Wakelam}, V. and {Gerin}, M.},
        title = "{Abundances of sulphur molecules in the Horsehead nebula. First NS$^{+}$ detection in a photodissociation region}",
      journal = {\aap},
         year = 2019,
        month = aug,
       volume = {628},
          eid = {A16},
        pages = {A16},
          doi = {10.1051/0004-6361/201935354},
archivePrefix = {arXiv},
       eprint = {1906.09932},
 primaryClass = {astro-ph.GA},
       adsurl = {https://ui.adsabs.harvard.edu/abs/2019A&A...628A..16R}
}

@ARTICLE{Padovani2009,
       author = {{Padovani}, M. and {Galli}, D. and {Glassgold}, A.~E.},
        title = "{Cosmic-ray ionization of molecular clouds}",
      journal = {\aap},
         year = 2009,
        month = jul,
       volume = {501},
       number = {2},
        pages = {619-631},
          doi = {10.1051/0004-6361/200911794},
archivePrefix = {arXiv},
       eprint = {0904.4149},
 primaryClass = {astro-ph.SR},
       adsurl = {https://ui.adsabs.harvard.edu/abs/2009A&A...501..619P}
}

@ARTICLE{Galli2019,
       author = {{Galli}, P.~A.~B. and {Loinard}, L. and {Bouy}, H. and {Sarro}, L.~M. and {Ortiz-Le{\'o}n}, G.~N. and {Dzib}, S.~A. and {Olivares}, J. and {Heyer}, M. and {Hernandez}, J. and {Rom{\'a}n-Z{\'u}{\~n}iga}, C. and {Kounkel}, M. and {Covey}, K.},
        title = "{Structure and kinematics of the Taurus star-forming region from Gaia-DR2 and VLBI astrometry}",
      journal = {\aap},
         year = 2019,
        month = oct,
       volume = {630},
          eid = {A137},
        pages = {A137},
          doi = {10.1051/0004-6361/201935928},
archivePrefix = {arXiv},
       eprint = {1909.01118},
 primaryClass = {astro-ph.SR},
       adsurl = {https://ui.adsabs.harvard.edu/abs/2019A&A...630A.137G}
}

@ARTICLE{RodriguezBaras2023,
       author = {{Rodr{\'\i}guez-Baras}, M. and {Esplugues}, G. and {Fuente}, A. and {Spezzano}, S. and {Caselli}, P. and {Loison}, J.~C. and {Roueff}, E. and {Navarro-Almaida}, D. and {Bachiller}, R. and {Mart{\'\i}n-Dom{\'e}nech}, R. and {Jim{\'e}nez-Serra}, I. and {Beitia-Antero}, L. and {Le Gal}, R.},
        title = "{Gas phase Elemental abundances in Molecular cloudS (GEMS). IX. Deuterated compounds of H$_{2}$S in starless cores}",
      journal = {\aap},
         year = 2023,
        month = nov,
       volume = {679},
          eid = {A120},
        pages = {A120},
          doi = {10.1051/0004-6361/202346869},
archivePrefix = {arXiv},
       eprint = {2309.00318},
 primaryClass = {astro-ph.GA},
       adsurl = {https://ui.adsabs.harvard.edu/abs/2023A&A...679A.120R}
}

@ARTICLE{Guzman2015,
       author = {{Guzm{\'a}n}, V.~V. and {Pety}, J. and {Goicoechea}, J.~R. and {Gerin}, M. and {Roueff}, E. and {Gratier}, P. and {{\"O}berg}, K.~I.},
        title = "{Spatially Resolved L-C$_{3}$H$^{+}$ Emission in the Horsehead Photodissociation Region: Further Evidence for a Top-Down Hydrocarbon Chemistry}",
      journal = {\apjl},
         year = 2015,
        month = feb,
       volume = {800},
       number = {2},
          eid = {L33},
        pages = {L33},
          doi = {10.1088/2041-8205/800/2/L33},
archivePrefix = {arXiv},
       eprint = {1502.02325},
 primaryClass = {astro-ph.GA},
       adsurl = {https://ui.adsabs.harvard.edu/abs/2015ApJ...800L..33G}
}

@ARTICLE{Pety2012,
       author = {{Pety}, J. and {Gratier}, P. and {Guzm{\'a}n}, V. and {Roueff}, E. and {Gerin}, M. and {Goicoechea}, J.~R. and {Bardeau}, S. and {Sievers}, A. and {Le Petit}, F. and {Le Bourlot}, J. and {Belloche}, A. and {Talbi}, D.},
        title = "{The IRAM-30 m line survey of the Horsehead PDR. II. First detection of the l-C$_{3}$H$^{+}$ hydrocarbon cation}",
      journal = {\aap},
         year = 2012,
        month = dec,
       volume = {548},
          eid = {A68},
        pages = {A68},
          doi = {10.1051/0004-6361/201220062},
archivePrefix = {arXiv},
       eprint = {1210.8178},
 primaryClass = {astro-ph.GA},
       adsurl = {https://ui.adsabs.harvard.edu/abs/2012A&A...548A..68P}
}

@ARTICLE{Oya2017,
       author = {{Oya}, Yoko and {Sakai}, Nami and {Watanabe}, Yoshimasa and {Higuchi}, Aya E. and {Hirota}, Tomoya and {L{\'o}pez-Sepulcre}, Ana and {Sakai}, Takeshi and {Aikawa}, Yuri and {Ceccarelli}, Cecilia and {Lefloch}, Bertrand and {Caux}, Emmanuel and {Vastel}, Charlotte and {Kahane}, Claudine and {Yamamoto}, Satoshi},
        title = "{L483: Warm Carbon-chain Chemistry Source Harboring Hot Corino Activity}",
      journal = {\apj},
         year = 2017,
        month = mar,
       volume = {837},
       number = {2},
          eid = {174},
        pages = {174},
          doi = {10.3847/1538-4357/aa6300},
archivePrefix = {arXiv},
       eprint = {1703.03653},
 primaryClass = {astro-ph.SR},
       adsurl = {https://ui.adsabs.harvard.edu/abs/2017ApJ...837..174O}
}

@ARTICLE{Molinari2010,
       author = {{Molinari}, S. and {Swinyard}, B. and {Bally}, J. and {Barlow}, M. and {Bernard}, J. -P. and {Martin}, P. and {Moore}, T. and {Noriega-Crespo}, A. and {Plume}, R. and {Testi}, L. and {Zavagno}, A. and {Abergel}, A. and {Ali}, B. and {Anderson}, L. and {Andr{\'e}}, P. and {Baluteau}, J. -P. and {Battersby}, C. and {Beltr{\'a}n}, M.~T. and {Benedettini}, M. and {Billot}, N. and {Blommaert}, J. and {Bontemps}, S. and {Boulanger}, F. and {Brand}, J. and {Brunt}, C. and {Burton}, M. and {Calzoletti}, L. and {Carey}, S. and {Caselli}, P. and {Cesaroni}, R. and {Cernicharo}, J. and {Chakrabarti}, S. and {Chrysostomou}, A. and {Cohen}, M. and {Compiegne}, M. and {de Bernardis}, P. and {de Gasperis}, G. and {di Giorgio}, A.~M. and {Elia}, D. and {Faustini}, F. and {Flagey}, N. and {Fukui}, Y. and {Fuller}, G.~A. and {Ganga}, K. and {Garcia-Lario}, P. and {Glenn}, J. and {Goldsmith}, P.~F. and {Griffin}, M. and {Hoare}, M. and {Huang}, M. and {Ikhenaode}, D. and {Joblin}, C. and {Joncas}, G. and {Juvela}, M. and {Kirk}, J.~M. and {Lagache}, G. and {Li}, J.~Z. and {Lim}, T.~L. and {Lord}, S.~D. and {Marengo}, M. and {Marshall}, D.~J. and {Masi}, S. and {Massi}, F. and {Matsuura}, M. and {Minier}, V. and {Miville-Desch{\^e}nes}, M. -A. and {Montier}, L.~A. and {Morgan}, L. and {Motte}, F. and {Mottram}, J.~C. and {M{\"u}ller}, T.~G. and {Natoli}, P. and {Neves}, J. and {Olmi}, L. and {Paladini}, R. and {Paradis}, D. and {Parsons}, H. and {Peretto}, N. and {Pestalozzi}, M. and {Pezzuto}, S. and {Piacentini}, F. and {Piazzo}, L. and {Polychroni}, D. and {Pomar{\`e}s}, M. and {Popescu}, C.~C. and {Reach}, W.~T. and {Ristorcelli}, I. and {Robitaille}, J. -F. and {Robitaille}, T. and {Rod{\'o}n}, J.~A. and {Roy}, A. and {Royer}, P. and {Russeil}, D. and {Saraceno}, P. and {Sauvage}, M. and {Schilke}, P. and {Schisano}, E. and {Schneider}, N. and {Schuller}, F. and {Schulz}, B. and {Sibthorpe}, B. and {Smith}, H.~A. and {Smith}, M.~D. and {Spinoglio}, L. and {Stamatellos}, D. and {Strafella}, F. and {Stringfellow}, G.~S. and {Sturm}, E. and {Taylor}, R. and {Thompson}, M.~A. and {Traficante}, A. and {Tuffs}, R.~J. and {Umana}, G. and {Valenziano}, L. and {Vavrek}, R. and {Veneziani}, M. and {Viti}, S. and {Waelkens}, C. and {Ward-Thompson}, D. and {White}, G. and {Wilcock}, L.~A. and {Wyrowski}, F. and {Yorke}, H.~W. and {Zhang}, Q.},
        title = "{Clouds, filaments, and protostars: The Herschel Hi-GAL Milky Way}",
      journal = {\aap},
         year = 2010,
        month = jul,
       volume = {518},
          eid = {L100},
        pages = {L100},
          doi = {10.1051/0004-6361/201014659},
archivePrefix = {arXiv},
       eprint = {1005.3317},
 primaryClass = {astro-ph.GA},
       adsurl = {https://ui.adsabs.harvard.edu/abs/2010A&A...518L.100M}
}

@ARTICLE{Juvela2012,
       author = {{Juvela}, M. and {Ristorcelli}, I. and {Pagani}, L. and {Doi}, Y. and {Pelkonen}, V. -M. and {Marshall}, D.~J. and {Bernard}, J. -P. and {Falgarone}, E. and {Malinen}, J. and {Marton}, G. and {McGehee}, P. and {Montier}, L.~A. and {Motte}, F. and {Paladini}, R. and {T{\'o}th}, L.~V. and {Ysard}, N. and {Zahorecz}, S. and {Zavagno}, A.},
        title = "{Galactic cold cores. III. General cloud properties}",
      journal = {\aap},
         year = 2012,
        month = may,
       volume = {541},
          eid = {A12},
        pages = {A12},
          doi = {10.1051/0004-6361/201118640},
archivePrefix = {arXiv},
       eprint = {1202.1672},
 primaryClass = {astro-ph.GA},
       adsurl = {https://ui.adsabs.harvard.edu/abs/2012A&A...541A..12J}
}

@ARTICLE{GoldsmithLanger1978,
       author = {{Goldsmith}, P.~F. and {Langer}, W.~D.},
        title = "{Molecular cooling and thermal balance of dense interstellar clouds.}",
      journal = {\apj},
         year = 1978,
        month = jun,
       volume = {222},
        pages = {881-895},
          doi = {10.1086/156206},
       adsurl = {https://ui.adsabs.harvard.edu/abs/1978ApJ...222..881G}
}

@ARTICLE{Visser2009,
       author = {{Visser}, R. and {van Dishoeck}, E.~F. and {Doty}, S.~D. and {Dullemond}, C.~P.},
        title = "{The chemical history of molecules in circumstellar disks. I. Ices}",
      journal = {\aap},
         year = 2009,
        month = mar,
       volume = {495},
       number = {3},
        pages = {881-897},
          doi = {10.1051/0004-6361/200810846},
archivePrefix = {arXiv},
       eprint = {0901.1313},
 primaryClass = {astro-ph.SR},
       adsurl = {https://ui.adsabs.harvard.edu/abs/2009A&A...495..881V}
}

@ARTICLE{Oberg2011,
       author = {{{\"O}berg}, Karin I. and {Boogert}, A.~C. Adwin and {Pontoppidan}, Klaus M. and {van den Broek}, Saskia and {van Dishoeck}, Ewine F. and {Bottinelli}, Sandrine and {Blake}, Geoffrey A. and {Evans}, II, Neal J.},
        title = "{The Spitzer Ice Legacy: Ice Evolution from Cores to Protostars}",
      journal = {\apj},
         year = 2011,
        month = oct,
       volume = {740},
       number = {2},
          eid = {109},
        pages = {109},
          doi = {10.1088/0004-637X/740/2/109},
archivePrefix = {arXiv},
       eprint = {1107.5825},
 primaryClass = {astro-ph.GA},
       adsurl = {https://ui.adsabs.harvard.edu/abs/2011ApJ...740..109O}
}

@ARTICLE{Navarro2024,
       author = {{Navarro-Almaida}, D. and {Lebreuilly}, U. and {Hennebelle}, P. and {Fuente}, A. and {Commer{\c{c}}on}, B. and {Le Gal}, R. and {Wakelam}, V. and {Gerin}, M. and {Rivi{\'e}re-Marichalar}, P. and {Beitia-Antero}, L. and {Ascasibar}, Y.},
        title = "{Grain growth and its chemical impact in the first hydrostatic core phase}",
      journal = {\aap},
         year = 2024,
        month = may,
       volume = {685},
          eid = {A112},
        pages = {A112},
          doi = {10.1051/0004-6361/202347847},
archivePrefix = {arXiv},
       eprint = {2403.01905},
 primaryClass = {astro-ph.GA},
       adsurl = {https://ui.adsabs.harvard.edu/abs/2024A&A...685A.112N}
}

@ARTICLE{Mumma2011,
       author = {{Mumma}, Michael J. and {Charnley}, Steven B.},
        title = "{The Chemical Composition of Comets{\textemdash}Emerging Taxonomies and Natal Heritage}",
      journal = {\araa},
         year = 2011,
        month = sep,
       volume = {49},
       number = {1},
        pages = {471-524},
          doi = {10.1146/annurev-astro-081309-130811},
       adsurl = {https://ui.adsabs.harvard.edu/abs/2011ARA&A..49..471M}
}

@ARTICLE{Sakai2013,
       author = {{Sakai}, Nami and {Yamamoto}, Satoshi},
        title = "{Warm Carbon-Chain Chemistry}",
      journal = {Chemical Reviews},
         year = 2013,
        month = dec,
       volume = {113},
       number = {12},
        pages = {8981-9015},
          doi = {10.1021/cr4001308},
       adsurl = {https://ui.adsabs.harvard.edu/abs/2013ChRv..113.8981S}
}

@ARTICLE{Draine1978,
       author = {{Draine}, B.~T.},
        title = "{Photoelectric heating of interstellar gas.}",
      journal = {\apjs},
         year = 1978,
        month = apr,
       volume = {36},
        pages = {595-619},
          doi = {10.1086/190513},
       adsurl = {https://ui.adsabs.harvard.edu/abs/1978ApJS...36..595D}
}

@ARTICLE{Oberg2023,
       author = {{{\"O}berg}, Karin I. and {Facchini}, Stefano and {Anderson}, Dana E.},
        title = "{Protoplanetary Disk Chemistry}",
      journal = {\araa},
         year = 2023,
        month = aug,
       volume = {61},
        pages = {287-328},
          doi = {10.1146/annurev-astro-022823-040820},
archivePrefix = {arXiv},
       eprint = {2309.05685},
 primaryClass = {astro-ph.EP},
       adsurl = {https://ui.adsabs.harvard.edu/abs/2023ARA&A..61..287O}
}

@ARTICLE{Beitia2024,
       author = {{Beitia-Antero}, L. and {Fuente}, A. and {Navarro-Almaida}, D. and {G{\'o}mez de Castro}, A.~I. and {Wakelam}, V. and {Caselli}, P. and {Le Gal}, R. and {Esplugues}, G. and {Rivi{\`e}re-Marichalar}, P. and {Spezzano}, S. and {Pineda}, J.~E. and {Rodr{\'\i}guez-Baras}, M. and {Canet}, A. and {Mart{\'\i}n-Dom{\'e}nech}, R. and {Roncero}, O.},
        title = "{Gas phase Elemental abundances in Molecular cloudS (GEMS). X. Observational effects of turbulence on the chemistry of molecular clouds}",
      journal = {\aap},
         year = 2024,
        month = aug,
       volume = {688},
          eid = {A188},
        pages = {A188},
          doi = {10.1051/0004-6361/202346955},
archivePrefix = {arXiv},
       eprint = {2410.04226},
 primaryClass = {astro-ph.GA},
       adsurl = {https://ui.adsabs.harvard.edu/abs/2024A&A...688A.188B}
}

@ARTICLE{Taillard2025,
       author = {{Taillard}, A. and {Wakelam}, V. and {Gratier}, P. and {Dartois}, E. and {Chabot}, M. and {Noble}, J.~A. and {Chu}, L.},
        title = "{Chemical constraints on the dynamical evolution of the cold core L694}",
      journal = {\aap},
         year = 2025,
        month = jun,
       volume = {698},
          eid = {A278},
        pages = {A278},
          doi = {10.1051/0004-6361/202553796},
archivePrefix = {arXiv},
       eprint = {2507.00843},
 primaryClass = {astro-ph.GA},
       adsurl = {https://ui.adsabs.harvard.edu/abs/2025A&A...698A.278T}
}

@ARTICLE{Bohlin1978,
       author = {{Bohlin}, R.~C. and {Savage}, B.~D. and {Drake}, J.~F.},
        title = "{A survey of interstellar H I from Lalpha absorption measurements. II.}",
      journal = {\apj},
         year = 1978,
        month = aug,
       volume = {224},
        pages = {132-142},
          doi = {10.1086/156357},
       adsurl = {https://ui.adsabs.harvard.edu/abs/1978ApJ...224..132B}
}

@ARTICLE{Tasa2025,
       author = {{Tasa-Chaveli}, A. and {Fuente}, A. and {Esplugues}, G. and {Navarro-Almaida}, D. and {Majumdar}, L. and {Rayalacheruvu}, P. and {Rivi{\`e}re-Marichalar}, P. and {Rodr{\'\i}guez-Baras}, M.},
        title = "{Gas phase Elemental abundances in Molecular cloudS (GEMS): XI. The evolution of HCN, HNC, and N$_{2}$H$^{+}$ isotopic ratios in starless cores}",
      journal = {\aap},
         year = 2025,
        month = aug,
       volume = {700},
          eid = {A226},
        pages = {A226},
          doi = {10.1051/0004-6361/202554121},
       adsurl = {https://ui.adsabs.harvard.edu/abs/2025A&A...700A.226T}
}

\newpage
\onecolumn
\appendix 

\section{Additional tables}

\small
\begin{longtable}{l l l l l l l}
\caption{\label{Line parameters of the detected molecules in B213-C2}Line parameters of the detected molecules in B213-C2.}\\
\hline\hline
\\
 Species&Line & Frequency & v & $\Delta \text{v}$ & $T_\text{MB}$ & $\int T_\text{MB}\text{dv}$\\
            &                &\small{(MHz)} & \small{(km$ \ \text{s}^{-1}$) }   &    \small{(km$ \ \text{s}^{-1}$) }         & \small{($\text{mK}$)} &\small{(mK $  $\text{km}$ \ \text{s}^{-1}$)}\\
\hline
\endfirsthead
\caption{Continued.}\\
\hline\hline
\\
 Species&Line & Frequency & v & $\Delta \text{v}$ & $T_\text{MB}$ & $\int T_\text{MB}\text{dv}$\\
            &                &\small{(MHz)} & \small{(km$ \ \text{s}^{-1}$) }   &    \small{(km$ \ \text{s}^{-1}$) }         & \small{($\text{mK}$)} &\small{(mK $  $\text{km}$ \ \text{s}^{-1}$)}\\
\hline
\endhead
\hline
\endfoot
\hline
\endlastfoot
\multirow{6}{*}{C$\mathrm{H}_{3}$CHO}&2( 1, 2)- 1( 1, 1) A, vt=0 & 37464.20 &$6.94 \pm 0.06$&$0.34 \pm 1.56$ &$7.91 \pm 1.16$ & $2.86 \pm 0.63$\\

&\multirow{2}{*}{2( 1, 2)- 1( 1, 1) E, vt=0}& \multirow{2}{*}{37686.93}& $5.76 \pm 0.09$ &$0.73 \pm 0.17$         &$5.80 \pm 1.79$ &$4.48 \pm1.09$ \\
                                                                                     && & $7.00 \pm 0.09$        &$0.54 \pm 0.16$         &$5.17 \pm 1.79$ &$2.99 \pm 0.93$\\

&\small{2( 0, 2)- 1( 0, 1) E, vt=0}& 38506.03& $6.91 \pm 0.03$  & $0.49 \pm 0.05$&$16.34 \pm 1.76$ &$8.44 \pm0.96$\\

&\small{2( 0, 2)- 1( 0, 1) A, vt=0}& 38512.08&$6.84 \pm 0.06$   &$0.81\pm 0.27$    &$10.21 \pm 1.52$ &$8.76 \pm 1.70$\\

&\small{2( 1, 1)- 1( 1, 0) E, vt=0} & 39362.54&$7.01 \pm 0.05$ &$0.71 \pm 0.14$ &$8.25 \pm 1.36$ &$6.25 \pm 0.96$\\

&\small{2( 1, 1)- 1( 1, 0) A, vt=0 }& 39594.29& $6.92 \pm 0.04$ &$0.59 \pm 0.10$   &$10.02 \pm 1.42$ &$6.25 \pm 0.88$\\

\hline
\multirow{5}{*}{ C$\mathrm{H}_{3}$OH}&\small{4( 1, 4)- 3(- 0, 3) E, vt=0 }&36169.26&$7.07 \pm 0.02$ & $0.62 \pm 0.05$& $19.72 \pm 1.29$ &$12.93 \pm 0.89$\\

&\small{7( 2, 5)- 8( 1, 8) E, vt=0}&  37703.76 &--       &--    & $\leq 2.68$&$ \leq 1.48$ \\

 &\small{7( 0, 7)- 6( 1, 6) A, vt=0}&44069.37&  -- &--&$\leq 2.96 $ &$\leq 1.51$\\

&\multirow{2}{*}{\small{1( 0, 1)- 0( 0, 0) A, vt=0}}&\multirow{2}{*}{48372.46} &       $6.978 \pm 0.003$&$0.47 \pm 0.01$& $254.57 \pm 3.89$ &$126.77 \pm 2.01$\\
 & &  &$5.58 \pm 0.02$  &$0.60 \pm 0.04$ &$64.42 \pm 3.89$ &$40.90 \pm 2.25$\\
&\small{1(- 0, 1)- 0(- 0, 0) E, vt=0}&48376.89& $7.02 \pm 0.08$&$0.51 \pm 0.24$&$16.62 \pm 4.63$ &$9.06 \pm 2.86$\\
\hline
\multirow{2}{*}{ CS}&\multirow{2}{*}{1-0}&\multirow{2}{*}{48990.95}&$5.605 \pm 0.005$ &$0.68 \pm 0.01$ &$910.94 \pm 4.99$ &$654.76 \pm 7.75$  \\
&&&$6.904 \pm 0.003$&$0.581 \pm 0.005$& $1371.45 \pm 4.99$& $848.50 \pm 5.81$ \\
\hline
\multirow{1}{*}{$\mathrm{H}_{2}$CS}& 1( 0, 1)- 0( 0, 0)& 34351.43& $7.08 \pm 0.02$       &$0.63 \pm 0.04$& $41.13 \pm 2.05$&$27.77 \pm 1.40$\\
\hline
\multirow{2}{*}{ HCNS}  &3-2&36908.48&-- &-- &$\leq 2.66$&$\leq 1.48$\\
&4-3&49211.11&-- &-- &$\leq 7.47$&$\leq 3.60$\\
\hline
\multirow{10}{*}{ HSCN}  &3( 1, 3)- 2( 1, 2)&34228.19&-- &-- &$\leq 2.88$&$\leq 1.67$\\

 & 3( 2, 2)- 2( 2, 1)&34405.13 &-- &-- &$\leq 1.94$&$\leq 1.12$\\

&3( 2, 1)- 2( 2, 0)&34405.28&-- &-- &$\leq 1.96$ &$\leq 1.13$\\

 &3( 0, 3)- 2( 0, 2)&34408.63 &-- &-- &$\leq 2.37$ &$\leq 1.37$\\

& 3( 1, 2)- 2( 1, 1)&34587.50&-- &-- &$\leq 2.16$&$\leq 1.24$\\

 & 4( 1, 4)- 3( 1, 3)&45637.36 &-- &-- &$\leq 5.83$ &$\leq 2.92$\\

& 4( 2, 3)- 3( 2, 2)&45873.29&-- &-- &$\leq 4.84$&$\leq 2.42$\\

 & 4( 2, 2)- 3( 2, 1)&45873.66&-- &-- &$\leq 4.74$ &$\leq 2.37$\\

& 4( 0, 4)- 3( 0, 3)&45877.81&-- &-- &$\leq 5.92$ &$\leq 2.96$\\

& 4( 1, 3)- 3( 1, 2)& 46116.44&-- &-- &$\leq 5.35$&$ \leq 2.67$\\
\hline
HCS$^{+}$& 1-0&42674.20&$7.00 \pm 0.03$& $0.56 \pm 0.07$&$20.61 \pm 2.18$ &$12.33 \pm 1.27$\\
\hline
\multirow{2}{*}{ OCS}  &\multirow{2}{*}{3-2}&\multirow{2}{*}{36488.81 }&        $6.91 \pm 0.06$        &      $0.68 \pm 0.15$ &  $9.51 \pm 1.62$  & $6.89 \pm 1.20$\\
& & &   $5.62 \pm 0.08$  &      $0.63 \pm 0.42$ & $7.39 \pm 1.62$   & $4.95 \pm 0.19$ \\
&4-3&48651.60 & $7.03 \pm 0.07$  &      $0.62 \pm 0.16$ & $20.55 \pm 4.99$   &$13.55 \pm 2.93$\\
\hline
SO$_{2}$ & 6( 2, 4)- 7( 1, 7)& 44052.86 &$7.44 \pm 0.10$&$0.70 \pm 0.17$&$5.96 \pm 1.95$ & $4.45 \pm 1.15$\\
\hline
\multirow{4}{*}{ CCS} &N= 2- 1, J= 3- 2&33751.37&       $6.996 \pm 0.002$                &$0.569 \pm 0.005$              &$246.12 \pm 1.70$   & $149.14 \pm 1.13$\\

&N= 3- 2, J= 3- 2&38866.42&     $6.99 \pm 0.01$          &      $0.50 \pm 0.03$   &  $39.05 \pm 1.48$  & $20.84 \pm 0.87$\\

&N= 4- 3, J= 3- 2&43981.02&     $7.00 \pm 0.01$          &      $0.50 \pm 0.03$   &$49.22 \pm 2.35$  &$26.28 \pm 1.29$ \\

&N= 3- 2, J= 4- 3&      45379.05&       $7.133 \pm 0.002$        &      $0.439 \pm 0.006$              & $367.74 \pm 3.02$  &$171.80 \pm 1.64$\\
\hline
\multirow{3}{*}{C$_{3}$S}& 6-5 &34684.37 &$7.024 \pm 0.006$     &$0.53 \pm 0.02$   &$60.47 \pm 1.27$ & $33.89 \pm 0.80$\\

& 7-6&40465.02 &        $7.036 \pm 0.005$&      $0.51 \pm 0.02$&$63.00 \pm 1.31$ &$34.35 \pm 0.77$\\

& 8-7&46245.62 &        $7.041 \pm 0.019$       &       $0.44 \pm 0.04$&$71.48 \pm 5.42$ & $33.34 \pm 2.77$\\
\hline
CH$_{3}$CCH & 2( 0)- 1( 0)& 34183.41&$6.99 \pm 0.03$    &$0.66 \pm 0.08$        &$18.26 \pm 1.75$ & $12.73 \pm 1.31$\\
\hline
\multirow{2}{*}{ CH$_{3}$CN}  &2( 1)- 1( 1) &36794.77 &--       &--     &$\leq 4.21$&$\leq 2.35$\\
&2( 0)- 1( 0) &36795.47 &$6.27 \pm 0.07$        &       $0.88 \pm 0.19$ &$8.81 \pm 1.77$& $8.24 \pm1.44$\\
\hline
\multirow{7}{*}{ HC$_{5}$N} &J = 12 - 11 &31951.77 &$6.95 \pm 0.01$      & $0.45 \pm 0.01$       &$141.10 \pm 3.55$&$67.61 \pm 2.35$\\
&J = 13 - 12 &34614.39 &$7.019 \pm 0.003$ &     $0.515 \pm 0.007$       &$123.55 \pm 1.23$&$67.76 \pm 0.76$\\
&J = 14 - 13 &  37276.99&$7.056 \pm 0.004$& $0.52 \pm 0.01$     &$115.05 \pm 1.48$&$63.37 \pm 0.89$\\
& J = 15 - 14&39939.59 &$7.126\pm 0.004$         &$0.547 \pm 0.009$     &$102.26 \pm 1.53$&$59.54 \pm 0.89$\\
&J = 16 - 15 & 42602.15&$7.038 \pm 0.004$        & $0.43 \pm 0.02$      &$101.45 \pm 2.07$&$46.12 \pm 1.11$\\
&J = 17 - 16 &45264.72 & $7.011 \pm 0.007$& $0.40 \pm 0.02$     &$91.88 \pm 2.70$&$39.56\pm 1.30$\\
&J = 18 - 17 &          47927.27&$7.020 \pm 0.009$       & $0.35 \pm 0.03$      &$90.77 \pm 4.19$&$33.54 \pm 1.93$\\
\hline

\multirow{7}{*}{ HC$_{7}$N}   &J = 28 - 27 &31583.70    & $7.31 \pm 0.14$       & $0.83 \pm 0.29$&$115.56 \pm 4.20$ &$10.26 \pm 3.29$  \\

  &J = 29 - 28 &32711.68        &$7.15 \pm 0.07$         &$0.50 \pm 0.28$       &$7.83 \pm 1.26$& $4.17 \pm 0.95 $ \\

  &J = 30 - 29 &33839.63        &$6.85 \pm 0.04$         &      $0.64 \pm 0.01$ &$8.79 \pm 1.08$ &$6.00 \pm 0.77 $ \\

  &J = 31 - 30 &34967.59        &$6.99 \pm 0.03$         &$0.33 \pm 1.37$ &$8.88 \pm 1.34$ &$3.09 \pm 0.53$ \\

  &J = 32 - 31 &36095.53        &$6.98 \pm 0.07$        &$0.68 \pm 0.18$ &$5.83 \pm 1.30$ &$4.20 \pm 0.88$ \\

  &J = 33 - 32 &37223.49        &--      &--&$\leq 2.36$&$\leq 1.31$ \\

  &J = 34 - 33 &38351.45        &--      &--     &$\leq 2.46$ &$\leq 1.35$ \\

  &     J = 35 - 34 &39479.41   &$7.16 \pm 0.11$        &$0.74\pm 0.21$ &$4.12 \pm 1.42$ &$3.25 \pm 0.92$ \\

  &J = 36 - 35 &40607.33        &       -- &-- &$\leq 3.11$&$\leq 0.28$ \\

  &J = 37 - 36&41735.26 &--      &--&$\leq 2.73$ &$\leq 1.43$ \\

  & J = 38 - 37&42863.20        & $7.13\pm 0.08$        & $0.57 \pm 0.17$&$6.24 \pm 1.73$&$3.78 \pm 0.98$ \\

  &J = 39 - 38&43991.13 &       -- &-- &$\leq 4.27$ &$\leq 2.18$ \\

  &J = 40 - 39&45119.06 & --&-- &$\leq 4.08$&$\leq 2.05$\\

  &J = 41 - 40&46246.98 & --&-- &$ \leq 5.25$ & $\leq 2.61$ \\

  &     J = 42 - 41&47374.90    & --&-- &$\leq 5.87$& $ \leq 2.89$ \\

  &J = 43 - 42&48502.81 & --&-- &$\leq 6.82$ &$\leq 3.31$\\

  &J = 44 - 43&49630.72 & --&-- &$\leq 7.76$& $\leq 3.73$ \\
\hline
PO$^{+}$ &1-0&47024.25& --&-- &$\leq 5.07$ &$\leq 2.50$ \\
\hline
\multirow{6}{*}{ CP}  &         \small{N= 1- 0, J=1/2-1/2, F= 0- 1}&    47011.71& --&-- &$\leq 5.21$&$\leq 2.57$ \\

  &\small{N= 1- 0, J=1/2-1/2, F= 1- 0 }&47124.24        & --&-- &$\leq 6.32$&$\leq 3.11$ \\

  &\small{N= 1- 0, J=1/2-1/2, F= 1- 1 }&47256.93        & --&-- &$\leq 5.70$&$\leq 2.80$ \\

  &\small{N= 1- 0, J=3/2-1/2, F= 1- 0 }&47979.86        & --&-- &$\leq 7.74$&$\leq 3.78$\\

  &\small{N= 1- 0, J=3/2-1/2, F= 2- 1} &47982.88        & --&-- &$\leq 6.23$&$\leq 3.04$\\

  &\small{N= 1- 0, J=3/2-1/2, F= 1- 1 }&48112.5 & --&-- &$\leq 6.15$ &$\leq 3.00$ \\
\hline
HCP &1-0&39951.90& --&-- &$\leq 2.35$&$\leq 1.26$ \\
\hline
\multirow{3}{*}{ CCP}  &J=5/2-3/2, $\Omega$=1/2, F= 3- 2, l=e &31765.86 & --&-- &$\leq 6.45$ &$ \leq 3.87$ \\
 &J=5/2-3/2, $\Omega$=1/2, F= 3- 2, l=f &31832.45       & --&-- &$\leq 8.05$&$\leq 4.82$ \\
 &J=7/2-5/2, $\Omega$=1/2, F= 4- 3, l=f &       44558.92        & --&-- &$\leq 3.48$&$\leq 1.76$ \\
\hline
\multirow{7}{*}{c-C$_{3}$H$_{2}$ }  &4( 4, 0)- 4( 3, 1) &       35360.93        & --&-- &$\leq 2.73$&$\leq 1.55$\\
&       5( 4, 1)- 5( 3, 2)&     42139.19        & --    & --     & $\leq 3.04$ & $\leq 1.58$\\
&4( 3, 1)- 4( 2, 2)&    42231.25& --&-- &$\leq 3.21$ &$\leq 1.67$ \\
&       3( 2, 1)- 3( 1, 2)&     44104.78        &        $6.98 \pm 0.01$        &  $0.44\pm 0.02$ &$53.56 \pm 2.50$&$25.09 \pm 1.23$ \\
&6( 5, 1)- 6( 4, 2)&    44624.90        & --&-- &$\leq 3.69$ &$\leq 1.87$ \\
&       5( 5, 0)- 5( 4, 1)&     46645.05        & --&-- &$\leq 5.30$ &$ \leq 2.63$\\
&       2( 1, 1)- 2( 0, 2)&     46755.61        & $6.980\pm 0.004$      & $0.41\pm 0.01$&$316.88 \pm 3.45$& $136.67 \pm 3.11$\\
\hline
\multirow{4}{*}{c-C$_{3}$H }  &\small{2( 1, 1)- 2( 1, 2), J=3/2-3/2, F= 2- 2}&44272.60     & --&-- &$ \leq 4.06$&$ \leq 2.06$ \\
&\small{2( 1, 1)- 2( 1, 2), J=3/2-3/2, F= 1- 1}&        44281.61        & --&-- &$\leq 3.73$ &$\leq 1.89$ \\
&       \small{2( 1, 1)- 2( 1, 2), J=5/2-5/2, F= 3- 3}&44610.39 & --&-- &$\leq 3.69$ &$\leq 1.87$ \\
&       \small{2( 1, 1)- 2( 1, 2), J=5/2-5/2, F= 2- 2}&         44610.87& --&-- &$\leq 3.92$ &$\leq 1.99$ \\
\hline
\multirow{6}{*}{l-C$_{3}$H }  &J=3/2-1/2, $\Omega$=1/2, F= 1- 1, l=f&32617.02   & --&-- &$\leq 6.99$ &$\leq 4.14$ \\
  &\small{J=3/2-1/2, $\Omega$=1/2, F= 2- 1, l=f}&32627.30       &       $7.073\pm 0.006$  &$0.56\pm 0.02$          &$63.93 \pm 1.39$&$38.37 \pm 0.98$ \\
  &\small{J=3/2-1/2, $\Omega$=1/2, F= 1- 0, l=f}&32634.39       &       $7.07\pm 0.01$           &$0.54\pm 0.06$          &$27.49 \pm 1.34$&$15.88 \pm 0.97$ \\
  &\small{J=3/2-1/2, $\Omega$=1/2, F= 2- 1, l=e}&       32660.65        &       $6.974\pm 0.007$          &       $0.58\pm 0.01$ &$61.76 \pm 1.33$&$38.13 \pm 0.88$ \\
  &\small{J=3/2-1/2, $\Omega$=1/2, F= 1- 0, l=e}&32663.36       &       $6.97\pm 0.02$   &       $0.51\pm 0.05$   &$25.61 \pm 1.67$&$13.79 \pm 1.07$ \\
  &\small{J=3/2-1/2, $\Omega$=1/2, F= 1- 1, l=e}&32667.67       &       $6.91\pm 0.03$           &$0.67\pm 0.09$          &$15.21 \pm 1.45$& $10.87 \pm 1.09$ \\
\hline
\multirow{3}{*}{HOCO$^{+}$}  &2( 1, 2)- 1( 1, 1)        &42598.19       &       $6.73\pm 0.10$&$0.72\pm 0.20$             &$6.18 \pm 1.92$&$4.76 \pm 1.24$ \\
 &      2( 0, 2)- 1( 0, 1)&42766.20     &               $7.04\pm 0.03$          &       $0.56\pm 0.07$    &$18.02 \pm 1.78$&$10.79 \pm 1.06$ \\
 &2( 1, 1)- 1( 1, 0)&42926.80   &               --      &--                      &$\leq 3.02$&$\leq 1.56$\\
\hline
\multirow{5}{*}{CCO}  & N= 1- 0, J= 1- 1&       32623.45&               --      &--                      &$ \leq 3.77$&$\leq2.23$ \\
 &      N= 2- 1, J= 1- 1        &       32738.61&               --      &--                      &$\leq 2.73$&$\leq 1.62$ \\
 &      N= 2- 1, J= 1- 2        &       43103.88&               --      &--                      &$\leq 3.22$&$\leq 1.66$ \\
 &N= 2- 1, J= 3- 2      &45826.71&              --      &--                      &$\leq 4.37 $&$\leq 2.18$\\
 &N= 2- 1, J= 2- 1      &46182.19&              --      &--                      &$\leq 4.19 $&$\leq 2.08$\\
\hline
\multirow{2}{*}{C$_{3}$O}  &4-3&38486.86        &$6.77 \pm 0.02$        &       $0.50 \pm 0.04$                &$11.63 \pm 1.36$ & $6.20 \pm 0.46$\\
 &5-4   &48108.50       &               $7.19\pm 0.02$          &$0.42\pm 0.05$                    &$37.66 \pm 3.75$&$16.73 \pm 1.76$\\
\hline

\multirow{17}{*}{CH$_{2}$CHCN }  &4( 1, 4)- 3( 1, 3)    &       37018.92        &               --      &--                      &$\leq 2.67 $&$\leq 1.48$\\
 &      4( 0, 4)- 3( 0, 3)&37904.85     &               --      &--                      &$\leq 2.91$&$\leq 1.60$ \\
 &4( 2, 3)- 3( 2, 2)&37939.62   &               --      &--                      &$\leq 2.63$&$\leq 1.44 $ \\
 &4( 3, 2)- 3( 3, 1)&37952.63   &               --      &--                      &$\leq 2.56$&$\leq 1.40$\\
 &4( 3, 1)- 3( 3, 0)&37952.73   &               --      &--                      &$\leq 2.47$&$\leq 1.36$\\
 &      4( 2, 2)- 3( 2, 1)      &37974.37               &               --      &--                      &$\leq 2.53$&$\leq 1.39$ \\
 &4( 1, 3)- 3( 1, 2)&38847.74   &               $6.61\pm 0.10$  &       $0.54\pm 0.17$            &$3.81 \pm 1.30$&$2.19 \pm 0.72$\\
 &      5( 1, 5)- 4( 1, 4)&46266.93     &               --      &       --               & $\leq 4.88$&$\leq 2.43$\\
 &      5( 0, 5)- 4( 0, 4)&47354.65     &               --      &--                      &$\leq 6.14$&$\leq 3.02$ \\
 &              5( 2, 4)- 4( 2, 3)&47419.79     &               --      &--                      &$\leq 5.72$&$\leq 2.81$\\
 &      4( 1, 3)- 4( 0, 4)&47436.36     &               --      &--                      &$\leq 5.72$&$\leq 2.81$ \\
 &      5( 3, 3)- 4( 3, 2)&47443.53     &               --      &--                      &$\leq 5.63$&$\leq 2.76$ \\
 &5( 3, 2)- 4( 3, 1)&47443.88   &               --      &--                      &$\leq 5.32 $&$\leq 2.61$\\
 &5( 4, 2)- 4( 4, 1)&47445.43   &               --      &--                      &$\leq 5.15$&$\leq 2.53$\\
 &      5( 4, 1)- 4( 4, 0)&47445.43&            --      &--                      &$\leq 5.18$&$\leq 2.54$\\
 &      5( 2, 3)- 4( 2, 2)&47489.23     &               --      &--                      &$\leq 5.23$&$\leq 2.57$ \\
 &      9( 0, 9)- 8( 1, 8)&47499.58&            --      &--                      &$\leq 4.92$&$\leq 2.41$ \\
 &      5( 1, 4)- 4( 1, 3)&48552.56     &               --      &--                      &$\leq 6.94$&$\leq 3.37$ \\
 &      5( 1, 4)- 5( 0, 5)&48634.28&            --      &--                      &$\leq 7.45$&$\leq 3.61$ \\
\hline
\multirow{3}{*}{H$_{2}$CCS}  &3( 1, 3)- 2( 1, 2)        &       33438.37&               --      &--                      &$\leq 2.82 $&$\leq 1.65 $\\
 &3( 0, 3)- 2( 0, 2     &       33611.70&               --      &--                      &$\leq 2.57$&$\leq 1.50$ \\
 &3( 1, 2)- 2( 1, 1)    &33783.23       &               --      &--                      &$\leq 5.71$&$\leq 3.32$ \\
  &     4( 1, 4)- 3( 1, 3)      &44584.34               &               --      &--                      &$\leq 3.49$&$\leq 1.77$ \\
 &4( 2, 3)- 3( 2, 2)    &44810.21               &               --      &--                      &$\leq 3.98$&$\leq 2.01$ \\
 &4( 2, 2)- 3( 2, 1)    &       44810.56        &               --      &--                      &$\leq 3.85$&$\leq 1.94$ \\
 &      4( 0, 4)- 3( 0, 3)      &       44815.32        &               --      &--                      &$\leq 4.03$&$\leq 2.04$ \\
 &4( 1, 3)- 3( 1, 2)    &       45044.18        &               --      &--                      &$\leq 4.29$&$\leq 2.16$ \\
\hline
\multirow{4}{*}{HCO}  &\small{\small{5( 1, 4)- 5( 1, 5), J=$\frac{9}{2}$-$\frac{9}{2}$, F= 4- 4}}&      42222.60        &               --      &--                      &$\leq 3.26$&$\leq 1.69$ \\
 &\small{\small{5( 1, 4)- 5( 1, 5), J=$\frac{9}{2}$-$\frac{9}{2}$, F= 5- 5}}&    42229.19        &               --      &--                      &$\leq 3.12$&$\leq 1.62$ \\
 &\small{\small{5( 1, 4)- 5( 1, 5), J=$\frac{11}{2}$-$\frac{11}{2}$, F= 5- 5}}&    42825.36&               --      &--                      &$\leq 2.96$&$\leq 1.53$ \\
 &\small{\small{5( 1, 4)- 5( 1, 5), J=$\frac{11}{2}$-$\frac{11}{2}$, F= 6- 6}}&    42840.62&               --      &--                      &$\leq 3.80$&$\leq 1.96$\\
\hline
\multirow{6}{*}{t-CH$_{3}$CH$_{2}$OH} &4( 1, 3)- 4( 0, 4)&32742.82&             --      &--                      &$\leq 2.38$&$\leq 1.40$ \\
 &5( 1, 4)- 5( 0, 5)&   36417.24&               --      &--                      &$\leq 2.22$&$\leq 1.24$\\
 &6( 1, 5)- 6( 0, 6)&   41124.95&               --      &--                      &$\leq 2.52$&$\leq 1.33$\\
 &4( 0, 4)- 3( 1, 3)&   46832.80&               --      &--                      &$\leq 4.90$&$\leq 2.42$\\
 &7( 1, 6)- 7( 0, 7)&   46980.15&               --      &--                      &$\leq 5.96 $&$\leq 2.94$\\
 &10( 1,10)- 9( 2, 7)&48079.03&         --      &--                      &$\leq 4.92$&$\leq 2.45$ \\
\hline
\multirow{4}{*}{CH$_{3}$OCH$_{3}$} &5( 1, 4)- 5( 0, 5) EE&39047.30&             --      &--                      &$\leq 2.89$&$\leq 1.56$\\
&       6( 1, 5)- 6( 0, 6) EE&43447.57&         --      &--                      &$\leq 3.55$&$\leq 1.82$ \\
&       7( 1, 6)- 7( 0, 7) EE&48901.45&         --      &--                      &$\leq 7.90$&$\leq 3.82$ \\
&7( 1, 6)- 7( 0, 7) AA&48902.63&                --      &--                      &$\leq 7.68$&$\leq 3.71$\\
\hline
\multirow{3}{*}{t-HCOOH} &2( 1, 2)- 1( 1, 1)&43303.71&          --      &--                      &$\leq 3.59$&$\leq 1.85$ \\

&2( 0, 2)- 1( 0, 1)&44911.74&           --      &--                      &$\leq 3.87$&$\leq 1.95$ \\

&2( 1, 1)- 1( 1, 0)&46581.23&           --      &--                      &$\leq 6.05$&$\leq 3.00$ \\
\hline
\multirow{5}{*}{c-HCOOH} &7( 1, 6)- 7( 1, 7)&39295.95&          --      &--                      &$\leq 2.50 $&$\leq 1.35$ \\

&5( 0, 5)- 4( 1, 4)&40778.14&           --      &--                      &$\leq 2.83$&$\leq 1.50$\\

&2( 1, 2)- 1( 1, 1)&42541.36&           --      &--                      &$\leq 3.04$&$\leq 1.58$ \\

&2( 0, 2)- 1( 0, 1)&    43926.44&               --      &--                      &$\leq 4.12$&$\leq 2.10 $\\

&2( 1, 1)- 1( 1, 0)&45351.35&           --      &--                      &$\leq 4.70$&$\leq 2.36$ \\
\hline
\multirow{3}{*}{l-C$_{3}$H$_{2}$} &2(1, 2)-1(1, 1)&41198.34     &       $6.999\pm 0.007$          &$0.41\pm 0.02$                 & $45.91 \pm 1.60$& $19.88 \pm 0.07$\\
&2(0, 2)-1(0,1)& 41584.68       &       $7.04\pm 0.01$          &       $0.49\pm 0.03$           & $39.20 \pm 1.85$& $20.52 \pm 1.72$\\

&2(1, 1)-1(1,0)& 41967.67       &       $7.027\pm 0.008$                &       $0.41\pm 0.03$           & $47.78 \pm 1.70$& $20.63 \pm 0.08$\\
\hline
HONC&2(0, 2)-1(0,1) &43813.37 &         --      &--                      &$\leq 3.02$&$\leq 1.54$\\
\hline
$^{13}$CS&1- 0&46247.56&$6.95\pm 0.06$  &$0.60\pm 0.13$         &$39.98 \pm 6.96$& $25.32 \pm 4.87$\\
\hline

\multirow{2}{*}{C$^{34}$S} &\multirow{2}{*}{1- 0 }&\multirow{2}{*}{48206.94} & $5.69\pm 0.01$                &$0.47\pm 0.03$  & $75.82 \pm 4.25$ & $38.29 \pm 2.14$  \\
 &&&$6.939\pm 0.005$            &$0.53\pm 0.01$  & $212.88 \pm 4.25$ & $120.00 \pm 4.87$  \\
\hline
C$^{33}$S & 1- 0&48585.89&              --      &--                      &$\leq 8.60$&$\leq 4.17$\\
\hline
\multirow{4}{*}{CC$^{34}$S} &   N= 2- 1, J= 3- 2& 33111.84&     $7.06\pm 0.05$ &$0.48\pm 0.09$    & $9.57 \pm 1.30$& $4.88 \pm 0.79$ \\
&       N= 3- 2, J= 3- 2&       38015.23 &              --      &--                      &$\leq 2.49$&$\leq 1.37$ \\
&       N= 4- 3, J= 3- 2 &42918.18 &--&--& $\leq 4.20$ & $\leq 2.17$  \\
& N= 3- 2, J= 4- 3&     44497.60 & --&--&$\leq 4.8$  & $\leq 2.43$  \\
\hline
\multirow{3}{*}{CCC$^{34}$S} &J=6-5 & 33844.24&         --      &--                      &$\leq 2.71$&$\leq 1.58$ \\
&J=7-6 &39484.87 &              --      &--                      &$\leq 2.73 $&$\leq 1.57$ \\
&J=8-7 &45125.46&               --      &--                      &$\leq 5.27$&$\leq 2.65$ \\
\hline
HC$^{34}$S$^{+}$ & 1-0&41983.06 &               --      &--                      &$\leq 3.00$&$\leq 1.57$ \\
\hline
\multirow{2}{*}{OC$^{34}$S} &   3-2 & 35596.87&         --      &--                      &$\leq 2.38$&$\leq 1.35$ \\
&4-3 & 47462.35&                --      &--                      &$\leq 6.08$&$\leq 2.98$ \\
\hline
\multirow{1}{*}{H$_{2}$C$^{34}$S }&1( 0, 1)- 0( 0, 0) &         33765.80&               --      &--                      &$\leq 2.51$&$\leq 1.46$\\
\hline

\end{longtable}

\newpage
\small
\begin{longtable}{l l l l l l l}
\caption{\label{Line parameters of the detected molecules in B213-C16}Line parameters of the detected molecules in B213-C16.}\\
\hline\hline
\\
 Species&Line & Frequency & v & $\Delta \text{v}$ & $T_\text{MB}$ & $\int T_\text{MB}\text{dv}$\\
            &                &\small{(MHz)} & \small{(km$ \ \text{s}^{-1}$) }   &    \small{(km$ \ \text{s}^{-1}$) }         & \small{($\text{mK}$)} &\small{(mK $  $\text{km}$ \ \text{s}^{-1}$)}\\
\hline
\endfirsthead
\caption{Continued.}\\
\hline\hline
\\
 Species&Line & Frequency & v & $\Delta \text{v}$ & $T_\text{MB}$ & $\int T_\text{MB}\text{dv}$\\
            &                &\small{(MHz)} & \small{(km$ \ \text{s}^{-1}$) }   &    \small{(km$ \ \text{s}^{-1}$) }         & \small{($\text{mK}$)} &\small{(mK $  $\text{km}$ \ \text{s}^{-1}$)}\\
\hline
\endhead
\hline
\endfoot
\hline
\endlastfoot
\multirow{6}{*}{C$\mathrm{H}_{3}$CHO}&\small{2( 1, 2)- 1( 1, 1) A, vt=0 }& 37464.20 &$ 6.42\pm 0.10 $&$0.38 \pm 0.76$ &$16.74 \pm 4.55$ & $6.69 \pm 2.62$\\

&\small{2( 1, 2)- 1( 1, 1) E, vt=0}& 37686.93&--         &--    & $\leq 7.93$&$\leq 4.37$ \\

&\small{2( 0, 2)- 1( 0, 1) E, vt=0}& 38506.03& $6.59 \pm 0.03 $ & $0.53\pm 0.06 $&$59.89 \pm 5.03$ &$33.87 \pm 3.55$\\

&\small{2( 0, 2)- 1( 0, 1) A, vt=0}& 38512.08& $ 6.64\pm 0.03$  & $0.51 \pm 0.05 $&$49.15 \pm 4.27$ &$26.66 \pm 2.39$\\

&\small{2( 1, 1)- 1( 1, 0) E, vt=0} & 39362.54&$ 6.65\pm 0.07$ &$0.67 \pm 0.16 $ &$19.32 \pm 4.21$ &$13.87 \pm 2.79 $\\

&\small{2( 1, 1)- 1( 1, 0) A, vt=0 }& 39594.29& $6.71 \pm 0.05$ &$0.55 \pm0.10 $       &$26.69 \pm 4.65$ &$15.48 \pm 2.66$\\

\hline
\multirow{5}{*}{ C$\mathrm{H}_{3}$OH}&\small{4( 1, 4)- 3(- 0, 3) E, vt=0 }&36169.26&$6.74 \pm 0.03 $ & $0.53 \pm0.09 $& $35.93 \pm 3.49 $ &$20.43 \pm  2.35$\\

&\small{7( 2, 5)- 8( 1, 8) E, vt=0}&  37703.76 &--       &--    & $\leq 8.66$&$\leq 4.77$\\

 &\small{7( 0, 7)- 6( 1, 6) A, vt=0}&44069.37&  -- &--&$\leq 13.52$ &$\leq 6.89$\\

&\small{1( 0, 1)- 0( 0, 0) A, vt=0}&48372.46 &  $6.647 \pm 0.007$&$0.53 \pm 0.02$& $438.14 \pm 12.31$ &$248.06 \pm 6.82 $\\
&\small{1(- 0, 1)- 0(- 0, 0) E, vt=0}&48376.89& $6.72\pm 0.08 $&$0.50 \pm 0.19 $&$43.19 \pm 13.56$ &$23.01 \pm 7.11$\\
\hline
CS&1-0&48990.95&$6.597 \pm 0.003 $ &$ 0.894 \pm 0.007 $ &$1733.52 \pm 15.76$ &$1649.90 \pm 10.73 $  \\
\hline
\multirow{1}{*}{$\mathrm{H}_{2}$CS}& 1( 0, 1)- 0( 0, 0)& 34351.43& $6.766 \pm 0.004 $     &$0.61 \pm 0.01 $& $282.06 \pm 3.47$&$184.28 \pm 2.40$\\
\hline
\multirow{2}{*}{ HCNS}  &3-2&36908.48&-- &-- &$\leq 8.99$  &$\leq 5.01$\\
&4-3&49211.11&  $6.81 \pm 0.08 $&$0.52 \pm 0.19 $& $46.00 \pm 13.06$ &$25.52 \pm 7.16 $\\
\hline
\multirow{10}{*}{ HSCN}  &3( 1, 3)- 2( 1, 2)&34228.19&-- &-- &$\leq 6.89$&$\leq 3.98$\\

 & 3( 2, 2)- 2( 2, 1)&34405.13 &-- &-- &$\leq 8.02$  &$\leq 4.62$\\

&3( 2, 1)- 2( 2, 0)&34405.28&-- &-- &$\leq 8.38$ &$\leq 4.83$\\

 &3( 0, 3)- 2( 0, 2)&34408.63 & $6.55 \pm 0.07 $&$0.97 \pm 0.18 $& $18.62 \pm 3.40$ &$19.18 \pm 2.94 $\\

& 3( 1, 2)- 2( 1, 1)&34587.50&-- &-- &$\leq 7.76$&$ \leq 4.46$\\

 & 4( 1, 4)- 3( 1, 3)&45637.36 &-- &-- &$\leq 14.06$ &$\leq 7.04$\\

& 4( 2, 3)- 3( 2, 2)&45873.29&-- &-- &$\leq 14.37$ &$\leq 7.17$\\

 & 4( 2, 2)- 3( 2, 1)&45873.66&-- &-- &$\leq 15.63$ &$\leq 7.80$\\

& 4( 0, 4)- 3( 0, 3)&45877.81&  $6.69 \pm 0.11 $&$0.77 \pm 0.24 $& $28.11 \pm 9.01$ &$23.05 \pm 5.97 $\\

& 4( 1, 3)- 3( 1, 2)& 46116.44&-- &-- &$\leq 14.50$&$\leq 7.22$\\
\hline
HCS$^{+}$& 1-0&42674.20&        $6.68 \pm 0.01 $&$0.59 \pm 0.03 $& $97.43 \pm 4.72$ &$ 61.17\pm 2.83 $\\
\hline
\multirow{2}{*}{ OCS}  &3-2&36488.81 &  $6.68 \pm0.05 $&$1.06 \pm 0.13 $& $32.49 \pm 4.08$ &$36.69 \pm 3.67 $\\

&4-3&48651.60 & $6.46 \pm 0.13$&$0.79 \pm 0.20 $& $31.69 \pm 12.19$ &$26.80 \pm 7.33 $\\
\hline
SO$_{2}$ & 6( 2, 4)- 7( 1, 7)& 44052.86 &-- &-- &$\leq 14.60$ &$\leq 7.44$\\
\hline
\multirow{4}{*}{ CCS}  &N= 2- 1, J= 3- 2&33751.37&      $6.683 \pm 0.001 $&$0.611 \pm 0.003 $& $993.88 \pm 3.84$ &$646.11 \pm 2.62 $\\

&N= 3- 2, J= 3- 2&38866.42&     $6.696 \pm 0.008 $&$0.62 \pm 0.02 $& $163.66 \pm 4.82$ &$108.66 \pm 3.03 $\\

&N= 4- 3, J= 3- 2&43981.02&     $6.69 \pm 0.01 $&$0.52 \pm 0.03 $& $180.73 \pm 8.35$ &$99.70 \pm 4.66 $\\

&N= 3- 2, J= 4- 3&      45379.05&       $6.811 \pm 0.002 $&$0.530 \pm 0.003 $& $1319.99 \pm 7.59$ &$745.30 \pm 4.19 $\\
\hline
\multirow{3}{*}{C$_{3}$S}& 6-5 &34684.37 &      $6.751 \pm 0.005$&$0.59 \pm 0.01$& $318.28 \pm 4.70$ &$200.77 \pm 3.09 $\\

& 7-6&40465.02 &        $6.759 \pm 0.004 $&$ 0.592 \pm 0.009$& $343.10 \pm 5.17$ &$216.03 \pm 3.12 $\\

& 8-7&46245.62 &        $6.76 \pm 0.02 $&$ 0.60\pm 0.04$& $356.55 \pm 8.42$ &$226.45 \pm 13.28 $\\
\hline
CH$_{3}$CCH & 2( 0)- 1( 0)& 34183.41&   $6.75 \pm 0.03 $&$0.58 \pm 0.09 $& $197.05 \pm 4.37$ &$120.87 \pm 12.43 $\\
\hline
\multirow{2}{*}{ CH$_{3}$CN}  &2( 1)- 1( 1) &36794.77 &--       &--     &$\leq 9.89$&$\leq 5.12$\\
&2( 0)- 1( 0) &36795.47 &       $6.05 \pm 0.05 $&$1.13 \pm 0.13$& $53.10 \pm 4.71$ &$63.59 \pm 6.10$\\
\hline
\multirow{7}{*}{ HC$_{5}$N} &J = 12 - 11 &31951.77 &    $6.727 \pm 0.002 $&$0.638 \pm 0.003 $& $868.24 \pm 4.08$ &$589.92 \pm 0.98 $\\
&J = 13 - 12 &34614.39& $6.752 \pm 0.001 $&$0.555 \pm 0.003 $& $954.20 \pm 3.64$ &$563.80 \pm 2.32 $\\
&J = 14 - 13 &  37276.99&       $6.822 \pm 0.001 $&$0.541 \pm 0.003 $& $908.68 \pm 4.05$ &$523.78 \pm 2.49 $\\
& J = 15 - 14&39939.59 &        $6.889 \pm 0.002 $&$0.534 \pm 0.003 $& $822.36 \pm 4.53$ &$467.15 \pm 2.66 $\\
&J = 16 - 15 & 42602.15&        $6.777 \pm 0.003 $&$0.506 \pm 0.007 $& $475.58 \pm 5.59$ &$256.00 \pm 3.06 $\\
&J = 17 - 16 &45264.72 &        $6.747 \pm 0.003 $&$0.537 \pm 0.007 $& $626.89 \pm 7.28$ &$358.13 \pm 4.04$\\
&J = 18 - 17 &          47927.27&       $6.770 \pm 0.004 $&$0.468 \pm 0.009 $& $617.10 \pm 9.65$ &$307.36 \pm 4.93 $\\
\hline
\multirow{13}{*}{ HC$_{7}$N}   &J = 28 - 27 &31583.70   &       $6.69 \pm 0.02 $&$0.62 \pm 0.04 $& $92.76 \pm 4.26$ &$61.10 \pm 3.14 $\\

  &J = 29 - 28 &32711.68        &       $ 6.84\pm 0.01 $&$0.57 \pm 0.05 $& $83.94 \pm 3.75$ &$55.00 \pm 2.70 $\\

  &J = 30 - 29 &33839.63        &       $6.64 \pm 0.02 $&$ 0.55 \pm 0.04 $& $72.69 \pm 3.50$ &$42.23 \pm 2.27 $\\

  &J = 31 - 30 &34967.59        &       $6.78 \pm 0.01 $&$0.41 \pm 0.10 $& $82.07 \pm 3.66$ &$35.55 \pm 1.86 $\\

  &J = 32 - 31 &36095.53&       $ 6.67\pm 0.02 $&$0.49 \pm 0.05$& $61.80 \pm 4.05$ &$32.24 \pm 2.47$\\

  &J = 33 - 32 &37223.49        &       $6.77 \pm 0.02 $&$ 0.53\pm 0.07$& $54.06 \pm 4.61$ &$30.36 \pm 2.79 $\\

  &J = 34 - 33 &38351.45        &       $ 6.84\pm 0.02 $&$0.55 \pm 0.06 $& $48.61 \pm 4.13$ &$28.53 \pm 2.56 $\\

  &     J = 35 - 34 &39479.41   &       $6.97 \pm 0.03 $&$ 0.48 \pm 0.06 $& $43.22 \pm 4.49$ &$22.12 \pm 2.49 $\\

  &J = 36 - 35 &40607.33        &       $ 6.69 \pm 0.05 $&$0.52 \pm 0.10 $& $30.92 \pm 5.18$ &$17.19 \pm 2.91 $\\

  &J = 37 - 36&41735.26 &       $6.75 \pm 0.04 $&$0.53 \pm 0.10$& $35.14 \pm 5.33$ &$19.82 \pm 3.11$\\

  & J = 38 - 37&42863.20        &       $6.09 \pm 0.08 $&$0.47 \pm 0.14 $& $20.80 \pm 6.26$ &$10.32 \pm 3.16 $\\

  &J = 39 - 38&43991.13 &       -- &-- &$\leq 17.83$&$\leq 9.10$ \\

  &J = 40 - 39&45119.06         &       -- &-- &$\leq 13.69$&$\leq 6.89$ \\

  &J = 41 - 40&46246.98 & --&-- &$\leq 14.57$&$\leq 7.25$\\

  &     J = 42 - 41&47374.90    & --&-- &$\leq 17.65$&$\leq 8.67$ \\

  &J = 43 - 42&48502.81 & --&-- &$\leq 19.37$&$\leq 9.41$ \\

  &J = 44 - 43&49630.72 & --&-- &$\leq 35.28$&$\leq 16.93$ \\
\hline
PO$^{+}$ &1-0&47024.25& --&-- &$\leq 15.39 $&$\leq 7.59$ \\
\hline
\multirow{6}{*}{ CP}  &         \small{N= 1- 0, J=1/2-1/2, F= 0- 1}&    47011.71& --&-- &$\leq 16.36$ &$\leq 8.07$\\

  &\small{N= 1- 0, J=1/2-1/2, F= 1- 0 }&47124.24        & --&-- &$\leq 17.47 $&$\leq 8.61$ \\

  &\small{N= 1- 0, J=1/2-1/2, F= 1- 1 }&47256.93        & --&-- &$\leq 18.35$&$\leq 9.03$ \\

  &\small{N= 1- 0, J=3/2-1/2, F= 1- 0 }&47979.86        & --&-- &$\leq 28.77$ &$\leq 14.05$ \\

  &\small{N= 1- 0, J=3/2-1/2, F= 2- 1} &47982.88        & --&-- &$\leq 16.8$ &$\leq 8.20$\\

  &\small{N= 1- 0, J=3/2-1/2, F= 1- 1 }&48112.5 & --&-- &$\leq 21.87$&$\leq 10.66$ \\
\hline
HCP &1-0&39951.90& --&-- &$\leq 9.08$&$\leq 4.86$ \\
\hline
\multirow{3}{*}{ CCP}  &J=5/2-3/2, $\Omega$=1/2, F= 3- 2, l=e &31765.86 & --&-- &$\leq 9.21$ &$\leq 5.53$ \\
 &J=5/2-3/2, $\Omega$=1/2, F= 3- 2, l=f &31832.45       & --&-- &$\leq 7.86$&$\leq 4.71$ \\
 &J=7/2-5/2, $\Omega$=1/2, F= 4- 3, l=f &       44558.92        & --&-- &$\leq 14.30$&$\leq 7.25$ \\
\hline
\multirow{7}{*}{c-C$_{3}$H$_{2}$ }  &4( 4, 0)- 4( 3, 1) &       35360.93        & --&-- &$\leq 6.59$ &$\leq 3.74$\\
&       5( 4, 1)- 5( 3, 2)&     42139.19        & --&-- &$\leq 11.01$ &$\leq 5.73$\\
&4( 3, 1)- 4( 2, 2)&    42231.25& --&-- &$\leq 8.90$ &$\leq 4.63$ \\
&       3( 2, 1)- 3( 1, 2)&     44104.78        & --&-- &$\leq 20.97$ &$\leq 10.68$\\
&6( 5, 1)- 6( 4, 2)&    44624.90        & --&-- &$\leq 12.86$ &$\leq 6.51$ \\
&       5( 5, 0)- 5( 4, 1)&     46645.05        & --&-- &$\leq 13.50$ &$\leq 6.68$ \\
&       2( 1, 1)- 2( 0, 2)&     46755.61        &       $6.669 \pm 0.004$&$0.490 \pm 0.009 $& $679.77 \pm 10.01$ &$354.63 \pm 5.27 $\\
\hline
\multirow{4}{*}{c-C$_{3}$H }  &\small{2( 1, 1)- 2( 1, 2), J=3/2-3/2, F= 2- 2}&44272.60     & --&-- &$\leq 19.91$&$\leq 10.12$ \\
&\small{2( 1, 1)- 2( 1, 2), J=3/2-3/2, F= 1- 1}&        44281.61        &       --&--&$\leq 19.85$ &$\leq 10.09$\\
&       \small{2( 1, 1)- 2( 1, 2), J=5/2-5/2, F= 3- 3}&44610.39 & --&-- &$\leq 11.96$ &$\leq 6.06$\\
&       \small{2( 1, 1)- 2( 1, 2), J=5/2-5/2, F= 2- 2}&         44610.87& --&-- &$\leq 11.80$ &$\leq 5.98$ \\
\hline
\multirow{6}{*}{l-C$_{3}$H }  &J=3/2-1/2, $\Omega$=1/2, F= 1- 1, l=f&32617.02   &       $6.81 \pm 0.08 $&$0.67 \pm 0.19 $& $23.80 \pm 3.67$ &$17.05 \pm 4.04 $\\
  &\small{J=3/2-1/2, $\Omega$=1/2, F= 2- 1, l=f}&32627.30       &       $6.75 \pm 0.02$&$0.62 \pm 0.04 $& $67.94 \pm 3.82$ &$44.69 \pm 2.58$\\
  &\small{J=3/2-1/2, $\Omega$=1/2, F= 1- 0, l=f}&32634.39       &       $6.74 \pm 0.04 $&$0.76 \pm 0.09 $& $30.43 \pm 3.21 $ &$24.50 \pm 2.52 $\\
  &\small{J=3/2-1/2, $\Omega$=1/2, F= 2- 1, l=e}&       32660.65        &       $6.66 \pm 0.02$&$0.64 \pm 0.05 $& $67.62 \pm 4.04$ &$46.35 \pm 2.84 $\\
  &\small{J=3/2-1/2, $\Omega$=1/2, F= 1- 0, l=e}&32663.36       &       $6.69 \pm 0.05 $&$ 0.54 \pm 0.11 $& $28.21 \pm 4.02$ &$16.12 \pm 2.59$\\
  &\small{J=3/2-1/2, $\Omega$=1/2, F= 1- 1, l=e}&32667.67       &               --      &--                      &$\leq 7.31 $&$\leq 4.32$ \\
\hline
\multirow{3}{*}{HOCO$^{+}$}  &2( 1, 2)- 1( 1, 1)        &42598.19       &               --      &--                      &$\leq 8.41$&$\leq 4.36$\\
 &      2( 0, 2)- 1( 0, 1)&42766.20     &       $6.91 \pm 0.06 $&$0.51 \pm 0.12 $& $26.89 \pm 6.02$ &$14.61 \pm 3.25 $\\
 &2( 1, 1)- 1( 1, 0)&42926.80   &               --      &--                      &$\leq 10.14$&$\leq 5.23$ \\
\hline
\multirow{5}{*}{CCO}  & N= 1- 0, J= 1- 1&       32623.45&               --      &--                      &$\leq 7.61 $&$\leq 4.51$ \\
 &      N= 2- 1, J= 1- 1        &       32738.61&               --      &--                      &$\leq 6.83$&$\leq 4.04$ \\
 &      N= 2- 1, J= 1- 2        &       43103.88&               --      &--                      &$\leq 10.15$&$\leq 5.23$\\
 &N= 2- 1, J= 3- 2      &45826.71&              --      &--                      &$\leq 15.01$&$\leq 7.50$ \\
 &N= 2- 1, J= 2- 1      &46182.19&              --      &--                      &$\leq 13.72$&$\leq 6.83$ \\
\hline
\multirow{2}{*}{C$_{3}$O}  &4-3&38486.86        &       $6.49 \pm 0.08 $&$0.89 \pm 0.25 $& $21.47 \pm 4.63$ &$21.27 \pm 4.14 $\\
 &5-4   &48108.50       &       $ 6.96 \pm 0.11$&$ 0.80 \pm 0.21 $& $40.16 \pm 13.94$ &$34.12 \pm 8.74 $\\
\hline
\multirow{20}{*}{CH$_{2}$CHCN }  &4( 1, 4)- 3( 1, 3)    &       37018.92        &               --      &--                      &$\leq 10.26$&$\leq 5.70$\\
 &      4( 0, 4)- 3( 0, 3)&37904.85     &       $6.58 \pm 0.03 $&$0.61 \pm 0.08 $& $47.64 \pm 5.11$ &$30.81 \pm 3.39 $\\
 &4( 2, 3)- 3( 2, 2)&37939.62   &               --      &--                      &$\leq 7.77$&$\leq 4.26$\\\
 &4( 3, 2)- 3( 3, 1)&37952.63   &               --      &--                      &$\leq 7.59$&$\leq 4.17$\\
 &4( 3, 1)- 3( 3, 0)&37952.73   &               --      &--                      &$\leq 7.59$&$\leq 4.17$\\
 &      4( 2, 2)- 3( 2, 1)      &37974.37               &               --      &--                      &$\leq 7.54$&$\leq 4.14$ \\
 &4( 1, 3)- 3( 1, 2)&38847.74   &       $6.41 \pm 0.08 $&$0.81 \pm 0.17 $& $12.27 \pm 4.61$ &$10.53 \pm 2.01 $\\
 &      5( 1, 5)- 4( 1, 4)&46266.93     &       $6.83 \pm 0.11 $&$1.15 \pm 0.23 $& $31.44 \pm 8.97$ &$38.78 \pm 7.07 $\\
 &      5( 0, 5)- 4( 0, 4)&47354.65     &       $6.75 \pm 0.06$&$0.65 \pm 0.12 $& $45.31 \pm 10.20$ &$31.50 \pm 5.77$\\
 &              5( 2, 4)- 4( 2, 3)&47419.79     &               --      &--                      &$\leq 16.71$&$\leq 8.21$\\
 &      4( 1, 3)- 4( 0, 4)&47436.36     &               --      &--                      &$\leq 18.37$&$\leq 9.02 $\\
 &      5( 3, 3)- 4( 3, 2)&47443.53     &               --      &--                      &$\leq 18.56$&$\leq 9.11$\\
 &5( 3, 2)- 4( 3, 1)&47443.88   &               --      &--                      &$\leq 17.43$&$\leq 8.56$\\
 &5( 4, 2)- 4( 4, 1)&47445.43   &               --      &--                      &$\leq 18.40$&$\leq 9.04$ \\
 &      5( 4, 1)- 4( 4, 0)&47445.43&            --      &--                      &$\leq 18.40$&$\leq 9.04$\\
 &      5( 2, 3)- 4( 2, 2)&47489.23     &               --      &--                      &$\leq 15.38$&$\leq 7.55$ \\
 &      9( 0, 9)- 8( 1, 8)&47499.58&            --      &--                      &$\leq 14.61$&$\leq 7.17$\\
 &      5( 1, 4)- 4( 1, 3)&48552.56     &       $6.61 \pm 0.11$&$0.87 \pm 0.20 $& $38.43 \pm 12.18$ &$35.70 \pm 7.86$\\
 &      5( 1, 4)- 5( 0, 5)&48634.28&            --      &--                      &$\leq 14.67$&$\leq 7.11$ \\
\hline
\multirow{4}{*}{H$_{2}$CCS}  &3( 1, 3)- 2( 1, 2)        &       33438.37&               --      &--                      &$\leq 6.12$&$\leq 3.58$\\
 &3( 0, 3)- 2( 0, 2)    &       33611.70&               --      &--                      &$\leq 7.20$&$\leq 4.20$ \\
 &3( 1, 2)- 2( 1, 1)    &33783.23       &               --      &--                      &$\leq 18.86$&$\leq 10.98 $ \\
 &      4( 1, 4)- 3( 1, 3)      &44584.34               &               --      &--                      &$\leq 12.78$&$\leq 6.47$ \\
 &4( 2, 3)- 3( 2, 2)    &44810.21               &               --      &--                      &$\leq 11.41$&$\leq 5.77$ \\
 &4( 2, 2)- 3( 2, 1)    &       44810.56        &               --      &--                      &$\leq 11.80$&$\leq 5.96$ \\
 &      4( 0, 4)- 3( 0, 3)      &       44815.32        &               --      &--                      &$\leq 12.85$&$\leq 6.49$ \\
 &4( 1, 3)- 3( 1, 2)    &       45044.18        &               --      &--                      &$\leq 10.98$&$\leq 5.53$\\
\hline
\multirow{1}{*}{HCO}  &\small{\small{5( 1, 4)- 5( 1, 5), J=$\frac{9}{2}$-$\frac{9}{2}$, F= 4- 4}}&      42222.60        &               --      &--                      &$\leq 10.37$&$\leq 5.40$ \\
 &\small{\small{5( 1, 4)- 5( 1, 5), J=$\frac{9}{2}$-$\frac{9}{2}$, F= 5- 5}}&    42229.19        &               --      &--                      &$\leq 9.37$&$\leq 4.88$ \\
 &\small{\small{5( 1, 4)- 5( 1, 5), J=$\frac{11}{2}$-$\frac{11}{2}$, F= 5- 5}}&    42825.36&               --      &--                      &$\leq 12.29$&$\leq 6.35$\\
 &\small{\small{5( 1, 4)- 5( 1, 5), J=$\frac{11}{2}$-$\frac{11}{2}$, F= 6- 6}}&    42840.62&               --      &--                      &$\leq 10.31$&$\leq 5.33$ \\
\hline
\multirow{6}{*}{t-CH$_{3}$CH$_{2}$OH} &4( 1, 3)- 4( 0, 4)&32742.82&             --      &--                      &$\leq 7.27$&$\leq 4.30$\\
 &5( 1, 4)- 5( 0, 5)&   36417.24&               --      &--                      &$\leq 6.82$&$\leq 3.82$\\
 &6( 1, 5)- 6( 0, 6)&   41124.95&               --      &--                      &$\leq 8.58$&$\leq 4.52$\\
 &4( 0, 4)- 3( 1, 3)&   46832.80&               --      &--                      &$\leq 13.80$&$\leq 6.82$\\
 &7( 1, 6)- 7( 0, 7)&   46980.15&               --      &--                      &$\leq 15.9$&$\leq 7.84$ \\
 &10( 1,10)- 9( 2, 7)&48079.03&         --      &--                      &$\leq 17.29$&$\leq 8.43$ \\
\hline
\multirow{4}{*}{CH$_{3}$OCH$_{3}$} &5( 1, 4)- 5( 0, 5) EE&39047.30&             --      &--                      &$\leq 9.79$&$\leq 5.30$\\
&       6( 1, 5)- 6( 0, 6) EE&43447.57&         --      &--                      &$\leq 11.84$&$\leq 6.07$ \\
&       7( 1, 6)- 7( 0, 7) EE&48901.45&         --      &--                      &$\leq 19.69$&$\leq 9.52$ \\
&7( 1, 6)- 7( 0, 7) AA&48902.63&        -- &--& $\leq 18.99$ &$\leq 9.19$\\
\hline
\multirow{3}{*}{t-HCOOH} &2( 1, 2)- 1( 1, 1)&43303.71&          --      &--                      &$\leq 10.02$&$\leq 5.15$ \\

&2( 0, 2)- 1( 0, 1)&44911.74&           --      &--                      &$\leq 13.12$&$\leq 6.62$ \\

&2( 1, 1)- 1( 1, 0)&46581.23&           --      &--                      &$\leq 13.01$&$\leq 6.44$ \\
\hline
\multirow{7}{*}{c-HCOOH} &7( 1, 6)- 7( 1, 7)&39295.95&          --      &--                      &$\leq 8.86$&$\leq 4.78$ \\

&5( 0, 5)- 4( 1, 4)&40778.14&           --      &--                      &$\leq 8.38$&$\leq 4.44$\\

&2( 1, 2)- 1( 1, 1)&42541.36&           --      &--                      &$\leq 10.93$&$\leq 5.67$\\

&2( 0, 2)- 1( 0, 1)&    43926.44&               --      &--                      &$\leq 12.42$&$\leq 6.34$\\

&2( 1, 1)- 1( 1, 0)&45351.35&           --      &--                      &$\leq 12.54$&$\leq 6.30$\\
\hline
\multirow{3}{*}{l-C$_{3}$H$_{2}$} &2(1, 2)-1(1, 1)&41198.34&    $6.68 \pm 0.02 $&$0.51 \pm 0.05$& $69.31 \pm 5.25$ &$37.64 \pm 3.02 $\\
&2(0, 2)-1(0,1)& 41584.68       &       $6.71 \pm 0.02$&$0.41 \pm 0.08$& $59.51 \pm 5.71$ &$26.03 \pm 3.22$\\

&2(1, 1)-1(1,0)& 41967.67       &       $6.70 \pm 0.02 $&$0.42 \pm 0.05$& $78.47 \pm 4.92$ &$35.29 \pm 2.62$\\
\hline
HONC&2(0, 2)-1(0,1) &43813.37 &         --      &--                      &$\leq 12.92$&$\leq 6.60$ \\
\hline
$^{13}$CS&1- 0&46247.56&        $6.65 \pm 0.08$&$0.68 \pm 0.18 $& $164.95 \pm 9.66$ &$119.50 \pm 27.67 $\\
\hline
C$^{34}$S &1- 0 &48206.94 &     $6.671 \pm 0.006  $&$0.616 \pm 0.02 $& $542.63 \pm 12.11$ &$355.58 \pm 7.08$\\
\hline
C$^{33}$S & 1- 0&48585.89&      $ 6.70 \pm 0.04 $&$0.42 \pm 0.10$& $65.88 \pm 12.57$ &$29.29 \pm 5.96$\\
\hline
\multirow{4}{*}{CC$^{34}$S} &   N= 2- 1, J= 3- 2& 33111.84&     $6.72 \pm 0.03$&$0.69 \pm 0.06$& $49.44 \pm 3.84$ &$36.33 \pm 2.83$\\
&       N= 3- 2, J= 3- 2&       38015.23 &              --      &--                      &$\leq 8.47$&$\leq 4.64$ \\
&       N= 4- 3, J= 3- 2 &42918.18&             --      &--                      &$\leq 10.34$&$\leq 5.34$\\
& N= 3- 2, J= 4- 3&     44497.60 &      $6.72 \pm 0.03$&$0.56 \pm 0.07$& $69.02 \pm 7.25$ &$41.30 \pm 4.14$\\
\hline
\multirow{3}{*}{CCC$^{34}$S} &J=6-5 & 33844.24& $6.53 \pm 0.03$&$0.34 \pm 0.70 $& $23.18 \pm 3.82$ &$8.34 \pm 1.95$\\
&J=7-6 &39484.87 &              --      &--                      &$\leq 8.14$&$\leq 4.38$ \\
&J=8-7 &45125.46&               --      &--                      &$\leq 11.30$&$\leq 5.69$ \\
\hline
HC$^{34}$S$^{+}$ & 1-0&41983.06 &               --      &--                      &$\leq 9.55$&$\leq 4.98$ \\
\hline
\multirow{2}{*}{OC$^{34}$S} &   3-2 & 3596.87&          --      &--                      &$\leq 7.57$&$\leq 4.29$\\
&4-3 & 47462.35&                --      &--                      &$\leq 15.26$&$\leq 7.49$ \\
\hline
\multirow{1}{*}{H$_{2}$C$^{34}$S }&1( 0, 1)- 0( 0, 0) &         33765.80&               --      &--                      &$\leq 7.08$&$\leq 4.12$\\

\hline
\end{longtable}

\small
\begin{longtable}{l l l l l l}

\caption{\label{Table: Column densities and abundances}Values of $T_{rot}$, $N_{tot}$ and fractional abundances relative to H$_{2}$ calculated for both sources.}\\
\hline\hline
\\
Species& Source & $T_{rot}$ (K) & $\text{N}_{tot} \ (\text{cm}^{-2})$ & [X/H$_{2}$] \\
\hline
\endfirsthead
\caption{Continued.}\\
\hline\hline
\\
Species& Source & $T_{rot}$ (K) & $\text{N}_{tot} \ (\text{cm}^{-2})$ & [X/H$_{2}$] \\
\hline
\endhead
\hline
\endfoot
\hline
\endlastfoot

\hline
\multirow{2}{*}{CH$_{3}$CHO}& C2&8  & $(4.80 \pm 0.55) \times 10^{11}$ & $(2.297 \pm 0.26) \times 10^{-11}$ \\
& C16 &8  & $(14.97 \pm 1.34 )\times 10^{11}$ & $(6.04 \pm 0.54) \times 10^{-11}$\\
\hline
\multirow{4}{*}{CH$_{3}$OH} &\multirow{2}{*}{C2} & $8.48 \pm 3.41$  & $(7.73 \pm 6.93  )\times 10^{12}$ & $(3.70 \pm 3.32) \times 10^{-10}$ \\
 & & $8$& $(37.73 \pm 0.60  )\times 10^{11}$ & $(18.05 \pm 0.28) \times 10^{-11}$ \\
&\multirow{2}{*}{C16} &$7.93 \pm 1.98$& $(15.24 \pm  9.06 )\times 10^{12}$ & $(6.15 \pm 3.65) \times 10^{-10}$ \\
 & & $8$& $(7.38 \pm  0.20 )\times 10^{12}$  & $(29.80 \pm 0.81) \times 10^{-11}$\\
\hline
\multirow{2}{*}{CS} &C2  & 8 & $(69.98 \pm 0.48) \times 10^{12}$& $(334.83 \pm 2.30) \times 10^{-11}$\\
& C16 & 8&$(136.08 \pm 0.89) \times 10^{11}$ &$(5.488 \pm 0.036) \times 10^{-10}$  \\
\hline
\multirow{2}{*}{H$_{2}$CS}  & C2&8 &$ (11.70 \pm 0.59 )\times 10^{11}$ & $(55.98 \pm 2.82) \times 10^{-12}$\\
 & C16 &8  & $(7.76 \pm 0.10) \times 10^{12}$ & $(31.29 \pm 0.40)  \times 10^{-11}$ \\
\hline
\multirow{2}{*}{HCNS} & C2&8&$\leq 6.16 \times 10^{9}$&$\leq 2.95 \times 10^{-13}$\\
 &C16&8& $(8.06 \pm 2.26) \times 10^{10}$& $ (3.25 \pm 0.91) \times 10^{-12}$ \\
\hline
\multirow{3}{*}{HSCN}& C2&8&$\leq 1.30 \times 10^{10}$ & $\leq 6.22 \times 10^{-13}$ \\
& \multirow{2}{*}{C16}& $5.62 \pm 4.32$& $(1.10 \pm 0.56 )\times 10^{11}$ & $(4.44 \pm 2.26) \times 10^{-12}$\\
&&$8$& $(1.60 \pm 0.24 )\times 10^{11}
$& $(6.45 \pm 0.97) \times 10^{-12}$ \\
\hline
\multirow{2}{*}{HCS$^{+}$}&C2&8& $(1.40 \pm 0.14) \times 10^{11}$& $(66.99 \pm 6.70) \times 10^{-13}$ \\
& C16&8 & $(6.92 \pm 0.32) \times 10^{11}$&  $(2.79 \pm 0.13) \times 10^{-11}$\\
\hline
OCS&C2& 8&$(9.32 \pm 1.62) \times 10^{11}$ & $(44.59 \pm 7.75) \times 10^{-12}$ \\
\multirow{2}{*}{OCS}&\multirow{2}{*}{C16}&$2.62 \pm 0.86$&$(3.87 \pm 1.73 )\times 10^{12} $& $(1.56 \pm 0.70) \times 10^{-10}$\\
&&$8$&$(4.99 \pm 0.50 )\times 10^{12}$& $(2.01 \pm 0.20) \times 10^{-10}$\\
\hline
\multirow{2}{*}{SO$_{2}$}& C2&8&$(7.42 \pm 1.92) \times 10^{12}$ &  $(35.50 \pm 9.19) \times 10^{-11}$\\
& C16&8&$ \leq 12.34 \times 10^{12}$&  $\leq 4.98 \times 10^{-10}$\\
\hline
\multirow{4}{*}{CCS} & \multirow{2}{*}{C2} &$4.88 \pm 0.29 $ & $(1.37 \pm 0.16 )\times 10^{12}$&   $(6.56 \pm 0.77) \times 10^{-11}$\\
& & $8 $ &$(20.30 \pm 0.15) \times 10^{11}$ & $(97.13 \pm 0.72) \times 10^{-12}$ \\
&\multirow{2}{*}{C16} & $4.92 \pm 0.15 $&$(5.94 \pm 0.36  )\times 10^{12}$ & $(2.39 \pm 0.15) \times 10^{-10}$ \\
& & $8 $& $(87.95 \pm 0.36 )\times 10^{11}$& $(35.46 \pm 0.15) \times 10^{-11}$\\
\hline
\multirow{4}{*}{C$_{3}$S}&\multirow{2}{*}{ C2}& $7.04 \pm 0.26 $&$(21.89 \pm 0.91) \times 10^{10}$ & $(104.78 \pm 4.35) \times 10^{-13}$ \\
& & $8 $& $(2.17 \pm 0.10) \times 10^{11}$ & $(104.00 \pm 4.78) \times 10^{-13}$ \\
&\multirow{2}{*}{ C16} &$9.17 \pm 0.53 $& $(14.47 \pm 0.73 )\times 10^{11} $ & $(5.85 \pm 0.29) \times 10^{-11}$\\
& & $8 $&$(13.37 \pm 0.20 )\times 10^{11}$ & $(53.91 \pm 0.81) \times 10^{-12}$ \\
\hline
\multirow{2}{*}{CH$_{3}$CCH}& C2& 8& $(6.76 \pm 0.70) \times 10^{12}$ & $(32.34 \pm 3.35) \times 10^{-11}$\\
& C16& 8& $(64.14 \pm 6.60) \times 10^{12}$ & $(2.59 \pm 0.27) \times 10^{-9}$ \\
\hline
\multirow{2}{*}{CH$_{3}$CN}&C2&8& $(6.09 \pm 1.06) \times 10^{10}$ & $(2.91 \pm 0.51)\times 10^{-12}$ \\
&C16&8 & $(4.70 \pm 0.45) \times 10^{11}$ & $(1.90 \pm 0.18) \times 10^{-11}$ \\
\hline
\multirow{4}{*}{HC$_{5}$N} &\multirow{2}{*}{C2}&$7.61 \pm 0.27 $ &$(69.90 \pm 5.22 )\times  10^{10}$ & $(33.44 \pm 2.50) \times 10^{-12}$\\
&&$8$& $(6.92 \pm 0.10 )\times  10^{11}$& $(3.31 \pm 0.11) \times 10^{-11}$\\
&\multirow{2}{*}{C16} &$7.60 \pm 1.10$ &$(5.63 \pm 1.72  )\times 10^{12}$ & $(2.27 \pm 0.69) \times 10^{-10}$ \\
&&$8$& $(57.54 \pm 0.10  )\times 10^{11}$&$(232.02 \pm 0.40) \times 10^{-12}$\\
\hline
\multirow{4}{*}{HC$_{7}$N}&\multirow{2}{*}{C2} & $14.15 \pm 4.97$ &$(1.08 \pm 0.78) \times 10^{11}$ &  $(5.16 \pm 3.73) \times 10^{-12}$\\
&&$8$&$(2.81 \pm 0.36) \times 10^{11}$& $(1.34 \pm 0.17) \times 10^{-11}$\\
&\multirow{2}{*}{C16} & $8.51 \pm 0.50 $&   $(18.99 \pm 0.41) \times 10^{11}$ & $(7.66 \pm 0.17) \times 10^{-11}$ \\
&&$8$&$(2.25 \pm 0.11) \times 10^{12}$& $(9.07 \pm 0.44) \times 10^{-11}$ \\
\hline
\multirow{2}{*}{c-C$_{3}$H$_{2}$}&C2&$8 $ & $(6.87 \pm 0.16 ) \times 10^{12}$ & $(32.87 \pm 0.77) \times 10^{-11}$ \\
&C16&8&$(18.00 \pm 0.27) \times 10^{12}$ & $(7.26 \pm 0.11) \times 10^{-10}$ \\
\hline
\multirow{2}{*}{l-C$_{3}$H}&C2& 8 & $(8.31 \pm 0.19) \times 10^{11}$& $(397.61 \pm 9.09)\times 10^{-13}$ \\
&C16&8& $(9.74 \pm 0.56) \times 10^{11}$ & $(3.93 \pm 0.23) \times 10^{-11}$ \\
\hline
\multirow{2}{*}{HOCO$^{+}$}&C2&8& $(1.21 \pm 0.12) \times 10^{11}$ & $(57.89 \pm 5.74) \times 10^{-13}$ \\
&C16&8 & $(1.64 \pm 0.36) \times 10^{11}$ & $(6.61 \pm 1.45) \times 10^{-12}$\\
\hline
\multirow{2}{*}{C$_{3}$O}&C2&8&$(7.77 \pm 0.58) \times 10^{10}$ & $(3.71 \pm 0.28)  \times 10^{-12}$\\
&C16&8& $(2.65 \pm 0.52) \times 10^{11}$ & $(1.07 \pm 0.21) \times 10^{-11}$\\
\hline
\multirow{2}{*}{CH$_{2}$CHCN}&C2&8& $(5.00 \pm 1.65) \times 10^{10}$ & $(23.92 \pm 7.89) \times 10^{-13}$\\
& C16&8&$(5.15 \pm 0.57) \times 10^{11}$ & $(2.08 \pm 0.23) \times 10^{-11}$ \\
\hline
\multirow{2}{*}{l-C$_{3}$H$_{2}$}&C2&8&$(31.15 \pm 0.11)\times 10^{10}$ & $(148.804 \pm 0.53)\times 10^{-13}$ \\
&C16&8& $(5.53 \pm 0.41) \times 10^ {11}$ & $(2.23 \pm 0.17) \times 10^{-11}$ \\
\hline
\multirow{2}{*}{$^{13}$CS}&C2&8& $(2.49 \pm 0.48) \times 10^{11}$ & $(1.19 \pm 0.23) \times 10^{-11}$\\
& C16& 8& $(1.17 \pm 0.27) \times 10^{12}$ & $(4.72 \pm 1.09) \times 10^{-11}$\\
\hline
\multirow{2}{*}{C$^{34}$S}&C2&8& $(11.06 \pm 0.45)\times 10^{11}$  & $(53.11 \pm 2.15) \times 10^{-12}$ \\
&C16&8& $(32.77 \pm 0.65) \times 10^{11}$ & $(13.23 \pm 0.26)\times 10^{-11}$\\
\hline
\multirow{2}{*}{C$^{33}$S}&C2&8&$\leq 7.54 \times 10^{10}$ & $\leq 3.61 \times 10^{-12}$\\
&C16&8& $(5.30 \pm 1.10) \times 10^{11}$& $(2.13 \pm 0.44) \times 10^{-11}$\\
\hline
\multirow{3}{*}{CC$^{34}$S}& C2&8 & $(6.36 \pm 1.03) \times 10^{10}$& $(3.04 \pm 0.49) \times 10^{-12}$ \\
&\multirow{2}{*}{C16}&$4.68 \pm 1.30$&$(3.14 \pm 0.64)\times 10^{11}$ & $(1.27 \pm 0.26) \times 10^{-11}$ \\
&  &$8$& $(4.74 \pm 0.37)\times 10^{11}$& $(2.19 \pm 0.17) \times 10^{-11}$\\
\hline
\multirow{2}{*}{CCC$^{34}$S}&C2&8& $\leq 1.21 \times 10^{-10}$ & $\leq 5.79 \times 10^{-13}$ \\
&C16&8& $(6.39 \pm 1.49) \times 10^{10}$ & $(2.58 \pm 0.60) \times 10^{-12}$  \\
\hline

\end{longtable}

\end{document}